\documentclass[twoside, onecolumn,11pt]{article}
\usepackage{extsizes}
\usepackage[super,sort&compress,comma]{natbib} 
\usepackage[version=3]{mhchem}
\usepackage[left=1.5cm, right=1.5cm, top=1.785cm, bottom=2.0cm]{geometry}
\usepackage{balance}
\usepackage{times,mathptmx}
\usepackage{sectsty}
\usepackage{graphicx} 
\usepackage{lastpage}
\usepackage[format=plain,justification=justified,singlelinecheck=true,font={stretch=1.125,small,sf},labelfont=bf,labelsep=space]{caption}
\usepackage{float}
\usepackage{fancyhdr}
\usepackage{fnpos}
\usepackage[english]{babel}
\addto{\captionsenglish}{%
  
}
\usepackage{array}
\usepackage{droidsans}
\usepackage{charter}
\usepackage[T1]{fontenc}
\usepackage[usenames,dvipsnames]{xcolor}
\usepackage{setspace}
\usepackage[compact]{titlesec}
\usepackage{hyperref}

\usepackage{epstopdf}

\definecolor{cream}{RGB}{222,217,201}
\newcommand{\red}[1]{{\color{black} #1}}
\newcommand{\blue}[1]{{\color{black} #1}}

\usepackage{titlesec}
\titlelabel{\thetitle.\quad}

\begin{document}


\makeFNbottom
\makeatletter
\renewcommand\LARGE{\@setfontsize\LARGE{15pt}{17}}
\renewcommand\Large{\@setfontsize\Large{12pt}{14}}
\renewcommand\large{\@setfontsize\large{10pt}{12}}
\renewcommand\footnotesize{\@setfontsize\footnotesize{7pt}{10}}
\makeatother

\renewcommand{\thefootnote}{\fnsymbol{footnote}}
\renewcommand\footnoterule{\vspace*{1pt}%
\color{cream}\hrule width 3.5in height 0.4pt \color{black}\vspace*{5pt}} 
\setcounter{secnumdepth}{5}

\makeatletter 
\renewcommand\@biblabel[1]{#1}            
\renewcommand\@makefntext[1]%
{\noindent\makebox[0pt][r]{\@thefnmark\,}#1}
\makeatother 
\renewcommand{\figurename}{\small{Figure}}
\sectionfont{\sffamily\Large}
\subsectionfont{\normalsize}
\subsubsectionfont{\bf}
\setstretch{1.125} 
\setlength{\skip\footins}{0.8cm}
\setlength{\footnotesep}{0.25cm}
\setlength{\jot}{10pt}
\titlespacing*{\section}{0pt}{4pt}{4pt}
\titlespacing*{\subsection}{0pt}{15pt}{1pt}

\makeatletter 
\newlength{\figrulesep} 
\setlength{\figrulesep}{0.5\textfloatsep} 

\newcommand{\topfigrule}{\vspace*{-1pt}%
\noindent{\color{cream}\rule[-\figrulesep]{\columnwidth}{1.5pt}} }

\newcommand{\botfigrule}{\vspace*{-2pt}%
\noindent{\color{cream}\rule[\figrulesep]{\columnwidth}{1.5pt}} }

\newcommand{\dblfigrule}{\vspace*{-1pt}%
\noindent{\color{cream}\rule[-\figrulesep]{\textwidth}{1.5pt}} }

\makeatother

\begin{center}
\noindent\huge\textbf{Honeycomb Layered Oxides}\\
\noindent\LARGE\centering{Structure, Energy Storage, Transport, Topology and Relevant Insights}
\end{center}

\noindent\large{Godwill Mbiti Kanyolo,\textit{$^{a}$} Titus Masese,\textit{$^{b, c}$} Nami Matsubara,\textit{$^{d}$} Chih-Yao Chen,\textit{$^{b}$} Josef Rizell,\textit{$^{e}$} Ola Kenji Forslund,\textit{$^{d}$} Elisabetta Nocerino,\textit{$^{d}$}  Konstantinos Papadopoulos,\textit{$^{e}$} Anton Zubayer,\textit{$^{d}$} Minami Kato,\textit{$^{c}$} Kohei Tada,\textit{$^{c}$} Keigo Kubota,\textit{$^{b, c}$} Hiroshi Senoh,\textit{$^{c}$} Zhen-Dong Huang,\textit{$^{f}$},  Yasmine Sassa,\textit{$^{e}$} Martin M{\aa}nsson \textit{$^{d}$} and Hajime Matsumoto \textit{$^{c}$}}\\

\noindent{\textit{$^{a}$Department of Engineering Science, The University of Electro-Communications, 1-5-1 Chofugaoka, Chofu, Tokyo 182-8585, Japan. }\\
\textit{$^{b}$AIST-Kyoto University Chemical Energy Materials Open Innovation Laboratory (ChEM-OIL), Sakyo-ku, Kyoto 606-8501, Japan.}\\
\textit{$^{c}$Research Institute of Electrochemical Energy, National Institute of Advanced Industrial Science and Technology (AIST), 1-8-31 Midorigaoka, Ikeda, Osaka 563-8577, Japan.} Email: titus.masese@aist.go.jp\\
\textit{$^{d}$Department of Applied Physics, Sustainable Materials Research \& Technologies (SMaRT), KTH Royal Institute of Technology, SE-10691 Stockholm, Sweden.}\\
\textit{$^{e}$Department of Physics, Chalmers University of Technology, SE-412 96 G\"{o}teborg, Sweden.}\\
\textit{$^{f}$Key Laboratory for Organic Electronics and Information Displays and Institute of Advanced Materials (IAM), Nanjing University of Posts and Telecommunications (NUPT), Nanjing, 210023, China.}}\\

\noindent\normalsize{
\textbf{The advent of nanotechnology has hurtled the discovery and development of nanostructured materials with stellar chemical and physical functionalities in a bid to address issues in energy, environment, telecommunications and healthcare. In this quest, a class of two-dimensional layered materials consisting of alkali or coinage metal atoms sandwiched between slabs exclusively made of transition metal and chalcogen (or pnictogen) atoms arranged in a honeycomb fashion have emerged as materials exhibiting fascinatingly rich crystal chemistry, high-voltage electrochemistry, fast cation diffusion \blue{besides playing} host to varied exotic electromagnetic and topological phenomena. Currently, with a niche application in energy storage as high-voltage materials, this class of honeycomb layered oxides serves as ideal pedagogical exemplars of the innumerable capabilities of nanomaterials drawing immense interest in multiple fields ranging from materials science, solid-state chemistry, electrochemistry and condensed matter physics. In this review, we delineate the relevant chemistry and physics of honeycomb layered oxides, and discuss their functionalities for tunable electrochemistry, superfast ionic conduction, electromagnetism and topology. Moreover, we elucidate the unexplored albeit vastly promising crystal chemistry space whilst outlining effective ways to identify regions within this compositional space, particularly where interesting electromagnetic and topological properties could be lurking within the aforementioned alkali and coinage-metal honeycomb layered oxide structures. We conclude by pointing towards possible future research directions, particularly the prospective realisation of Kitaev-Heisenberg-Dzyaloshinskii-Moriya interactions with single crystals and Floquet theory in closely-related honeycomb layered oxide materials.}} \\




\renewcommand*\rmdefault{bch}\normalfont\upshape
\rmfamily
\section*{}
\vspace{-1cm}


\newpage

\section{Introduction}

\begin{figure*}[!htb]
 \centering
 \includegraphics[height=15cm]{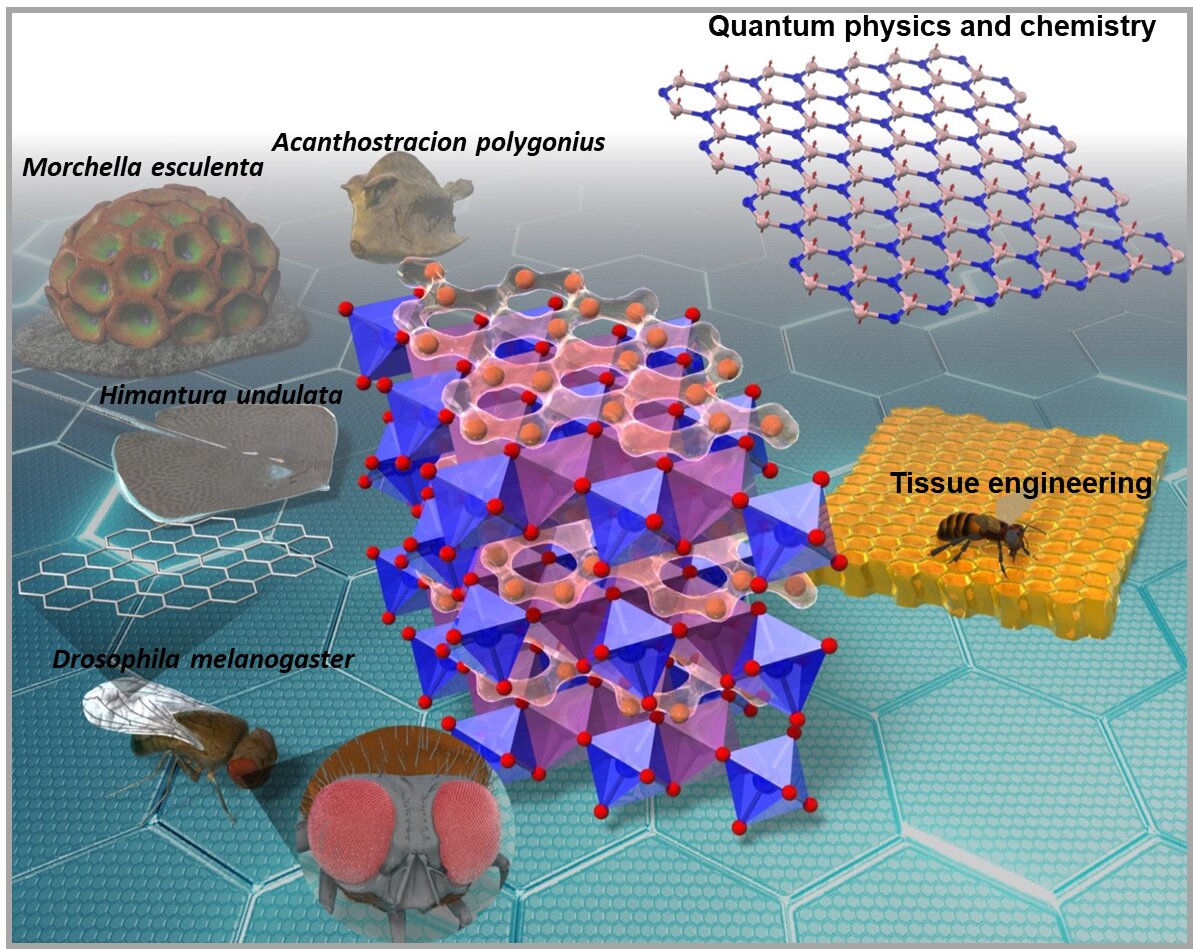}
 \caption{
 Schematic illustration of the various realisations of the honeycomb structure found not only in energy storage materials, but also as pedagogical models in condensed-matter physics, solid-state chemistry and extending to tissue engineering.\cite{ Zhang2015} Specific varieties of fungi ({\it videlicet}, {\it Morchella esculenta}) tend to adopt honeycomb-like structures,\cite{ garcia2006} whilst insects such as the fruit flies ({\it Drosophila melanogaster}) have their wing cells in honeycomb configuration; \cite{ sugimura2013mech, ikawa2018a, de2017} thus endowing them with excellent rigidity. In addition, the honeycomb whip ray ({\it Himantura undulata}) \cite{ yucel2017} and the honeycomb cowfish ({\it Acanthostracion polygonius})\cite{ garcia2018b} have honeycomb patterns on their body that are thought to aid in their facile movement and camouflage.}
 \label{Fig_1}
\end{figure*}

Charles Darwin famously described the honeycomb as an engineering masterpiece that is ``absolutely perfect in economising labour and wax''.\cite{Darwin1859} For over two millennia, scientists and philosophers alike have found a great deal of fascination in the honeycomb structures found in honeybee hives. These hexagonal prismatic wax cells built by honey bees to nest their larvae, store honey and preserve pollen are revered as a feat in precision engineering and admired for their elegance in geometry.\cite{Zhang2015} The honeycomb framework offers a rich tapestry of qualities adopted in myriads of fields such as mechanical engineering, architectural design biomedical engineering {\it et cetera} (as briefly outlined in \textbf{Fig. \ref{Fig_1}}).\cite{Zhang2015, Karihaloo2013}

\begin{figure*}[!htb]
\centering
  \includegraphics[height=14cm]{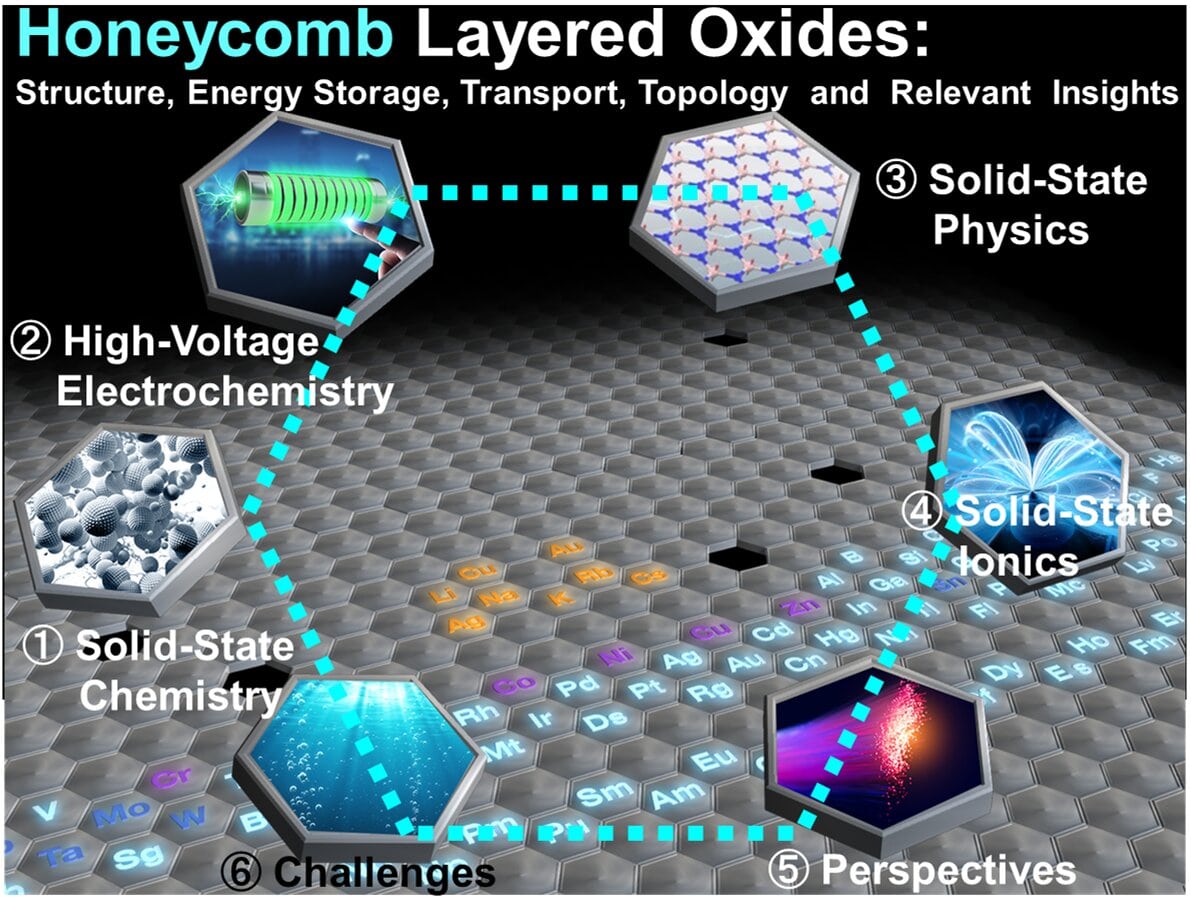}
  \caption{
  Illustration of the various sections to be covered in this review. This starts from solid-state chemistry, physics, electrochemistry to solid-state ionics. We finally adumbrate on the challenges and perspectives of these honeycomb layered oxides.}
  \label{Fig_2}
\end{figure*}

The discovery and isolation of graphene in 2004, not only revolutionised our understanding of two-dimensional materials but also unveiled a new platform for fabricating novel materials with customised functionalities.\cite{Novoselov2004, Kitagawa2018} Two-dimensional (2D) nanostructures are crystalline systems comprising covalently bonded atom cells in planar arrangement of mesoscopic thicknesses. Due to their small size, this class of materials exhibits highly controlled and unique optical, magnetic, or catalytic properties.\cite{Kadari2016, Pu2018, Roudebush2015, Kumar2012, Zvereva2013, Derakhshan2007, Viciu2007, Morimoto2007, Derakhshan2008, Koo2008, Zvereva2012, Schmitt2014, Sankar2014, Xiao2019, Zvereva2015a, Itoh2015, Zvereva2015b, Bera2017, Upadhyay2016, Koo2016, Zvereva2016, Kurbakov2017, Karna2017, Zvereva2017, Werner2019, Stavropoulos2019, Korshunov2020, Yao2020, Li2019a, Miura2006, Schmitt2006, Miura2007, Miura2008, Li2010, Kuo2012, Roudebush2013a, Zhang2014, Jeevanesan2014, Lefrancois2016, Wong2016, Werner2017, Scheie2019, Mao2020} By substituting the constituent atoms and manipulating the atomic cell configurations, materials with remarkable physicochemical properties such as high electron mobility, unique optical and chemical functionality can be tailor-made for various technological realms such as catalysts, superconductivity, sensor applications and energy storage. Assembling such materials into vertical stacks of layered combinations allows the insertion of atoms within the stacks creating a new class of multi-layered heterostructures. A unique characteristic of these structures is that inter-layer bonds holding together the thin 2D films are significantly weaker (Van der Waals bonds) than the covalent bonds within the 2D monolayers, which further augments the possibility for emergent properties exclusive to these materials.

In electrochemistry, layered frameworks composed of alkali or coinage metal atoms interposed between 2D sheets of hexagonal (honeycomb) transition metal and chalcogen (or pnictogen) oxide octahedra have found great utility as next-generation cathode materials for capacious rechargeable battery systems.\cite{Sathiya2013, Grundish2019, Yang2017, Yuan2014, Masese2018} The weak interlayer bonds between transition metal slabs facilitate facile mobility of intercalated atoms during the battery operation (de)insertion processes, endowing these heterostructures with ultrafast ionic diffusion rendering them exemplar high energy (and power) density cathode materials. Furthermore, the unique topological changes occurring during these electrochemical processes have been seen to induce new domains of physics entailing enigmatic optical, electromagnetic and quantum properties that promise to open new paradigms of computational techniques and theories quintessential in the field of quantum material science catapulting the discovery of materials with novel functionalities.\cite{Kadari2016, Pu2018, Roudebush2015, Kumar2012, Zvereva2013, Derakhshan2007, Viciu2007, Morimoto2007, Derakhshan2008, Koo2008, Zvereva2012, Schmitt2014, Sankar2014, Xiao2019, Zvereva2015a, Itoh2015, Zvereva2015b, Bera2017, Upadhyay2016, Koo2016, Zvereva2016, Kurbakov2017, Karna2017, Zvereva2017, Werner2019, Stavropoulos2019, Korshunov2020, Yao2020, Li2019a, Miura2006, Schmitt2006, Miura2007, Miura2008, Li2010, Kuo2012, Roudebush2013a, Zhang2014, Jeevanesan2014, Lefrancois2016, Wong2016, Werner2017, Scheie2019}This has created an entirely new platform of study, encompassing fields, {\it {\it inter alia}}, materials science, solid-state chemistry, electrochemistry and condensed matter physics.\cite{Kurbakov2020, motome2020, Nalbandyan2013, Zvereva2017, Bhardwaj2014, Yao2020, Kumar2013, Morimoto2007, Gupta2013, Berthelot2012, Laha2013, McCalla2015, Kumar2012, Taylor2019, Roudebush2013b, Berthelot2012a, Uma2016, He2017, He2018, Stratan2019, Xu2005, Ramlau2014, Bera2017, Smirnova2005, Schmidt2013, Seibel2013, Seibel2014, Liu2016, Bhange2017, Gyabeng2017, Yan2019, Yadav2019, Smaha2015, Brown2019, Heymann2017, Nalbandyan2013a, Schmidt2014}

Although layered oxides encompass a broad class of materials with diverse structural frameworks and varied emergent properties, in this review, we focus on the stellar properties innate in the aforementioned class of honeycomb layered oxides comprising alkali or coinage metal atoms sandwiched between slabs consisting of transition metal oxide octahedra surrounding chalcogen or pnictogen oxide octahedra in a honeycomb configuration. To provide deeper insights, we delineate the fundamental chemistry underlying their material design along with emergent domains of physics. We further highlight their functionalities for tunable electrochemistry, superfast ionic conduction, electromagnetism and topology, as illustrated in \textbf{Fig. \ref{Fig_2}}. Moreover, we highlight the unexplored albeit vastly promising crystal chemistry space whilst outlining effective ways to identify regions within this compositional space, particularly where potentially interesting electromagnetic and topological properties could be lurking within the aforementioned honeycomb layered oxides. The looming challenges are also discussed with respect to the governing chemistries surrounding honeycomb layered oxides. Finally, we conclude by pointing towards possible future research directions, particularly the prospective realisation of Kitaev-Heisenberg-Dzyaloshinskii-Moriya interactions in closely-related honeycomb layered oxide materials, and their connection to Floquet theory and fabrication efforts that are not limited to single crystals.

\newpage

\section{Materials chemistry of honeycomb layered oxides}

\subsection{Chemical composition}

\begin{figure*}[!b]
\centering
  \includegraphics[height=19cm]{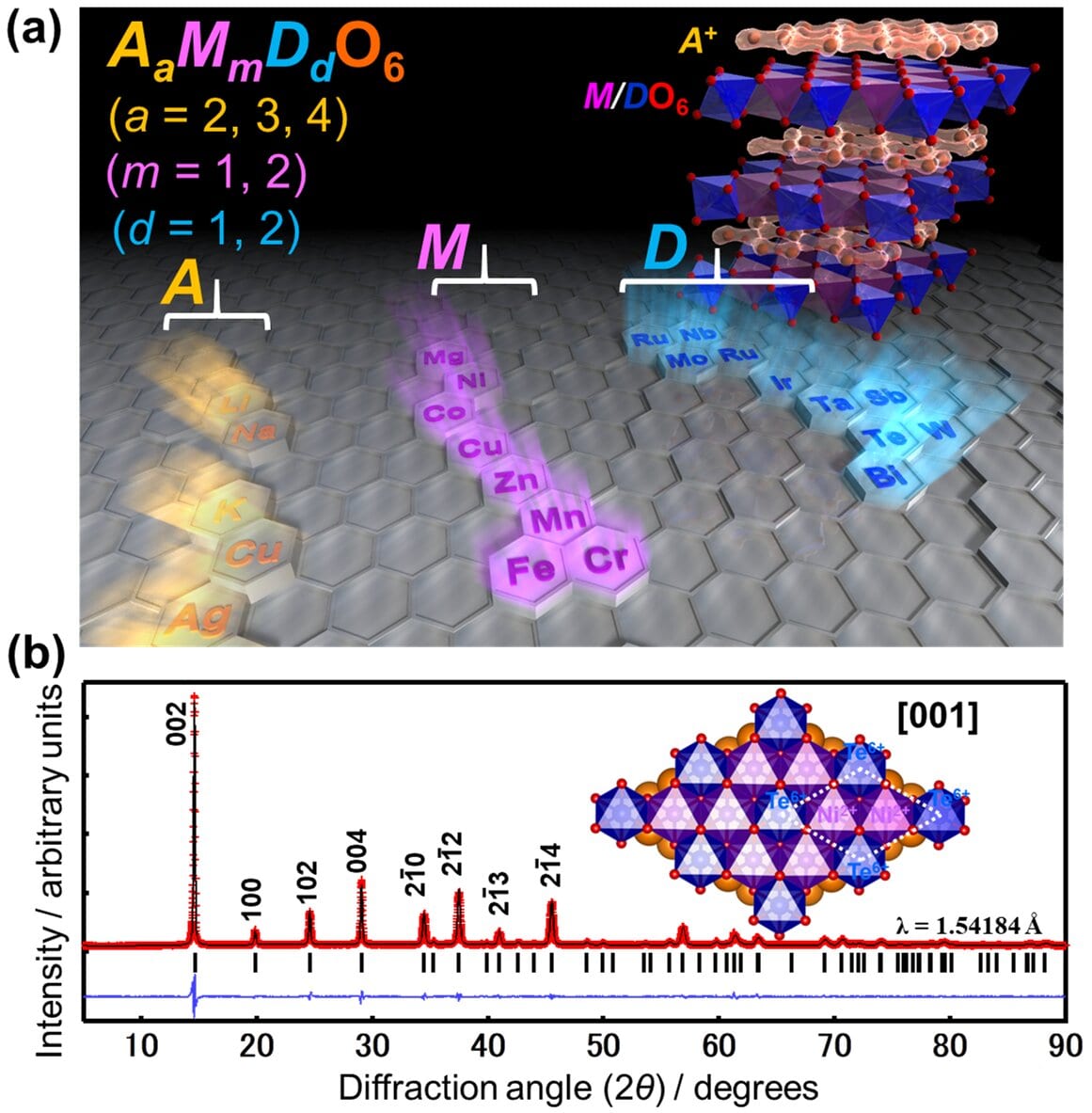}
  \caption{Combination of elements that constitute materials exhibiting the honeycomb layered structure. (a) Choice of elements for layered oxide compositions (such as $\rm {\it A}^{+}_{2}{\it M}^{2+}_{2}{\it D}^{6+}O_{6}$ ($\rm {\it A}^{+}_{2/3}{\it M}^{2+}_{2/3}{\it D}^{6+}_{1/3}O_{2}$), $\rm {\it A}^{+}_{3}{\it M}^{2+}_{2}{\it D}^{5+}O_{6}$ ($\rm {\it A}^{+}{\it M}^{2+}_{2/3}{\it D}^{5+}_{1/3}O_{2}$), {\it et cetera}.) that can adopt honeycomb configuration of transition metal atoms.  Inset shows a polyhedral view of the crystal structure of layered honeycomb oxides, with the alkali atoms (shown as brown spheres) sandwiched between honeycomb slabs (blue). (b) X-ray diffraction (XRD) pattern of $\rm K_2Ni_2TeO_6$ (12.5\% cobalt-doped) honeycomb layered oxide.\cite {Yoshii2019} Inset: Slab of layered oxide showing the honeycomb arrangement of magnetic nickel (Ni) atoms around non-magnetic tellurium (Te) atoms. Dashed line highlights the unit cell. (b) adapted from ref. \citenum{Yoshii2019} 
  with permission \red{(Creative Commons licence 4.0)}.}
  \label{Fig_3_0}
\end{figure*}

\begin{figure*}[!b]
\centering
  \includegraphics[height=12.5cm]{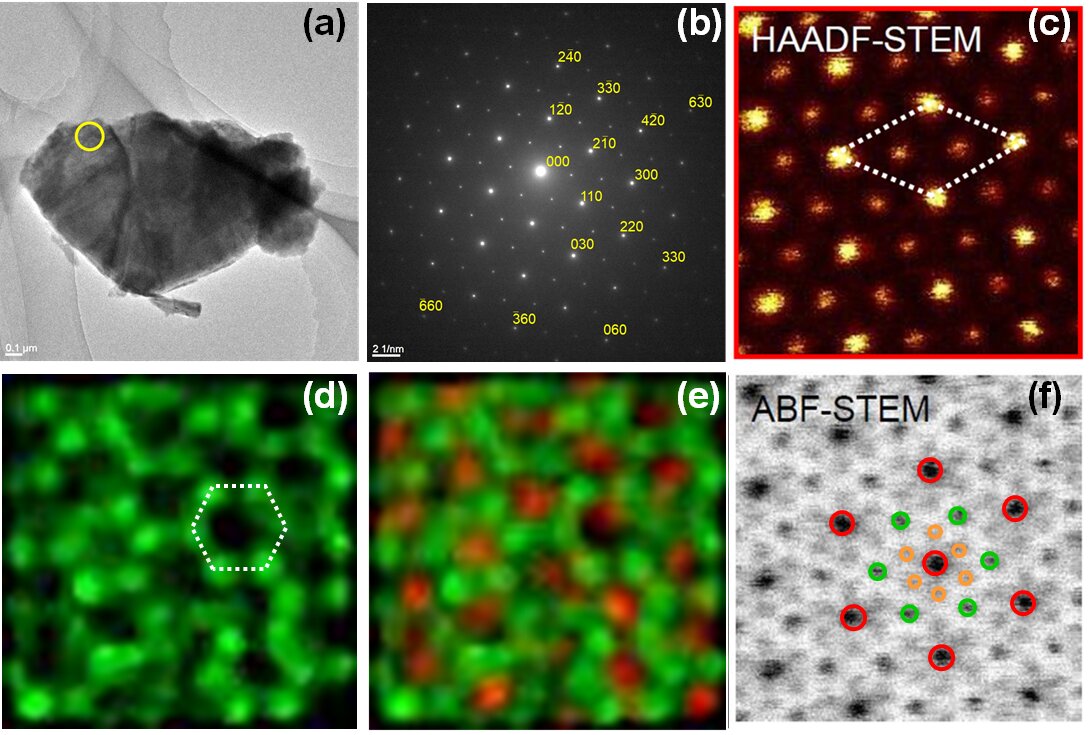}
  \caption{(a) High-resolution transmission electron microscopy (TEM) image of a crystallite of $\rm K_2Ni_2TeO_6$ (12.5\% cobalt-doped) honeycomb layered oxide and (b) Corresponding electron diffractograms taken along the [001] zone axis.\cite {Yoshii2019} (c) Visualisation (along the $c$-axis [001]) of the honeycomb configuration of Ni atoms around Te atoms (in brighter contrast) using High-Angle Annular Dark\red{-}Field Scanning TEM (HAADF-STEM). Dashed lines indicate the unit cell. (d) STEM imaging with $\rm Ni$ atoms (partially with $\rm Co$) (in green) assuming a honeycomb fashion (as highlighted in dashed lines) and (e) STEM imaging showing Te atoms (in red) surrounded by transition metal atoms. (f) Annular Bright\red{-}Field TEM (ABF-TEM) of segments manifesting potassium atoms (in brown) assuming a honeycomb fashion and overlapped with oxygen atoms. Note that some portions of the honeycomb ordering of transition metal atoms slightly appear obfuscated, owing to sensitivity of the samples to long-time beam exposure.(a, b) \red{reproduced and adapted from ref. \citenum{Yoshii2019} under Creative Commons licence 4.0.}}
  \label{Fig_3_2}
\end{figure*}

\begin{figure*}[!t]
\centering
  \includegraphics[width=\textwidth]{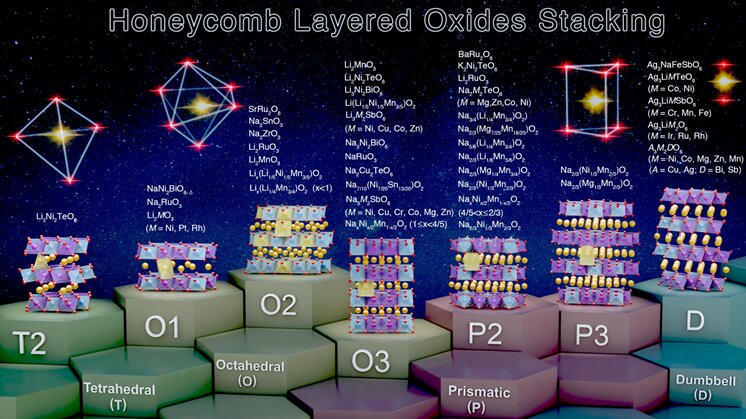}
  \caption{Summary of the various stacking sequences adopted by representative honeycomb layered oxides \red{including those that entail pnictogen or chalcogen atoms in the slab layers}. Note here that \red{`}T\red{'}, \red{`}O\red{'} and \red{`}P\red{'} denote the coordination of the alkali or coinage \red{metal} atoms (sandwiched between the honeycomb slabs) with the adjacent oxygen atoms of the honeycomb slab, {\it id est}, tetrahedral, octahedral and prismatic coordination, respectively. The numbers or digits (\red{`}1\red{'}, \red{`}2\red{'} and \red{`}3\red{'}) indicate the repetitive alkali-atom layers per unit cell, as denoted in Hagenmuller\red{-Delmas'} notation.\cite{Delmas1976} \red{Note that the predominant stacking sequences have been shown, for ease of explanation. A more elaborate list of the stacking sequences for other honeycomb layered oxides reported thus far are provided in Table 
  \red{\ref{Table_2}}.\cite {navaratnarajah2019, zheng2018, koch2015, tamaru2013, gazizova2018, mogare2004, lu2019, bastow1994, wang2013, robertson2003, boulineau2009, zuo2018, strobel1988, ma2014, shinova2005, okada1999, asakura1999, omalley2008, hermann2019, luo2013, Todorova2011b, kimber2014, yu2019, jang2020, james1988, house2020, eum2020, maitra2018, yabuuchi2014, cabana2013, song2019, Zvereva2013, Kurbakov2020, motome2020, Nalbandyan2013, Zvereva2017, Bhardwaj2014, Yao2020, Kumar2013, Morimoto2007, Gupta2013, Berthelot2012, Laha2013, McCalla2015, Kumar2012, Taylor2019, Roudebush2013b, Berthelot2012a, Uma2016, He2017, He2018, Stratan2019, Xu2005, Ramlau2014, Bera2017, Smirnova2005, Schmidt2013, Seibel2013, Seibel2014, Liu2016, Bhange2017, Gyabeng2017, Yan2019, Yadav2019, Smaha2015, Brown2019,Szillat1995, Greaves1990, Skakle1997, Nagarajan2002, Gupta2015, zvereva2016d, Mather2000, Mather1995, nguyen2020, wang2018a, marchandier2020}}}
  \label{Fig_3_3}
\end{figure*}

The aforementioned honeycomb layered oxides generally adopt the following chemical compositions, taking into account that charge electro-neutrality is maintained and assuming ordered structures: $A^{+}_{2}M^{2+}_{2}D^{6+}\rm O^{2-}_{6}$ (or equivalently as $\rm {\it A}^{+}_{2/3}{\it M}^{2+}_{2/3}{\it D}^{6+}_{1/3}O^{2-}_{2}$), $\rm {\it A}^{+}_{3}{\it M}^{2+}_{2}{\it D}^{5+}O^{2-}_{6}$ ($\rm {\it A}^{+}{\it M}^{2+}_{2/3}{\it D}^{5+}_{1/3}O^{2-}_{2}$), $\rm {\it A}^{+}_{4.5}{\it M}^{3+}_{0.5}{\it D}^{6+}O^{2-}_{6}$ ($\rm {\it A}^{+}_{3/2}{\it M}^{3+}_{1/6}{\it D}^{6+}_{1/3}O^{2-}_{2}$), $\rm {\it A}^{+}_{4}{\it M}^{3+}{\it D}^{5+}O^{2-}_{6}$ ($\rm {\it A}^{+}_{4/3}{\it M}^{3+}_{1/3}{\it D}^{5+}_{1/3}O^{2-}_{2}$), $\rm {\it A}^{+}_{2}{\it D}^{4+}O^{2-}_{3}$
($\rm {\it A}^{+}_{4/3}{\it D}^{4+}_{2/3}O^{2-}_{2}$), $\rm {\it A}^{+}_{3}{\it A'}^{+}{\it M}^{3+}{\it D}^{5+}O^{2-}_{6}$ ($\rm {\it A}^{+}{\it A'}^{+}_{1/3}{\it M}^{3+}_{1/3}{\it D}^{5+}_{1/3}O^{2-}_{2}$), $\rm {\it A}^{+}_{4}{\it M}^{2+}$\\
$\rm {\it D}^{6+}O^{2-}_{6}$ ($\rm {\it A}^{+}_{4/3}{\it M}^{2+}_{1/3}{\it D}^{6+}_{1/3}O^{2-}_{2}$ or $\rm {\it A}^{+}_{8}{\it M}^{2+}_{2}{\it D}^{6+}_{2}O^{2-}_{12}$), amongst others (\textbf{Fig. \ref{Fig_3_0}a}).\cite{Kumar2012, Zvereva2013, Kurbakov2020, motome2020, Nalbandyan2013, Zvereva2017, Bhardwaj2014, Yao2020, Kumar2013, Morimoto2007, Gupta2013, Berthelot2012, Laha2013, McCalla2015, Taylor2019, Roudebush2013b, Berthelot2012a, Uma2016, He2017, He2018, Stratan2019, Xu2005, Ramlau2014, Bera2017, Smirnova2005, Schmidt2013, Seibel2013, Seibel2014, Liu2016, Bhange2017, Gyabeng2017, Yan2019, Yadav2019, Smaha2015, Brown2019,Szillat1995, Greaves1990, Skakle1997, Nagarajan2002, Gupta2015, zvereva2016d, Mather2000, Mather1995, nguyen2020, navaratnarajah2019, zheng2018, koch2015, tamaru2013, gazizova2018, mogare2004, lu2019, bastow1994, wang2013, robertson2003, boulineau2009, zuo2018, strobel1988, ma2014, shinova2005, okada1999, asakura1999, omalley2008, hermann2019, luo2013, Todorova2011b, kimber2014, yu2019, jang2020, james1988} Here $M$ denotes transition metal atoms such as $\rm Ni, Co, Mn,$ $\rm Fe, Cu, Zn, Cr$ (including $\rm Mg$); $\rm {\it D}$ denotes $\rm Te, Sb, Bi, Nb, Ta, W,$ $\rm Ru, Ir, Os$; $\rm {\it A}$ and $A'$ denote alkali atoms such as $\rm Li, Na, K$ or coinage atoms like $\rm Cu$ and $\rm Ag$ (with $A \neq A'$). \blue{It is worth mentioning that the honeycomb layered framework is not limited to compositions entailing only one species of alkali atoms. Oxides compositions comprising mixed-alkali atoms such as $\rm Na_3LiFeSbO_6$, $\rm Na_2LiFeTeO_6$, $\rm Ag_3LiRu_2O_6$, $\rm Ag_3NaFeSbO_6$, $\rm Ag_3LiIr_2O_6$, $\rm Ag_3Li{\it M}TeO_6$ ($\rm {\it M}  =  Co, Ni$), $\rm Ag_3Li{\it M}SbO_6$ ($M  = \rm  Cr, Mn, Fe$) and, more recently, $\rm Li_{3-{\it x}}Na_{\it x}Ni_2SbO_6$ as well as  oxides with off-stoichiometric compositions such as $\rm Li_3Co_{1.06}TeO_6$ have been explored} 
\red{with the aim of merging favourable attributes from multiple species to improve various material functionalities, for instance, battery performance.}\blue{\cite{Heymann2017, Nalbandyan2013a, Schmidt2014, Vallee2019, Bette2019, Kimber2010, Brown2020} Other atypical honeycomb layered oxide compositions are those encompassing alkali or mixed-alkali species atoms embedded within the transition metal slabs. Compositions such as ${\rm Li}_x{\rm {\it M}}_{y}{\rm Mn}_{1-y}\rm O_2$ ($0 < x < 1$; $0 < y < 1$; $M  = \rm  Li, LiNi$), $\rm Na_{3/4}(Li_{1/4}Mn_{3/4})O_2$, $\rm Na_{5/6}(Li_{1/4}Mn_{3/4})O_2$, $\rm Na_{2/3}(Li_{1/6}Mn_{5/6})O_2$, \textit{et cetera} have also been considered for use as \red{high-performance battery materials},\cite{house2020, eum2020, maitra2018, wang2018a, yabuuchi2014, cabana2013, song2019} \red{and can be envisaged to exhibit unique three-dimensional diffusion as they have some of their alkali atoms reside in the honeycomb layers.} Although three-dimensional diffusion is covered in a later section, it is pertinent to note that this review will mainly focus on two-dimensional diffusion of \red{cations} in relation to topotactic curvature evolutions. Other classes of honeycomb layered oxide frameworks with vastly different properties and applications are also beyond the scope of this review.}\blue{\cite{doudin2017, goniakowski2020, pomp2016, clark2019, sahoo2020, rani2015, asai2017, streltsov2018}} 

\subsection{Preparative methods}

High-temperature solid-state synthesis is often considered an expedient route to synthesise most of the above mentioned honeycomb layered oxides because their initial precursor materials usually require high temperatures to activate the diffusion of individual atoms.\cite{Uppuluri2018} In this technique, precursors are mixed in stoichiometric amounts and pelletised to increase the contact surface area of these reactants. Finally, they are fired at high temperatures (over 700$^{\circ}$ C) resulting in thermodynamically-stable honeycomb layered structures. The firing environment (argon, nitrogen, air, oxygen, carbon monoxide, \red{hydrogen}, {\it et cetera}) needs to be adequately controlled to obtain materials with the desired oxidation states of transition metals. For example, an inert firing environment is demanded for layered oxides that contain $\rm Mn^{2+}$ and $\rm Fe^{2+}$; otherwise oxidised samples containing $\rm Mn^{3+}$ and $\rm Fe^{3+}$ are essentially formed. \blue{Nonetheless, not all high\red{-}temperature synthesis processes are as restrictive, as compositions containing $\rm Ni^{2+}$, such as $\rm {\it A}_2Ni_2TeO_6$ ($A = \rm Li, Na, K$, {\it et cetera}.) can be synthesised under air  to obtain samples that contain $\rm Ni$ still in the divalent state. Moreover, varying the synthesis protocols (annealing temperature and time, types of precursors, \red{thermal ramp rate}, {\it et cetera}) during high-temperature synthesis not only aid in enhancing the scalability of the synthesis but also gives rise to the possibility of obtaining new polymorphs (or polytypes), as has been shown in the high-temperature synthesis of $\rm Na_3Ni_2BiO_6$, $\rm Na_4NiTeO_6$ and $\rm Na_3Ni_2SbO_6$. \cite {Liu2016, Ma2015, He2017, Yang2017}}

The topochemical ion-exchange synthesis route is also possible for honeycomb layered oxides with accessible kinetically-metastable phases.\cite{Bette2019, Hosogi2008, Martinolich2017, Todorova2011, heubner2020}  Despite the high binding strength amongst adjacent atoms within the honeycomb slab, the use of cations with higher charge-to-radius ratio such as $\rm Li^{+}$ in $\rm LiNO_3$ can drive out the $\rm Na^{+}$ atoms present in $\rm Na_2Cu_2TeO_6$ by lowering their electrostatic energy to create $\rm Li_2Cu_2TeO_6$.\cite{Kumar2013} Here, the two precursors are heated together at a moderate temperature (300$^{\circ}$ C) triggering the diffusion of $\rm Na^{+}$ and $\rm Li^{+}$. Other oxides that can be prepared via the ion-exchange route include $\rm Li_3Co_2SbO_6$, $\rm Ag_3{\it M}_2SbO_6$ ($\rm {\it M} = Ni, Co$ and $\rm Zn$), $\rm Li_2Ni_2TeO_6$, $\rm Ag_3LiGaSbO_6$, $\rm Ag_3LiAlSbO_6$, $\rm Ag_3Ni_2BiO_6$ \red{and $\rm Li_3{\it M}_2SbO_6$ ($\rm {\it M} = Fe$ and $\rm Mn$).}\cite{Berthelot2012, Zvereva2016, Bhardwaj2014, Grundish2019, Yadav2019} However, it is worthy to recapitulate that there exists exemplars of compounds that can be synthesised topochemically such as $\rm Ag_3Co_2SbO_6$ \red{and $\rm Ag_3Li{\it D}O_6$ ($\rm {\it D} = Ru, Ir$)} that are exceptions to the rule that ion exchange can only happen from ions with lower charge-to-radius ratio to those with higher ratios (taking into account $\rm Ag$ has a larger ionic radius than $\rm Li$\cite{shannon1970})\cite{Zvereva2016, Bette2019}. The syntheses of these honeycomb layered oxides are typically done at ambient pressures; however, high-pressure syntheses routes remain unexplored, a pursuit which may expand their material platforms. Equally important is the utilisation of low-temperature routes such as sol-gel and mechanochemical synthesis since they do not require {\it apriori} high-temperatures to attain thermodynamically-stable phases. \red{Although not delved in the scope of this review, honeycomb layered oxides \blue{entailing alkaline-metal atoms as the resident cations} such as $\rm SrRu_2O_6$ and $\rm BaRu_2O_6$ can be prepared via a low-temperature hydrothermal synthesis route. \cite {marchandier2019, wang2020, marchandier2020}} Finally, despite the scarce exploration of electrochemical ion-exchange \cite {bo2015, armstrong1996, capitaine1996, orikasa2014s, paulsen1999, robertson2001, gwon2014, baskar2017, naveen2018, hwang2018d} of honeycomb layered oxide materials, invaluable results on the $\rm K^{+}$/$\rm Na^{+}$ ion-exchange process in $\rm Na_3Ni_2SbO_6$ has been recently reported, demonstrating this process as a promising route to pursue.\cite{Kim2020}

\subsection{Crystallography}

\red{
\begin{table*}[!t]
\caption{Stacking sequences adopted by a smorgasbord of honeycomb layered oxides hitherto reported.\red{\cite {navaratnarajah2019, zheng2018, koch2015, tamaru2013, gazizova2018, mogare2004, lu2019, bastow1994, wang2013, robertson2003, boulineau2009, zuo2018, strobel1988, ma2014, shinova2005, okada1999, asakura1999, omalley2008, hermann2019, luo2013, Todorova2011b, kimber2014, yu2019, jang2020, james1988, house2020, eum2020, maitra2018, yabuuchi2014, cabana2013, song2019, Zvereva2013, Kurbakov2020, motome2020, Nalbandyan2013, Zvereva2017, Bhardwaj2014, Yao2020, Kumar2013, Morimoto2007, Gupta2013, Berthelot2012, Laha2013, McCalla2015, Kumar2012, Taylor2019, Roudebush2013b, Berthelot2012a, He2017, He2018, Stratan2019, Xu2005, Ramlau2014, Bera2017, Smirnova2005, Schmidt2013, Seibel2013, Seibel2014, Liu2016, Bhange2017, Gyabeng2017, Yan2019, Yadav2019, Smaha2015, Brown2019,Szillat1995, Greaves1990, Skakle1997, Nagarajan2002, Gupta2015, zvereva2016d, Mather2000, Mather1995, nguyen2020, wang2018a, Schmidt2014, Uma2016, Bhardwaj2014}}
}\label{Table_1}
\begin{center}
\scalebox{0.83}{
\begin{tabular}{ccl} 
\hline
\textbf{Coordination of alkali atoms} & \textbf{Hagenmuller-Delmas' notation} & \textbf{Honeycomb layered oxide compositions} \\
\textbf{with oxygen} & \textbf{(slab stacking sequence)} &  \\
\hline\hline
& & \\
Tetrahedral & T2 & $\rm Li_2Ni_2TeO_6$\\
& &\\
Octahedral & O1 & $\rm NaNi_2BiO_{6-\delta}$, $\rm Na_2RuO_3$, $\rm Li_2MO_3$ ($M = \rm Ni, Pt, Rh$), \red{$\rm BaRu_2O_6
$}\\
 & O2 & $\rm Li_2MnO_3$, ${\rm Li}_x\,(\rm Li_{1/5}Ni_{1/5}Mn_{3/5})O_2$ ($x < 1$), ${\rm Li}_x\,(\rm Li_{1/4}Mn_{3/4})O_2$,\\
 & & $\rm Na_2ZrO_3$, $\rm Na_2SnO_3$, $\rm Li_2RuO_3$, $\rm SrRu_2O_6$\\
 & O3 & $\rm Li_2Ni_2TeO_6$, $\rm Li_3Ni_2BiO_6$, $\rm Na_3Ni_2BiO_6$, $\rm Na_2Cu_2TeO_6$,\\
 & & ${\rm Na}_3M_2\rm SbO_6$ ($M = \rm Ni, Cu, Cr, Co, Mg, Zn$), \red{$\rm Li_2Cu_2TeO_6$}\\
 & & ${\rm Li}_3M_2\rm SbO_6$ ($M = \rm Ni, Cu, Co, Zn$), $\rm NaRuO_3$,\\
 & & $\rm Li_2MnO_3$, $\rm Li\,(Li_{1/5}Ni_{1/5}Mn_{3/5})O_2$, $\rm Na_{7/10}\,(Ni_{7/20}Sn_{13/20})O_2$,\\
 & & ${\rm Na}_x{\rm Ni}_{x/2}{\rm Mn}_{1-x/2}\rm O_2$ ($1 \red{\leq}$ $x < 4/5$)\red{, $\rm Na_3LiFeSbO_6$, $\rm Li_4MTeO_6$}\\ 
 & & \red{($M = \rm Co, Ni, Cu, Zn$), ${\rm Li_4}M \rm SbO_6$ ($M = \rm Cr, Fe, Al, Ga, Mn$)}\\
 & & \\
Prismatic & P2 & $\rm K_2Ni_2TeO_6$, $\rm Na_2M_2TeO_6$ ($M = \rm Mg, Zn, Co, Ni$)\\
 & & $\rm Na_{3/4}\,(Li_{1/4}Mn_{3/4})O_2$, $\rm Na_{2/3}\,(Mg_{7/25}Mn_{18/25})O_2$, $\rm BaRu_2O_6$,\\
 & & $\rm Na_{5/6}\,(Li_{1/4}Mn_{3/4})O_2$, $\rm Na_{2/3}\,(Li_{1/6}Mn_{5/6})O_2$,\\
 & & $\rm Na_{2/3}\,(Mg_{1/4}Mn_{3/4})O_2, Na_{2/3}\,(Ni_{1/3}Mn_{2/3})O_2$, $\rm Li_2RuO_3$,\\
 & & ${\rm Na}_x{\rm Ni}_{x/2}{\rm Mn}_{1-x/2}\rm O_2$ \red{($2/3 \leq x < 4/5$),} $\rm Na_{2/3}Ni_{1/3}Mn_{2/3}O_2$,\\
 & P3 & $\rm Na_{2/3}\,(Mg_{1/3}Mn_{2/3})O_2$, $\rm Na_{2/3}\,(Ni_{1/3}Mn_{2/3})O_2$\\ 
 & & \\
Dumbbell & D & $\rm Ag_3NaFeSbO_6$, ${\rm Ag_3Li}M\rm TeO_6$ ($M = \rm Co, Ni$),\\
(linear) & & ${\rm Ag_3Li}M\rm SbO_6$ ($M = \rm Cr, Mn, Fe$), ${\rm Ag_3Li}M_2\rm O_6$ ($M = \rm Ir, Ru, Rh$),\\
 & & $A_3M_2D\rm O_6$ ($M = {\rm Ni, Co, Mg, Zn, Mn}$; $A = \rm Cu, Ag$; $D = \rm Bi, Sb$)\\
 & & \\
\hline
\end{tabular}}
\end{center}
\end{table*}
}

To ascertain the crystal structure of honeycomb layered oxides and discern the precise location of the constituent atoms, transmission electron microscopy (TEM), neutron diffraction (ND) and X-ray diffraction (XRD) analyses can be performed on single-crystals or polycrystalline samples. Although the XRD is the most commonly used crystallography technique, it is ineffective in analysing oxides composed of lighter atoms such as $\rm Li, H,$ and $\rm B$ due to their low scattering intensity. Also, honeycomb layered oxides with elements of similar atomic number are difficult to distinguish using XRD because they diffract with similar intensity.

To distinguish light elements or elements with close atomic numbers on honeycomb layered oxides, the ND is used because the neutron beam interacts directly with the nucleus, hence the ability to observe light elements. In spite of the high accuracy, the equipment remain very expensive and ND experiments require the use of very large sample amounts to obtain high-resolution data- an impediment to materials that can only be prepared on a small scale. 

Although, XRD analys\red{e}s accurately validate the precise crystal structure of honeycomb oxides with heavy elements such as $\rm K_2Ni_2TeO_6$ (partially doped with $\rm Co$), as shown in \textbf{Fig. \ref{Fig_3_0}b}, TEM is used to obtain unequivocal information relating to the structure of materials at the atomic scale. A number of studies have reported the utilisation of TEM analyses on honeycomb layers of oxides to determine, with high-precision, the arrangement of atoms within the honeycomb lattice and the global order of atoms within the structure of materials in a honeycomb lattice.\cite{Masese2018, Roudebush2013b, Ramlau2014, Liu2016, Masese2019, Yoshii2019, Wang2019b, Xiao2020, Todorova2011, bette2019a} Likewise, the honeycomb lattice comprising $\rm Te$ surrounded by transition metals in $\rm K_2Ni_2TeO_6$ (\textbf{Fig. \ref{Fig_3_0}b}) can be seen from state-of-the-art TEM images, shown in \textbf{Fig. \textbf{\ref{Fig_3_2}}}. It is worth noting that TEM analyses are expensive to conduct and may lead to damage of samples because of the strong electron beam. We also note that the sensitivity of samples to the electron beam differ even within slightly the same honeycomb layered oxide composition. For instance, $\rm Cu_3Co_2SbO_6$ is more susceptible to electron\red{-}beam damage than $\rm Cu_3Ni_2SbO_6$,\cite {Roudebush2013b} which implies that tuning of the chemical composition of these materials can induce structural stability necessary to perform intensive TEM analyses.

\subsection{Nomenclature}

In a notation system promulgated by Hagenmuller\red{, Delmas} and co-workers,\red{\cite{Delmas1976, delmas1981}} honeycomb layered oxides can \red{also} be classified according to the arrangement of \red{their} honeycomb layers (stackings) 
The notation comprises a letter to represent the bond coordination of $A$ alkali, \blue{alkaline-earth} or coinage \red{metal} atoms with the surrounding oxygen atoms (generally, \red{`}$\rm T$\red{'} for tetrahedral, \red{`}$\rm O$\red{'} for octahedral, or \red{`}$\rm P$\red{'} for prismatic) and a numeral that indicates the number of repetitive honeycomb layers (slabs) per each unit cell (mainly, \red{`}1\red{'}, \red{`}2\red{'} or \red{`}3\red{'}) \red{as shown in \textbf{Table \ref{Table_1}}}. For instance, $\rm Na_2{\it M}_2TeO_6$ (with $\rm {\it M}$ being $\rm Mg, Zn, Co$ or $\rm Ni$) possess P2-type structures, the nomenclature arises from their repetitive two-honeycomb layers sequence in the unit cell with prismatic coordination of Na atoms with oxygen in the interlayer region.\cite{Sankar2014, Karna2017, Kurbakov2020, Berthelot2012a,Evstigneeva2011} Structures such as O3-type stackings can be found in $\rm Na_3{\it M}_2SbO_6$ (here $\rm {\it M} = Zn, Ni, Mg$ or $\rm Cu$) and $\rm Na_3LiFeTeO_6$, whereas $\rm Na_3Ni_2SbO_6$ and $\rm Na_3Ni_2BiO_6$ reveal O1-type and P3-type stackings during the electrochemical extraction of alkali $\rm Na$ atoms.\cite{Yuan2014, Seibel2013, Liu2016, Bhange2017, Schmidt2014, Zheng2016, Seibel2014} Note that the aforementioned oxide compositions are representative of the main stackings observed, and is by no means, an exhaustive summary. Ag- and Cu-based honeycomb layered oxides, prepared via topochemical ion-exchange, such as $A_3M_2D\rm O_6$ ($\rm {\it A} = Ag, Cu$; $\rm {\it M} = Ni, Mn, Co, Zn$; $\rm {\it D} = Bi, Sb$) and related oxides, adopt a linear \red{(dumbbell-like)} coordination of alkali or coinage \red{metal} atoms with the adjacent two oxygen atoms with an intricate multiple stacking sequence of the honeycomb slabs.\cite{zvereva2016d, Roudebush2013a, politaev2009, Bhardwaj2014} The various stacking sequences exhibited by representative honeycomb layered oxides are detailed in \textbf{Fig. \ref{Fig_3_3}} \red{and \textbf{Table \ref{Table_1}}}.

\subsection{Stacking sequences}

\blue{The stacking sequences of the honeycomb slabs as well as the emplacement patterns of the oxygen atoms play a crucial role in the nature of emergent properties; even minuscule differences in atomistic placements could result in distinct variations of crystal frameworks with an assortment of physochemical properties.\cite{breger2005, Evstigneeva2011, cabana2013, Ma2015, Liu2016}} In general, the various manner of stackings observed in honeycomb layered oxides is contingent on the synthesis procedure, the content of alkali $A$ atoms sandwiched between the honeycomb slabs and the nature of alkali $A$ cations (that is, $\rm Li, Na, K$ and so forth).\cite{Hosaka2020} Different stacking sequences of the honeycomb slabs are observed in, for example, honeycomb layered oxides that comprise $\rm Na$ and $\rm Li$ atoms. $\rm Na$ atoms, with larger radii, tend to have a strong affinity to coordinate with six oxygen atoms; adopting octahedral (O) or prismatic (P) coordination.\cite{Hosaka2020} $\rm Li$ atoms, {\it vide infra}, have been found to possess tetrahedral (T) and octahedral coordination, as recently observed in $\rm Li_2Ni_2TeO_6$.\cite{Grundish2019} Further, TEM analyses performed on oxides such as $\rm Na_3Ni_2BiO_6$, indicate assorted sequences of honeycomb ordering.\cite{Liu2016} Using high-angle annular dark-field scanning transmission electron microscopy (HAADF-STEM) imaging studies, Khalifah and co-workers have broached another labeling scheme to allow the indication of the number of repetitive honeycomb layers.\cite{Liu2016} Using their notation, they illustrated that $\rm Na_3Ni_2BiO_6$ had 6 layers (6L), 9 layers (9L) and 12 layers (12L) of stacking honeycomb ordering sequence (periodicity). Such sequence of honeycomb ordering (and even stacking disorder) can be influenced by the reaction kinetics during the use of various synthesis conditions and higher orders of stacking sequences (4L, 6L, 9L, 12L, {\it et cetera}.) can be anticipated in honeycomb layered oxides.

\newpage

\red{
\section{Magneto-spin models of honeycomb layered oxides}
}


\red{
\subsection{Crystal structure considerations}
}

\begin{figure*}[!t]
\centering
  \includegraphics[width=\textwidth]{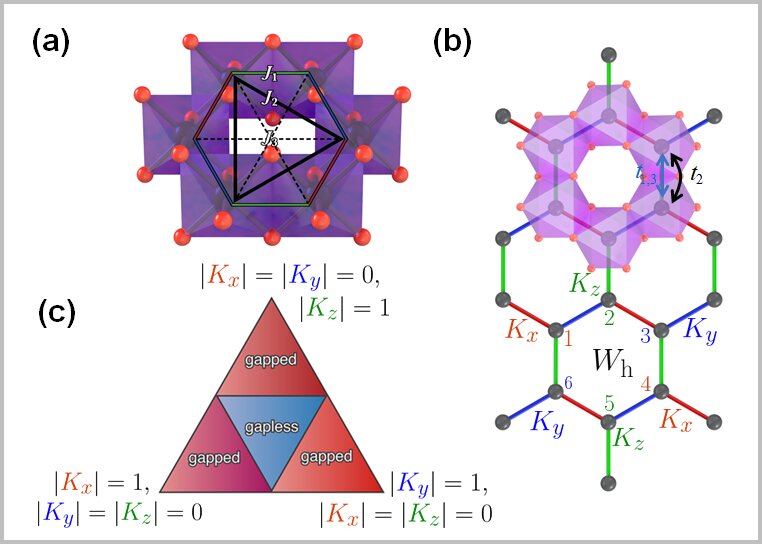}
  \caption{Nature of magnetic configurations adopted by honeycomb layered oxide materials. \red{(a) Fragment of the transition metal slab showing the honeycomb configuration of 
  transition metal atoms and the possible spin interactions with neighbouring magnetic atoms. Here $J_i$ (where $i = 1, 2$ and $3$) represents the magnetic exchange interactions (Kitaev, Haldane and higher-order interactions respectively) between an atom and its $i$-th neighbour. Transition metal atoms (in purple) are depicted  surrounded by oxygen atoms (in red) in octahedral coordinations. (b) A schematic of the realisation of the Kitaev model. $K_x$, $K_y$ and $K_z$ denote the Kitaev coupling constants for the next-neighbouring transition metal bonds in the axes (depicted in red, blue and green, respectively). $W_{\rm h}$ denotes the plaquette flux operator with links labelled as `1', `2', `3', `4', `5' and `6'. The blue ($t_{1,3}$) or black ($t_2$) double-headed arrow indicates the hopping path of an electron from one transition metal atom to an adjacent one \textit{directly} or \textit{indirectly} (via a shared oxygen atom) respectively along the spin-directional bonds of the honeycomb lattice. (c) The solution of the Kitaev model depicted as a phase diagram in the form of a triangle showing the spectrum constraints in the Brillouin zones as either gapped or gapless. The vertices of the triangle correspond to $|K_z| = 1$, $|K_x| = |K_y| = 0$, $|K_x| = 1$, $|K_y| = |K_z| = 0$ and $|K_y| = 1$, $|K_x| = |K_z| = 0$.}}
  \label{Fig_4_1}
\end{figure*}

As aforementioned, honeycomb layered oxides mainly comprise alkali cations $\rm {\it A}^{+}$ sandwiched in a framework containing layers or slabs of $\rm {\it M}$ and $\rm {\it D}$ atoms coordinated, octahedrally, with oxygen atoms. $\rm {\it M}$ atoms are essentially magnetic with a valency of $+2$ or $+3$, whilst $\rm {\it D}$ atoms are non-magnetic and generally possess valency states (oxidation states or oxidation numbers) of $+4, +5$ or $+6$. The $\rm {\it M}O_6$ and $\rm {\it D}O_6$ octahedra assume a honeycomb configuration within the layers; $\rm {\it D}O_6$ octahedra being surrounded by six $\rm {\it M}O_6$ octahedra, as shown in \red{\textbf{Fig. \ref{Fig_4_1}a}}. Note that such an ordered honeycomb configuration of magnetic $\rm {\it M}$ atoms around non-magnetic $\rm {\it D}$ atom is contingent on their ionic radii.\cite{Kumar2013, Berthelot2012a, Schmidt2013, Yadav2019} For instance, for honeycomb layered oxides with $\rm Te$ (or even $\rm Bi$), as the $\rm {\it D}$ atoms and transition metal atoms such as $\rm Ni$, ${\it M}$ atoms, typically form ordered honeycomb configurations in oxides such as $\rm Na_2Ni_2TeO_6$, $\rm Na_4NiTeO_6$, $\rm Na_3Ni_2BiO_6$ and, more recently, $\rm K_2Ni_2TeO_6$.\cite{Yang2017, Masese2018, Sankar2014, Karna2017, Kurbakov2020, Berthelot2012a, Liu2016, Evstigneeva2011, Zheng2016}. However there is slight disorder between Te and surrounding metal atoms within the honeycomb slab as noted in $\rm Na_2Zn_2TeO_6$.\cite{li2020b} On the other hand, in honeycomb layered oxides with $\rm Sb$ as the $\rm {\it D}$ metal atoms, such as $\rm Na_3Ni_2SbO_6$, disordered honeycomb configurations are often observed.\cite{Ma2015} This is due to the movement of $\rm Sb$ ($\rm {\it D}$) atoms to the sites of $\rm Ni$ ($\rm {\it M}$) atoms, which have similar ionic radii,\cite{shannon1970} a phenomenon commonly referred to as 
\red{`}cationic site mixing\red{'}. Also worthy of mention, is that the ionic radii of the sandwiched $\rm {\it A}$ atoms in honeycomb layered oxides has influence on the honeycomb ordering. This has been noted in $\rm Li$-based honeycomb \red{layered} oxides such as $\rm Li_4NiTeO_6$ \red{and $\rm Li_2Ni_2TeO_6$}, whereby $\rm Li$ atoms are located in the sites of $\rm Ni$ atoms.\red{\cite{Sathiya2013, Zvereva2015a, Grundish2019}} Hereafter, magnetism of honeycomb layered oxides with ordered honeycomb configurations of magnetic $\rm {\it M}$ atoms around $\rm {\it D}$ atoms shall be discussed, to serve as an entry point to 
\red{
some fundamental models of 
magneto-spin phenomena that have generated tremendous research interest in recent years.
}



The honeycomb arrangement of magnetic metal atoms ($\rm {\it M}$) within the slabs of honeycomb layered oxides often leads to fascinating magnetic behaviour. This is due to the interactions generated from the spins innate in the magnetic atoms (what is commonly termed as magnetic coupling). As is explicitly shown in \red{\textbf{Fig. \ref{Fig_4_1}a}}, such interactions primarily originate from spins from the adjacent magnetic atoms (Kitaev-type interactions (denoted as $J_1$)) within the honeycomb lattice, but they can also be influenced by spins of magnetic atoms from adjoining layers in the honeycomb configuration (\textit{id est}, Haldane-type interactions ($J_2$)). Spin interactions emanating from distant atoms may still occur and shall herein be classified as higher-order interactions ($J_3$).\cite{Zvereva2015b} Such magnetic interactions can be of varied fashion, spanning over short distances across the honeycomb lattice (what is termed as short-range interactions) or long distances extending to those of the adjacent honeycomb slabs (long-range interactions). 

\red{
\subsection{Kitaev model}

In the case of $J_1$ (type) interactions, the spin-spin interactions can either be anisotropic (Kitaev) or isotropic (Heisenberg) leading to an interaction Hamiltonian of the form, 
\begin{subequations}
\begin{align}
    H = 2\sum_{\langle j,k \rangle \in x} K_x S_j^x S_k^x + \sum_{\langle j,k \rangle \in x} K_y S_j^y S_k^y
    + \sum_{\langle j,k \rangle \in z} K_z S_j^z S_k^z
    + J_1\sum_{\langle j,k \rangle \in \gamma} \vec{S}_{j}\cdot\vec{S}_{k},
    \label{Kitaev_Heisenberg_eq}
\end{align}
where $\gamma = x,y,z$, $K_{x, y, z}$ are the Kitaev interaction terms shown in \textbf{Fig. \ref{Fig_4_1}b}, $J_1$ is the Heisenberg interaction term and $\vec{S}_j = (S_j^x, S_j^y, S_j^z)$ are the spin matrices. The sum i\blue{s} taken over next\blue{-}neighbour interaction \blue{s}ites corresponding to $J_1$ interactions. In a seminal paper, Kitaev showed that \textbf{eq. (\ref{Kitaev_Heisenberg_eq})} with $J_1 = 0$ is exactly soluble into a ground state of a topological superconductor in terms of a Kitaev-quantum spin liquid (K-QSL).\cite{Kitaev2006, zhou2017} K-QSL is a spin quantum state with long\red{-}range entanglement and short\red{-}range order of spin moments which continue to fluctuate coherently \blue{whilst still maintaining their disordered formation} even at low temperatures.  

In particular, the K-QSL Hamiltonian has a conserved quantity $W_{\rm h} = (S_1^xS_2^yS_3^zS_4^xS_5^yS_6^z)_{\rm h}$ for a given (honeycomb) plaquette $\rm h$, \textit{id est} as shown in \textbf{Fig. \ref{Fig_4_1}b}. Each plaquette satisfies $W_{\rm h}^2 = 1$ ($W_{\rm h} = \pm 1$) and the commutator $[H(J_1 = 0), W_{\rm h}] = i\partial W_{\rm h}/\partial t = 0$ in the Heisenberg picture of quantum mechanics, which immensely simplifies calculations of the ground state properties such as spin-spin correlation functions. For instance, since $W_{\rm h}$ and $H(J_1 = 0)$ can be simultaneously diagonalised, the ground state corresponds to $W_{\rm h} = 1$ for all plaquettes (known as the vortex\blue{-}free state) whereas flipping any one plaquette to $W_{\rm h} = -1$ corresponds to the lowest excited energy state (single vortex state) with a finite energy gap. Moreover, $W_{\rm h}$ commutes with both spin operators $S_j^{\gamma}$ and $S_k^{\gamma '}$ at different \blue{s}ites $j$ and $k$ if and only if $\gamma = \gamma '$ and $k = j \pm 1$ within a given plaquette. Otherwise, one can find $W_{\rm h}$ that commutes with either one of the operators but anti-commutes with the other \blue{one}. Thus, using the cyclic property of the trace $\langle abc \rangle = \langle bca \rangle = \langle cab \rangle$ where $a = S_j^\gamma S_k^{\gamma '}$, $b = W_{\rm h}$, $c = W_{\rm h}$, one can show that there is no long\red{-}range spin-spin correlation in the honeycomb lattice, \textit{id est} $\langle S_j^\gamma S_k^{\gamma '}\rangle = \langle S_j^\gamma S_k^{\gamma '}W_{\rm h}^2\rangle = \langle W_{\rm h} S_j^\gamma S_k^{\gamma '}W_{\rm h}\rangle = \langle S_j^\gamma W_{\rm h} S_k^{\gamma '}W_{\rm h}\rangle = -\langle S_j^\gamma W_{\rm h} S_k^{\gamma '}W_{\rm h}\rangle = -\langle S_j^\gamma S_k^{\gamma '}W_{\rm h}^2\rangle = -\langle S_j^\gamma S_k^{\gamma '} \rangle = 0$.\cite{Baskaran2007} This is the hallmark of a quantum spin liquid.\cite{savary2016, broholm2020} 

The energy spectrum of the model is exactly soluble by mapping the spin operators in $H(J = 0)$ to Majorana fermions operators by a Jordan-Wigner transformation.\cite{chen2008} (Majorana fermions are uncharged propagating degrees of freedom with quantum statistics described by anti-commuting self-adjoint quantum operators that square to $1$). This yields the phase diagram in the form of the triangle shown in \textbf{Fig. \ref{Fig_4_1}c} for half the Brillouin zone, where for $|K_x| < |K_y| + |K_z|$, $|K_y| < |K_x| + |K_z|$ and $|K_z| < |K_x| + |K_y|$ under the constraint $|K_x| + |K_y| + |K_z| = 1$, the spectrum is gapless whilst the rest of the Brillouin zone is gapped.\cite{Kitaev2006} The vertices of the triangle correspond to $|K_z| = 1$, $|K_x| = |K_y| = 0$, $|K_x| = 1$, $|K_y| = |K_z| = 0$ and $|K_y| = 1$, $|K_x| = |K_z| = 0$. 

In the gapped phase, spin correlations decay exponentially over a length scale inversely proportional to the gap. Thus, fermionic, vortex or quasi-particle (fermion $+$ vortex) excitations do not have long\red{-}range interactions. However, they can interact topologically as they move around each other (their worldlines in \red{three-dimensional (3D)} space satisfy specific braiding rules).\cite{Kitaev2006, freedman2003, kitaev2003} Thus, the energy gap favours topological interactions which reveal their Abelian anyonic statistics (\textit{id est}, the quantum phases acquired by particle exchange is additive). \blue{This carries major implications in topological quantum computing as it suggests the possibility of achieving the stabilisation of quantum bits (qubits), considering that any such uncharged system is extremely hard to disturb, \textit{exempli gratia} by a photon, because photons do not couple to uncharged quasi-particles.} 
 Moreover, the quantum state is topologically protected by the energy gap which cuts off any other non-topological interactions. Hence, quantum computation can be carried out topologically by particle exchange exploiting the braiding rules.\cite{Kitaev2006, freedman2003}

On the other hand, the quasi-particles in the gapless phase will have long\red{-}range interactions. Nonetheless, these can be suppressed by introducing a gap via a time-reversal symmetry breaking term $H_{\rm int} = -\sum_{\gamma}\sum_j h_{\gamma}S_j^{\gamma} $ in Hamiltonian $H(J_1 = 0)$ in \textbf{eq. (\ref{Kitaev_Heisenberg_eq})}, corresponding to the interaction of the spins with an external magnetic field $h_{\gamma} = (h_x, h_y, h_z)$. Since this term destroys the exact solubility of the Kitaev model, the model is solved perturbatively to yield a gap $\Delta \sim h_xh_yh_z/K^2$, where $K \equiv K_x = K_y = K_z$.\cite{Kitaev2006} Such a gapped phase admits non-Abelian quasi-particles, that are even more robust for quantum computing than the Abelian quasi-particles (\textit{id est}, the quantum phase acquired by the wavefunction under the exchange of quasi-particles \blue{is} non-additive). 




\subsection{Kitaev-Heisenberg model}

\begin{figure*}[!t]
\centering
  \includegraphics[width=\textwidth]{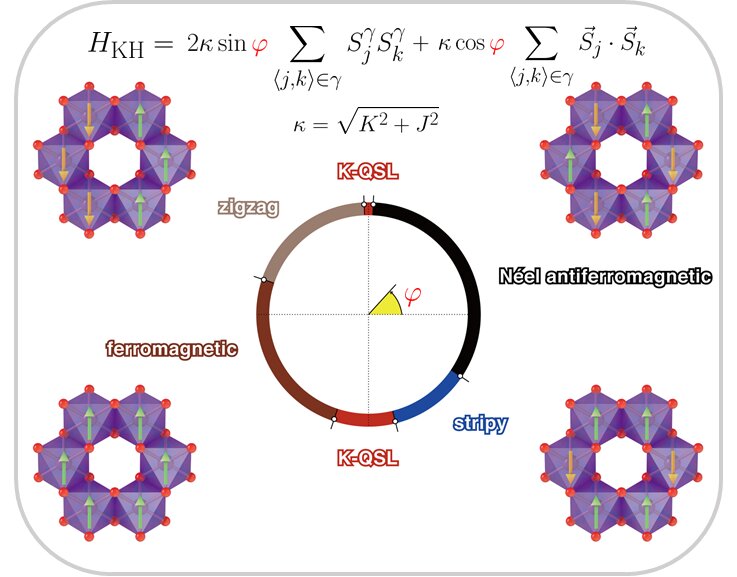}
  \caption{Various spin configurations that can be realised in honeycomb layered frameworks, described by the Kitaev-Heisenberg Hamiltonian ($H_{\rm KH}$) in \textbf{eq. (\ref{Kitaev_Heisenberg_eq2})} with coupling constants parametrised by angular variable $\varphi$,  entailing magnetic atoms (transition metals that are magnetic such as $\rm Ni, Fe, Co$ \textit{et cetera}), \red{based on the phase diagram of the Kitaev-Heisenberg model. Phases of the Kitaev-Heisenberg model (ferromagnetic, stripy, zigzag, N\'eel ferromagnetic, $et$ $cetera$) are shown with the direction of the electron spins in the honeycomb lattice.} The green arrows show the \blue{spin-up}, whereas the brown arrows show \blue{spin-down} alignment of the magnetic moment of the transition metal atoms.}
  \label{Fig_4_2}
\end{figure*}

Based on the robustness of the ungapped phase ($K \equiv K_x = K_y = K_z$) under a finite magnetic field ($\vec{h} \neq 0$) for quantum computing, it is illustrative to consider the behaviour of this ungapped phase in \textbf{eq. (\ref{Kitaev_Heisenberg_eq})} with a finite Heisenberg term ($J_1 \equiv J \neq 0$) and no magnetic field ($\vec{h} = 0$),
\begin{align}
     H_{\rm KH} = 2K\sum_{\langle j,k \rangle \in \gamma} S_j^{\gamma} S_k^{\gamma} + J\sum_{\langle j,k \rangle \in \gamma} \vec{S}_{j}\cdot\vec{S}_{k}. 
\label{Kitaev_Heisenberg_eq2}
\end{align}
\end{subequations}
This Hamiltonian has been diagonalised using a 24-site diagonalisation method by Chaloupka, Jackeli and Khaliullin,\cite{chaloupka2013} through a parametrisation $\kappa = \sqrt{K^2 + J^2}$ and $J = \kappa\cos\varphi$ and $K = \kappa\sin\varphi$, to yield 1) a phase diagram of the Kitaev-Heisenberg model with two distinct gapless K-QSL (disordered) phases around $\varphi = \pi/2$ and $\varphi = 3\pi/2$ where $J \simeq 0$, and 2) four ordered phases given by N\'eel antiferromagnetic, ferromagnetic, zigzag and stripy phases, as shown in \textbf{Fig. \ref{Fig_4_2}}. Distinct phase transitions between two phases occur at peaks corresponding to $-\partial^2 E_{0}/\partial \varphi^2$, where $E_{0}$ is the ground-state energy of the Kitaev-Heisenberg model, with the smallest peaks occurring at two ordered-disordered phase boundaries, signifying a relatively smoother transition compared to the typical ordered-ordered phase transitions.


Achieving the Hamiltonian given in \textbf{eq. (\ref{Kitaev_Heisenberg_eq2})} in a condensed matter system with $J = 0$ is \blue{considered to be} the Holy Grail of topological quantum computing.\cite{kitaev2003, Kitaev2006} Honeycomb layered oxides consisting of alkali or coinage metal atoms sandwiched between slabs exclusively made of transition metal and chalcogen (or pnictogen) atoms, which are the prime focus of this review, typically exhibit N\'eel antiferromagnetic properties,\cite{Zvereva2013, Derakhshan2007, Viciu2007, Morimoto2007, Derakhshan2008, Koo2008, 
Kumar2012, Schmidt2014, Sankar2014, Xiao2019, Zvereva2015a, Itoh2015, Zvereva2015b, Koo2016, Zvereva2016, Karna2017, Zvereva2017, Stavropoulos2019, Korshunov2020, Yao2020, Li2019a, Miura2006, Schmitt2006, Miura2007, Miura2008, Li2010, Kuo2012, Roudebush2013a, Zhang2014, Jeevanesan2014, Lefrancois2016, Wong2016, Werner2017, Scheie2019, Schmitt2014, Kurbakov2020, motome2020, Nalbandyan2013} which suggests that the Heisenberg term ($J$) dominates over the Kitaev term ($K$). Nonetheless,} observing the K-QSL phase in honeycomb layered oxides based on magnetic transition atoms with 3$d$ orbitals such as $M = \rm Co$, \red{has shown \blue{great} promise due to the localised nature of the magnetic electrons which favour the charge-transfer phase\cite{liu2020} of the Mott insulator -- a prerequisite for the realisation of the Kitaev-Heisenberg model.} For instance $\rm Na_2Co_2TeO_6$, has been shown to be favoured by the suppression of non-Kitaev interactions with rather moderate external magnetic fields, since external magnetic fields vastly suppress the Haldane-type and higher\blue{-}order interactions.\cite{zhong2020, liu2018, sano2018, xu2020} \red{Likewise, \blue{$\rm Na_3Co_2SbO_6$} has shown a rather promising route to the realisation of K-QSL via strain or pressure control.\cite{liu2020}} Conversely, it is also possible to suppress the Kitaev-type interactions instead by designing the honeycomb lattice to be composed of alternating magnetic and non-magnetic atoms, leaving only Haldane-type interactions corresponding to additional interaction terms in the Hamiltonian; \cite{Haldane1988} effectively attaining a quantum anomalous Hall insulator (or also referred to as a Chern insulator).\cite{Regnault2011, Thonhauser2006}

\red{
Based on the immense interest generated by the so-called Kitaev materials,\cite{trebst2017, ye2012, chaloupka2013, sizyuk2014, lee2015, choi2012, rau2016, knolle2014, glamazda2016, singh2012, liu2020, liu2018} we shall endeavour to briefly digress into this different class of honeycomb layered oxides, namely the honeycomb iridium oxides (iridates) embodied by $A_2 \rm IrO_3$ with $A = \rm Na, Li$ or $\rm Cu$\cite {ye2012, chaloupka2013, sizyuk2014, lee2015, choi2012, rau2016, clancy2012, knolle2014, glamazda2016, singh2012, wan2011, chaloupka2010, mazin2013, manni2014, chun2015, nishimoto2016, andrade2014, guo2020, takayama2015, wang2014, kimchi2011, chaloupka2015, yamaji2014, bhattacharjee2012, suzuki2015, hermann2018, cao2017, cao2018, rau2014} and other related compounds such as $\rm RuCl_3$, which also hold promise to realising the Kitaev-Heisenberg Hamiltonian given in \textbf{eq. (\ref{Kitaev_Heisenberg_eq2})}. These materials exist in different phases (polymorphs), often label\blue{l}ed with a prefix $\alpha -$, $\beta -$ or $\gamma -$ 
signifying the layering of the honeycomb sublattices which facilitate the existence of stacking faults. For instance, $\rm Li_2 \rm IrO_3$ is polymorphic with three distinct phases, $\alpha -\rm Li_2 \rm IrO_3$, $\beta -\rm Li_2 \rm IrO_3$ and $\gamma- \rm Li_2 \rm IrO_3$ which \blue{adopt the} honeycomb, hyper-honeycomb and stripy honeycomb crystal structure\blue{s}, respectively.\blue{\cite{majumder2018, nishimoto2016, katukuri2016, kimchi2014, pearce2020, glamazda2016, takayama2015, modic2014, singh2012}} $\alpha -\rm Li_2 \rm IrO_3$ exhibits N\'eel antiferromagnetic behaviour below their N\'eel temperature ($15$ K), and is paramagnetic above 15 K.\cite{freund2016} Its crystalline structure consists of $\rm IrO_6$ octahedra arranged in a honeycomb fashion and separated by $\rm Li^+$ cations, with other $\rm Li^+$ cations situated within the honeycomb slab. On the other hand, $\rm RuCl_3$ also exists in two main polymorphic phases, $\alpha - \rm RuCl_3$ and $\beta - \rm RuCl_3$. However, $\beta - \rm RuCl_3$ is prepared under specific conditions at low temperatures, and irreversibly reverts to the more stable polymorphic counterpart, $\alpha - \rm RuCl_3$ at approximately 500$^\circ$ C. The $\rm RuCl_3$ layers consist of $\rm RuCl_6$ octahedra li\red{n}ked by van der Waal forces forming interlayers devoid of any cations. The discovery of high-temperature K-QSL (half-integer thermal hall conductivity) in $\alpha -\rm RuCl_3$ has brought such materials into the forefront of the pursuit of the non-Abelian K-QSL\cite{kasahara2018, vinkler2018, kasahara2018_2, yokoi2020} and its related applications to topological quantum computing.\cite{aasen2020} In particular, the magnetic spins, originating from the $\rm Ir^{4+}$ ions or $\rm Ru^{3+}$ surrounded by the ligand \red{$\rm O^{2-}$} or $\rm Cl^{-}$ ions respectively, lead to a $K < 0$ (\blue{f}erromagnetic) Kitaev interaction via the celebrated Jackeli-Khaliullin mechanism.\cite{jackeli2009} 

\red{

\subsection{Realising the Kitaev interaction term}

To have a qualitative understanding of Jackeli-Khaliullin mechanism, we briefly offer an intuitive picture of the relevant crystal structure, energy scales and interactions that are predicted to give rise to \blue{the Kitaev coupling constant} $K$ in \textbf{eq. (\ref{Kitaev_Heisenberg_eq2})}. Whenever the valence
electron-electron \blue{Coulomb (Hubbard)} interaction strength $U$ in a (semi-)metal is greatly larger than their ion-to-ion hopping rate $t$ (\textit{id est}, $U \gg t \simeq 0$), the valence electrons tend to localise proximal to their parent ions forming a Mott insulator.\cite{georges1996, imada1998} On the other hand, their spins will transition from a disordered state \blue{(in this case, paramagnetic)} to an ordered configuration \blue{($viz.$, ferromagnetic, stripy, zigzag, N\'eel ferromagnetic, $et$ $cetera$)} below a transition temperature, with their magneto-spin dynamics governed to leading order by the Heisenberg term $J$ in \textbf{eq. (\ref{Kitaev_Heisenberg_eq2})} where $K \simeq 0$. However, \blue{the hopping rate $t$ is small but finite (\textit{id est} in h}oneycomb layered materials such as iridates and $\alpha - \rm RuCl_3$ , $U \gg t \neq 0$), which warrants the modification of the Heisenberg model.\blue{\cite{winter2017models, motome2020,liu2020, rau2014, cao2018, cao2017, hermann2018, suzuki2015, bhattacharjee2012, yamaji2014, chaloupka2015, kimchi2011, andrade2014, singh2012, rau2016, lee2015}}   

In irid\blue{a}tes, the Kitaev interaction term is predicted to arise from the \blue{$d$ orbitals of the ${\rm Ir^{4+}}$ ion containing 5 electrons ($5d^5$)}.\blue{\cite{winter2017models, motome2020,liu2020}}  In particular, \blue{from a solid-state chemistry perspective}, the $5d$ orbital of a free $\rm Ir^{4+}$ ion is 10-fold degenerate ($d_{x^2-y^2}$, $d_{z^2}$, $d_{xy}$, $d_{yz}$ and \blue{$d_{xz}$}, spin $1/2$). This degeneracy is lifted by crystal field splitting, into a \blue{4}-fold degenerate $e_g$ orbital ($d_{x^2-y^2}$ and $d_{z^2}$, spin $1/2$) and a lower energy 6-fold degenerate $t_{2g}$ orbital ($d_{xy}$, $d_{yz}$ and \blue{$d_{xz}$}, spin $1/2$) when a $\rm Ir^{4+}$ ion is bonded with the ligand $\rm O^{2-}$ ions forming the octahedral structure depicted in \textbf{Fig. \ref{Fig_4_1}b}. Since the ligand $\rm O^{2-}$ ions approach the $\rm Ir^{4+}$ ions at the centre of the \blue{o}ctahedra along the $x$, $y$ and $z$ axes, this finite energy difference $\Delta$ between the $e_g$ and $t_{2g}$ orbitals arises from the fact that the electrons in the $d_{x^2-y^2}$ and $d_{z^2}$ orbitals have a greater Coulomb repulsion compared to the electrons in the $d_{xy}$, $d_{yz}$ and \blue{$d_{xz}$} orbitals. The triplet $t_{2g}$ orbital experiences a further splitting into a ground $j_{\rm eff} = 3/2$ doublet state and an excited $j_{\rm eff} = 1/2$ singlet state due to spin-orbit coupling. The $j_{\rm eff} = 1/2$ state is composed of a linear combination of spin $1/2$ and orbital ($d_{xy}$, $d_{yz}$, \blue{$d_{xz}$}) entangled states. 

\blue{For brevity, s}pin-orbit coupling (SOC) is a relativistic effect where, in the inertial frame of an electron orbiting a stationary nucleus generating an electric field $\vec{E}$ at a distance $|\vec{r}| = r$ with momentum $\vec{p}$, there is a finite magnetic field $\vec{B} \propto \vec{p}\times\vec{E} = |\vec{E}|\vec{p}\times\vec{r}/r$, that couples to the spin of the electron, thus generating a magnetic moment proportional to $\vec{B}\cdot\vec{S} \propto \lambda \vec{L}\cdot\vec{S} = H_{\rm SOC}$ where $\vec{L} = \vec{r}\times\vec{p}$ is the angular momentum of the electron and $\lambda$ is the spin-orbit coupling strength.\blue{\cite{cao2018, cao2017, rau2016}}  This energy splitting can be written in terms of the total angular momentum $\vec{J} = \vec{L} + \vec{S}$ as $H_{\rm SOC} = \lambda(J^2 - L^2 - S^2)/2$, where we have used $J^2 = (\vec{L} + \vec{S})^2 = L^2 + 2\vec{L}\cdot\vec{S} + S^2$. Note that the $p$ orbital splitting (states label\blue{l}ed by total \blue{quantum} number $j$) will differ from the $t_{2g}$ orbital splitting (states label\blue{l}ed by effective total quantum number $j_{\rm eff}$) by the sign of $\lambda$. Thus, taking the spin and angular momentum quantum numbers respectively as $s = 1/2$ and $l = 1$ and using $\langle L^2 \rangle = l(l + 1)$, $\langle S^2 \rangle = s(s + 1/2)$ and $\langle J^2 \rangle = j(j + 1)$ where $j = l \pm s$, the energy splitting between the ground $j = 1/2$ singlet state and the excited $j = 3/2$ doublet state becomes $\langle H_{\rm SOC} \rangle = 3\lambda/2$.\cite{burns1993}

Thus, $4$ electrons ($2$ spin\blue{-}up and $2$ spin\blue{-}down \blue{states}) occupy the $j_{\rm eff} = 3/2$ doublet state and the remaining 1 electron (spin\blue{-}up or spin\blue{-}down) occupies the $j_{\rm eff} = 1/2$ singlet state, leaving a spin\blue{-}down or spin\blue{-}up singlet state unoccupied.\cite{rau2014} It is this unoccupied state (hole) which has a finite $U > t \neq 0$ hopping rate from one $\rm Ir^{4+}$ ion to an adjacent one. \blue{Thus, one condition \textit{sine qua non} for the realisation} of the Jackeli-Khaliullin mechanism is the existence of the $j_{\rm eff} = 1/2$ singlet state in a $5d$ transition metal ion such as $\rm Ir^{4+}$ and $\rm Ru^{3+}$, bonded with ligand ions such as \blue{$\rm O^{2-}$} and \blue{$\rm Cl^{2-}$} forming an octahedron.\blue{\cite{motome2020}} Since hopping occurs from the $j_{\rm eff} = 1/2$ state of a transition metal ion \textit{directly} to an adjacent one (labelled as $t_{1,3}$ in \textbf{Fig. \ref{Fig_4_1}b}) or \textit{indirectly} via the $p$ orbital\blue{s} of a ligand ion (labelled as $t_2$ in \textbf{Fig. \ref{Fig_4_1}b}), the second condition for the realisation of the Jackeli-Khaliullin mechanism concerns the geometric orientation of adjacent octahedra. In particular, the adjacent octahedra in honeycomb layered materials can either share a vertex (180$^\circ$ bond) or an edge (90$^\circ$ bond), which leads to either a single \textit{indirect} $t_2$ hopping path or a pair of equivalent $t_2$ hopping paths respectively. It is the pair of $t_2$ hopping paths in the 90$^\circ$ bond which results in the Jackeli-Khaliullin mechanism. This is because the Kanamori Hamiltonian\cite{rau2014} describing the Mott insulator\cite{georges1996, imada1998} is perturbed by the hopping Hamiltonian with terms resulting from the pair of $t_2$ paths, leading to destructive interference of the symmetric Heisenberg exchange terms but constructive interference of the asymmetric Kitaev exchange terms. Such a calculation results in a ferromagnetic ($K < 0$) Kitaev term,\cite{trebst2017, rau2014} 
\begin{align*}
    K = -\frac{4t_2^2}{3}\left (\frac{1}{U - 3J_{\rm H}} - \frac{1}{U - J_{\rm H}} \right )\simeq -\frac{8t_2^2}{3U^2}J_{\rm H},
\end{align*}
where $J_{\rm H} \ll U$ is the so-called Hund's coupling originating from the repulsion of electrons occupying the same orbital due to Pauli exclusion principle.

\begin{figure*}[!b]
\centering
 \includegraphics[width=\textwidth]{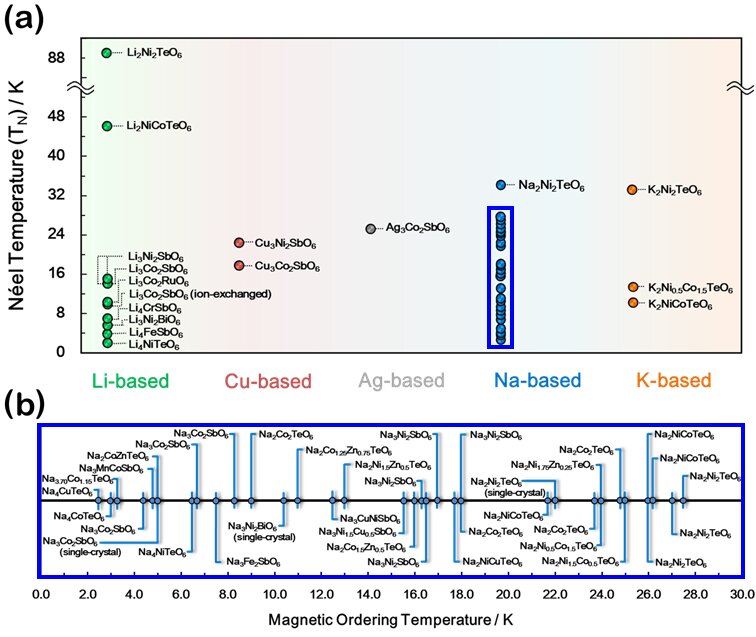}
  \caption{\red{Magnetic transition temperatures of representative honeycomb layered oxide materials.
  (a) Various magnetic transition temperatures attained in honeycomb layered oxides that entail a change in spin configuration to antiferromagnetic states ({\it videlicet}., N\'eel temperature). 
  (b) The magnetic transition temperatures of $\rm Na$-based honeycomb layered oxides (that have mostly been subject of passionate research owing to their intriguing magnetism) has been highlighted for clarity to readers.\cite{Zvereva2013, Derakhshan2007, Viciu2007, Morimoto2007, Derakhshan2008, Koo2008, 
  Kumar2012, Schmidt2014, Sankar2014, Xiao2019, Zvereva2015a, Itoh2015, Zvereva2015b, Koo2016, Zvereva2016, Karna2017, Zvereva2017, Stavropoulos2019, Korshunov2020, Yao2020, Li2019a, Miura2006, Schmitt2006, Miura2007, Miura2008, Li2010, Kuo2012, Roudebush2013a, Zhang2014, Jeevanesan2014, Lefrancois2016, Wong2016, Werner2017, Scheie2019, Schmitt2014, Kurbakov2020, motome2020, Nalbandyan2013} 
  }}
  \label{Fig_5}
 \end{figure*}

Typically, since spin-orbit coupling is greater for large transition metal atoms with $4d$ or $5d$ orbitals, much of the search for K-QSL has focused on their $j_{\rm eff} = 1/2$ state. Based on the Jackeli-Khaliullin mechanism, calculations are often conducted via complementary spin-wave analysis and exact diagonalisation, which indeed reveal that \blue{Kitaev coupling constant is ferromagnetic} ($K < 0$) for $A_2 \rm IrO_3$ when ($A = \rm Li, Na$).\cite{chaloupka2010} Unfortunately, the finite \textit{direct} $t_{1,3}$ hopping from the $j_{\rm eff} = 1/2$ state have been shown to restore the Heisenberg term as well as introduce another bond\blue{-}independent coupling, often labelled by $\Gamma$.\cite{rau2014} Thus, these non-Kitaev terms hinder the prospects of realising K-QSL with the conventional Jackeli-Khaliullin mechanism. Moreover, crystalline distortions of the octahedra, \textit{exempli gratia} Jahn-Teller distortions introduce long-range order that smears out the K-QSL phase, hence hindering its experimental realisation.\blue{\cite{Winter2016, jiang2010}} 

This has \blue{lent further impetus to} the consideration of realisation of honeycomb layered materials beyond the Jackeli-Khaliullin mechanism, which rely not only on $f$ orbitals but also on $d$ orbitals.
\cite{jang2019, liu2018, liu2020, motome2020} In particular, honeycomb layered oxides entailing \blue{for instance high-spin 3}$d^7$ ions such as $\rm Co^{2+}$, $\rm Fe^+$ and $\rm Ni^{3+}$ are considered as apposite models.\cite{motome2020, liu2020} Since $\rm Fe^+$-based honeycomb layered oxides are difficult to stabilise in the \blue{chemical compositions enumerated in Section 2} , it leaves \blue{$\rm Co^{2+}$}-based honeycomb layered oxides entailing pnictogen or chalcogen atoms such as $A_3 \rm Co_2SbO_6$, $A_3 \rm Co_2BiO_6$, $A_2 \rm Co_2TeO_6$, $A_4 \rm CoTeO_6$ (where $A$ can be $\rm Li, Na, K, Ag, Cu, Rb,$ \textit{et cetera}.), amongst others as promising candidates. As for $\rm Ni^{3+}$ compounds, they can be synthesised, for instance via the (electro)chemical oxidation of $\rm Ni^{2+}$\blue{-}based honeycomb layered oxides such as $A_2 \rm Ni_2TeO_6$, $A_3 \rm Ni_2SbO_6$, $A_3 \rm Ni_2BiO_6$, $A_4 \rm NiTeO_6$ ($A = \rm Li, Na, K, Ag, Cu,$ \textit{et cetera}) and so forth. Indeed, honeycomb layered oxides such as $\rm Na_2Co_2TeO_6$, $\rm Na_3Co_2SbO_6$ and $\rm NaNi_2BiO_{6-\delta}$ have generated traction in the search for K-QSL state.\cite{motome2020,liu2020} 



} 


\subsection{Beyond the Kitaev-Heisenberg model with higher\blue{-}order interactions}

For instance, the necessity to include the $J_2$ and $J_3$ Heisenberg couplings in the case of the Mott\blue{-}insulating layered iridates $A_2 \rm IrO_3$ ($A = \rm Na, Li$\blue{)} into \textbf{eq. (\ref{Kitaev_Heisenberg_eq2})}, thus further generalising the Kitaev-Heisenberg model has been tackled, \textit{exempli gratia}, by \citeauthor{kimchi2011} within the so-called Kitaev-Heisenberg-$J_2$-$J_3$ model.\cite{kimchi2011} Generally,}
higher\red{-}order interactions are best observable when these layered oxides are cooled down to extremely low temperatures, where the thermal motion of the spins is suppressed or negligibly small. At a unique magnetic ordering (transition) temperature, the spins align themselves in specific directions along the honeycomb configuration in various manners signifying a phase transition into new states of matter, as shown in \red{\textbf{Fig. \ref{Fig_4_2}}}. For example, a paramagnetic material transitions into antiferromagnetic when spins align in the same direction (parallel) or the opposite directions (antiparallel). Depending on the magnetic phase of matter they transition into, transition temperatures can be termed as N\'eel temperature or Curie temperature.\cite{Spaldin2006, Kittel2005} N\'eel temperature is the transition temperature where antiferromagnetic materials become paramagnetic and {\it vice versa}. We shall focus on N\'eel temperature since a vast majority of the honeycomb layered oxides display antiferromagnetic transitions at low temperatures. \textbf{Figure \ref{Fig_5}} shows the N\'eel temperatures for most honeycomb layered oxides \red{(incorporating pnictogen or chalcogen atoms),} which tend to be at lower temperatures (below 40 K).\cite{Zvereva2013, Derakhshan2007, Viciu2007, Morimoto2007, Derakhshan2008, Koo2008, 
Kumar2012, Schmidt2014, Sankar2014, Xiao2019, Zvereva2015a, Itoh2015, Zvereva2015b, Koo2016, Zvereva2016, Karna2017, Zvereva2017, Stavropoulos2019, Korshunov2020, Yao2020, Li2019a, Miura2006, Schmitt2006, Miura2007, Miura2008, Li2010, Kuo2012, Roudebush2013a, Zhang2014, Jeevanesan2014, Lefrancois2016, Wong2016, Werner2017, Scheie2019, Schmitt2014, Kurbakov2020, motome2020, Nalbandyan2013} Another intriguing manifestation of antiferromagnetism is the manner in which the antiparallel spins align in the honeycomb configuration. The antiparallel spins may assume various conformations such as zigzag ordering or alternating stripe-like (stripy) patterns within the honeycomb slab. Zigzag spin structure has been observed in honeycomb layered oxides, such as $\rm Li_3Ni_2SbO_6$, $\rm Na_3Co_2SbO_6$, $\rm Na_2Co_2TeO_6$, amongst others.\cite{Zvereva2012, Bera2017, Kurbakov2017, Werner2019, Werner2017, Ramlau2014}

Correspondingly, competing magnetic interactions on honeycomb lattices may induce both antiferromagnetic and ferromagnetic spin re-ordering, with the latter dominating when an external magnetic field is applied; a process referred to as spin-flop magnetism, observed in oxides such as $\rm Na_3Co_2SbO_6$ and $\rm Li_3Co_2SbO_6$.\cite{Viciu2007} Moreover, depending on the distance between the spins, spiral-like or helical spin arrangements may result. Competing interactions or `frustrations' may also cause the spins in a honeycomb lattice to orient haphazardly (magnetic disorder), even at low temperatures, leading to a plenitude of exotic magnetic states such as spin-glasses and spin-flop behaviour as has been noted in oxides such as $\rm Li_3Co_2SbO_6$.\cite{Stratan2019, Bieringer2000} Complex magnetic phase diagrams as well as enigmatic interactions (Heisenberg-Kitaev interactions, Dzyaloshinskii-Moriya (DM) interactions, {\it et cetera} are discussed in the last section of this review) can be envisaged in the honeycomb layered oxides.\cite{Winter2016} For instance, the Heisenberg-Kitaev model describes the magnetism in honeycomb lattice Mott insulators with strong spin-orbit coupling. An asymmetric (DM) spin interaction term in the Heisenberg-Kitaev model can be shown to lead to (anti-)vortices-like magnetic nanostructures commonly referred to as magnetic skyrmions that act as one of possible solutions describing equilibrium spin configurations in ferromagnetic/antiferromagnetic materials.\cite{Roessler2006} The binding of these vortex/anti-vortex pairs over long distances in 2D constitutes a higher-order interaction that becomes finite at a certain temperature when these materials undergo a Berezinskii, Kosterlitz and Thouless (BKT) transition - an example of a topological phase transition.\cite{Kosterlitz1973, Korshunov2020} The possibility of these (and more) higher\red{-}order interactions demonstrates that there is room for both experimentalists and theorists in physics and chemistry to expand the pedagogical scope of honeycomb layered oxides. 

\newpage

\section{Solid-state ion diffusion in honeycomb layered oxides}

\begin{figure*}[!b]
\centering
  \includegraphics[height=12cm]{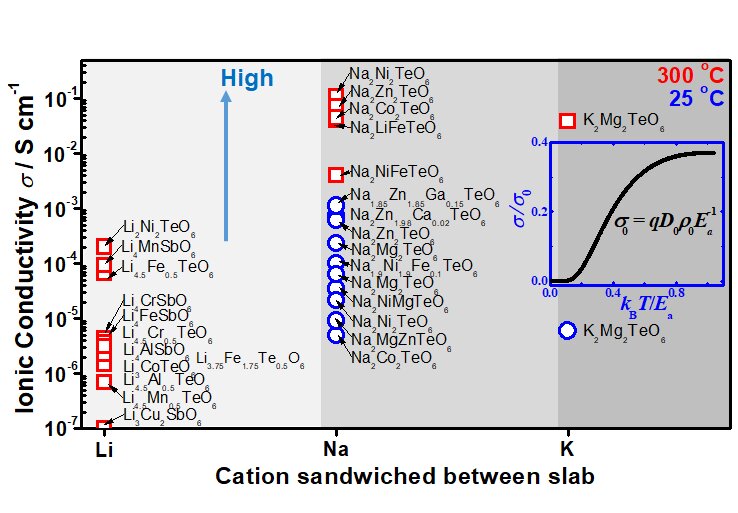}
  \caption{Solid-state diffusive properties of typical cations sandwiched between honeycomb slabs of various layered oxides showing values of ionic conductivity attained in honeycomb layered oxides at room temperature and also at high temperature (300$^{\circ}$ C).\cite{Masese2018, Nalbandyan2013a, Evstigneeva2011, Li2018Chemistry, Li2018ACSApplMater, Wu2018, Deng2019, Bianchini2019, Wu2020, Dubey2020} Honeycomb layered oxides based on tellurates generally tend to show high ionic conduction, owing to the partial occupancy of alkali atoms in distinct crystallographic sites that facilitate rapid hopping diffusion mechanism. Inset shows the plot of the dependency of (normalised) conductance $\sigma$ to (normalised) thermal energy $k_{\rm B}T$ of the cations determined by linear response of the diffusion current to low amplitude, slowly oscillating voltage by electrochemical impedance spectroscopy (EIS).}
  \label{Fig_6}
\end{figure*}

\begin{figure*}[!t]
\centering
  \includegraphics[width=\textwidth]{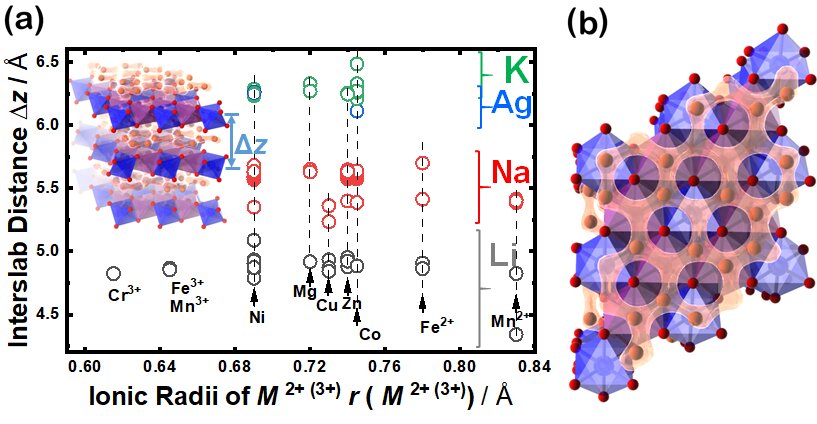}
  \caption{Regarding the honeycomb interslab distance and the size of sandwiched alkali or coinage \red{metal} atoms. 
  (a) Correlation of the \red{average} interslab distance ($\Delta z$) and the size ({\it id est}, the Shannon-Prewitt ionic radius) of the sandwiched alkali ion ($\it A$) in honeycomb layered oxides adopting the following compositions: $A^{+}_2M^{2+}_2D^{6+}\rm O_6$ ($A^{+}_{2/3}M^{2+}_{2/3}D^{6+}_{1/3}\rm O_2$), $A^{+}_{3}M^{2+}_2D^{5+}\rm O_6$ ($A^{+}M^{2+}_{2/3}D^{5+}_{1/3}\rm O_2$), $A^{+}_{4}M^{3+}D^{5+}\rm O_6$ ($A^{+}_{4/3}M^{3+}_{1/3}D^{5+}_{1/3}\rm O_2$), $A^{+}_4M^{2+}D^{6+}\rm O_6$ ($A^{+}_{4/3}M^{2+}_{1/3}D^{6+}_{1/3}\rm O_2$), {\it et cetera} where $M$ = Fe, Mn, Co, Ni, Cu, Zn, Mg \red{or a combination of at least two transition or alkaline-earth metal atoms}; $D$ = Te, Sb, Bi, Nb ; $A$ = Cu, Ag, Li, Na, K.\red{\cite {Masese2018, Masese2019, Yoshii2019, Berthelot2012a, Berthelot2012, Schmidt2013, Stratan2019, zvereva2016d, Zvereva2013, Zvereva2012, Zvereva2016, Yadav2019, chen2020}} \red{For clarity's sake, the error bars associated with each data point are smaller than the size of the data markers}. (b) A fragment of a honeycomb layered oxide such as $\rm K_2Ni_2TeO_6$ in the $ab$ plane, showing the two-dimensional \red{(2D)} diffusion channels of potassium ions. This figure also shows that apart from the type of alkali or coinage \red{metal} atom, where profound change in the interlayer distance can be expected, the nature of the transition metal atom $M$ also influences the interlayer distance albeit to a smaller extent \red{in some instances}.}
  \label{Fig_8}
\end{figure*}

\begin{table*}[!t]
\caption{
Values of ionic conductivity measured using electrochemical impedance spectroscopy (EIS) along with the activation energy ($E_{\rm a}$) attained in representative honeycomb layered oxides at room temperature and also at high temperature (300$^{\circ}$ C).\red{\cite{Masese2018, Nalbandyan2013a, Evstigneeva2011, Li2018Chemistry, Li2018ACSApplMater, Wu2018, Deng2019, Wu2020, Dubey2020, Kumar2013, Greaves1990}} The pellet compactness, amongst other factors, do influence the conductivity of ceramics and thus have been furnished (where possible).}\label{Table_2}
\begin{center}
\scalebox{0.75}{
\begin{tabular}{llllc} 
\hline
\textbf{Compound} & $\sigma_{573\,\rm K}/$S cm$^{-1}$ & $\sigma_{300\,\rm K}/$S cm$^{-1}$ & $E_{\rm a}/$ eV & \textbf{Pellet compactness}\\ 
 & ($300^\circ$ C ($573$ K)) & ($25 \pm 3^\circ$ C ($298 \pm 3$ K)) & & \% \\ 
\hline\hline
& & & &\\
$\rm K_2Mg_2TeO_6$ &	$3.8 \times 10^{-2}$  &	$\sim 10^{-5}$ & $0.92$ & $\sim 70$\\
& & & &\\
$\rm Li_2Ni_2TeO_6$ & $2.0 \times 10^{-4}$ & & $0.80$ ($333 \sim 573$ K) & \\	
$\rm Li_3Co_{1.06}TeO_6$ & $\rm 1.6 \times 10^{-6}$ & & &\\		
$\rm Li_3Cu_2SbO_6$ & $1.0 \times 10^{-7}$ & & &		\\
$\rm Li_{3.5}Zn_{1.5}BiO_{5.75}$ & $1.13 \times 10^{-5}$ & & $0.37$ ($373 \sim 573$ K) & \\
$\rm Li_{3.75}Fe_{1.75}Te_{0.5}O_6$ & $2.21 \times 10^{-6}$ & & $0.60$ &\\ 
$\rm Li_4CrSbO_6$	& $4.31 \times 10^{-6}$ & & $0.66$ ($300 \sim 573$ K) & \\
$\rm Li_4FeSbO_6$ &	$3.66 \times 10^{-6}$ &	& $0.57$ ($300 \sim 573$ K) &\\
$\rm Li_4MnSbO_6$ &	$9.33\times 10^{-5}$ & & $0.57$ ($300 \sim 573$ K) &\\
$\rm Li_4AlSbO_6$ &	$3.05 \times 10^{-6}$ & & $0.91$ ($300 \sim 573$ K) &\\
$\rm Li_{4.5}Cr_{0.5}TeO_6$ & $3.24 \times 10^{-6}$ & & $0.53$ ($373 \sim 573$ K) & \\	
$\rm Li_{4.5}Mn_{0.5}TeO_6$ & $6.88 \times 10^{-7}$ & & $0.73$ ($373 \sim 573$ K) &	\\
$\rm Li_{4.5}Al_{0.5}TeO_6$ & $1.49 \times 10^{-6}$	& & $0.66$ ($373 \sim 573$ K) &	\\	
$\rm Li_{4.5}Fe_{0.5}TeO_6$ & $6.76 \times 10^{-5}$	& & $0.70$ &\\ 
& & & &\\
$\rm Na_{1.9}Ni_{1.9}Fe_{0.1}TeO_6$	& $4.7 \times 10^{-2}$ & $1 \times 10^{-4}$ & $0.38$ ($373 \sim 623$ K) & $72$\\
$\rm Na_2Zn_2TeO_6$ & $(5.1 \sim 7.0) \times 10^{-2}$ & $9 \times 10^{-5}$ & &	$55 \sim 68$\\
$\rm Na_2Co_2TeO_6$	& $4.4 \times 10^{-2}$ & & & \\			
$\rm Na_2Co_2TeO_6$	& $(3.1 \sim 4.4)\times 10^{-2}$ &	$(3.8 \sim 4.9) \times 10^{-6}$ & $0.52$ ($373 \sim 623$ K)	& $56 \sim 82$\\
$\rm Na_2Mg_2TeO_6$	& $2.3 \times 10^{-2}$ & $6.3 \times 10^{-5}$ & & $74$ \\
$\rm Na_2Mg_2TeO_6$	& & $2.3 \times 10^{-4}$ & $0.341$ ($323 \sim 393$ K) &	$87.2$\\ 	
$\rm Na_2NiFeTeO_6$	& $\sim 4.0 \times 10^{-3}$ & & $0.46 \sim 0.49$ ($343 \sim 663$ K) & \\	
$\rm Na_2NiFeTeO_6$	& $4.0 \times 10^{-3}$ & & & \\			
$\rm Na_2NiMgTeO_6$	& & $2.13 \times 10^{-5}$ & $0.59$ ($T < 303$ K) & \\	
$\rm Na_2MgZnTeO_6$	& &	$9 \times 10^{-6}$ & $0.36$ ($T < 303$ K) & \\
$\rm Na_2Zn_2TeO_6$	& &	$(6.29 \sim 7.54) \times 10^{-4}$ & & \\
$\rm Na_2Zn_2TeO_6$	& &	$\sim 6 \times 10^{-4}$ & & \\
$\rm Na_2Zn_2TeO_6$	& &	$5.7 \times 10^{-4}$ & & \\
$\rm Na_{2-\it x}Zn_{2-\it x}Ga_{\it x}TeO_6$ $(x = 0.15)$ & & $(6.29 \sim 10.9) \times 10^{-4}$ & &	\\
$\rm Na_2Zn_{2-\it x}Ca_{\it x}TeO_6$ $(x = 0 \sim 0.05)$	& &	$7.54 \times 10^{-4} (x = 0.02)$ & & \\
$\rm Na_2Zn_2TeO_6$ (Ga-doped)	& &	$8.3 \times 10^{-4}$ & & \\
$\rm Na_2Ni_2TeO_6$ & $(1.01 \sim 1.08) \times 10^{-1}$ &	$(8 \sim 34)\times10^{-4}$ & $0.55$ $(373 \sim 623\,\,\rm K)$ &	$79.6 \sim 80.3$\\
$\rm Na_2Ni_2TeO_6$ & & $2 \times 10^{-6}\,\,(323\,\,\rm K)$ & $\sim 0.58 (3)$ ($T \geq 383$ K), $0.39$ ($T < 353$ K) & 90\\
$\rm Na_2LiFeTeO_6$ & $4 \times 10^{-2}$ & & $0.44 \sim 0.49$ ($343 \sim 663$ K)&\\
& & & &\\
\hline
\end{tabular}}
\end{center}
\end{table*}

High ionic conductivity is a prerequisite for superfast ionic conductors that may serve as solid electrolytes for energy storage devices. The presence of mobile alkali atoms sandwiched in honeycomb slabs, as is present in honeycomb layered oxides, endows them with fast ionic conduction not only at high temperatures but also at room temperature. \textbf{Figure \ref{Fig_6}} shows the ionic conductivity of honeycomb layered oxides reported to date, with the tellurate-based honeycomb layered oxides exhibiting the highest conductivity so far.\cite{Masese2018, Kumar2012, Evstigneeva2011, Nalbandyan2013a, Sau2015, Sau2016, Li2018Chemistry, Li2018ACSApplMater, Wu2018, Deng2019, Bianchini2019, Wu2020, Dubey2020} Experimentally, the measurement of ionic conductivity is conducted via linear response techniques with polarised electromagnetic fields such as electrochemical impedance spectroscopy (EIS).\cite{Lasia1999} EIS entails applying a low amplitude low frequency oscillating voltage (current) and measuring the current (voltage) response. The current-to-voltage ratio determines the inverse of the impedance (admittance) of the material, where the real part of the admittance $Y(\omega) = \sigma(\omega) + i\omega\varepsilon$ is proportional to the conductivity $\sigma(\omega)$ and the imaginary part is proportional to the permittivity $\varepsilon$ of the material.

\subsection{Heuristics of cationic diffusion}

In order to rationalise the heuristics behind the high ionic conductivity of the honeycomb layered oxides and predict associated outcomes, it is imperative to introduce a detailed theoretical approach that incorporates the thermodynamics of the cations. In particular, the connection between the ionic conductivity of honeycomb layered oxides and other physical measurable quantities such as the diffusion of solid-state alkali cations at thermal equilibrium and very low frequencies undergoing Brownian motion satisfies the fluctuation-dissipation theorem.\cite{Weber1956} Here, we showcase this approach based on heuristic arguments that captures the diffusion aspects of ionic conductivity of the (honeycomb) layered materials.\cite{Kanyolo2020}

Ionic conductivity of {\it A} cations can be heuristically modeled under a Langevin-Fick framework of equations,\cite{Fick1995, Langevin1908}
\begin{subequations}\label{Langevin-Fick_eq}
\begin{align}
 -D\Vec{\nabla}\rho (t,\vec{x}) = \Vec{j} \equiv \rho(t,\vec{x}) \Vec{v},\\
 \frac{d\Vec{p}}{dt} = -\frac{1}{\mu}\Vec{v} - q\Vec{\nabla}V(t,\vec{x}),
\end{align}
\end{subequations}
where $q$ is the unit charge of the cation, $D = D_{0}\exp (-\beta E_{\rm a})$ is the Arrhenius equation relating the diffusion coefficient to the activation energy (per mole) of diffusion ($E_{\rm a}$), $D_{0}$ is the maximal diffusion coefficient, $\beta = 1/k_{\rm B}T$ is the inverse temperature, $T$ is the temperature at equilibrium, $k_{\rm B}$ is the Boltzmann constant, $\rho(t,\vec{x})$  is the concentration of alkali cations, $\Vec{v}$ is their velocity vector and $V(t,\vec{x})$ is a time-dependent voltage distribution over the material. Imposing charge and momentum conservation, $-\vec{\nabla}\cdot\vec{j} = \partial \rho/\partial t = 0$ and $d\vec{p}/dt = 0$ respectively, and assuming the ionic concentration satisfies the Boltzmann distribution at thermal equilibrium, $\rho(T,t,\vec{x}) = \rho_{0}\exp(-\beta qV(t,\vec{x}))$ (where $\rho_{0}$ is the ionic density at zero voltage) leads to the ionic conductivity $\sigma = q\mu\rho$ proportional to the mobility $\mu$ of the alkali cations, which satisfies the fluctuation-dissipation relation $\mu = \beta D$ first derived by Einstein and Smoluchowski to describe particles undergoing Brownian motion (diffusion).\cite{Einstein1905, Smoluchowski1906, Feynman2005} Based solely on \textbf{eq. (\ref{Langevin-Fick_eq})}, the ionic conductivity of the alkali cations of honeycomb layered oxides, as summarised in \textbf{Fig. \ref{Fig_6}} and \red{\textbf{Table \ref{Table_2}}}, is related to the equilibrium temperature of the materials. 

In particular, the ionic conductivity computes to $\sigma(T,t,\vec{x}) = q\mu\rho(T,t,\vec{x}) = qD_{0}\rho_{0}\beta\exp(-\beta\left \{E_{\rm a} + qV(t,\vec{x})\right \})$. Plotting the ionic conductivity versus the normalised temperature $k_{\rm B}T/(E_{\rm a} + qV)$, we find that the ionic conductivity scales with the equilibrium temperature in the regime $k_{\rm B}T/(E_{\rm a} + qV) \sim k_{\rm B}T/E_{\rm a}$, which is always satisfied in EIS measurements. For $k_{\rm B}T/E_{\rm a} < 1$, raising the temperature increases the thermal motion of the cations, which in turn raises the ionic conductivity. \textbf{Figure \ref{Fig_6}} displays the ionic conductivity attained in honeycomb layered oxides at room temperature (25$^\circ$ C) and also at high temperature (300$^\circ$ C), which showcases the increase of ionic conductivity with temperature as expected. 

\subsection{Correlation between interslab distances and cationic diffusion}

Moreover, classical motion of the alkali cations and other electromagnetic interactions along the $z$ direction are precluded, since these materials often satisfy the condition $\Delta z \gg r_{\rm ion}$, where $\Delta z$ is the interlayer/interslab separation distance and $r_{\rm ion}$ is the ionic radius of the alkali cations. This condition effectively restricts the electrodynamics in these layered materials to two dimensions (2D), and is almost always satisfied since the ionic radius of the alkali cations is correlated with interslab distance, as shown in \textbf{Fig. \ref{Fig_8}a}. For instance, alkali cations with large ionic radii such as K in the layered oxide $\rm K_2Ni_2TeO_6$ are restricted to the two-dimensional (2D) honeycomb diffusion channels in the $ab$ plane (\textbf{Fig. \ref{Fig_8}b}), as has also been shown in $\rm Na_2Ni_2TeO_6$.\cite{Bera2020, sau2016ion1, sau2015ion1} Thus, the large interslab separation, together with the $\rm TeO_6$ octahedra acts as a barrier preventing inter-channel exchange of the alkali cations. 

Other factors that affect the ionic conductivity of the cations include the ionic radius of the $\rm {\it A}$ cations in relation to the $\rm {\it M}$ atoms. For instance, in the case of $\rm Li_2Ni_2TeO_6$ (where $\rm {\it M} = Ni$) in comparison to $\rm Na_2Ni_2TeO_6$ and $\rm K_2Ni_2TeO_6$, $\rm Ni$ atoms act as impurities in the diffusion dynamics of $\rm {\it A} = Li$ since the interlayer separation distance in $\rm Li_2Ni_2TeO_6$ is vastly smaller compared to $\rm Na_2Ni_2TeO_6$ and $\rm K_2Ni_2TeO_6$. Electrochemically, collisions with such impurities in these honeycomb layered oxides are suppressed by the larger interslab distance in conjunction with the greater sizes of $\rm Na$ and $\rm K$ atoms relative to $\rm Li$, which ensures their facile mobility within the two dimensional planes. In contrast, the smaller interslab distance and the equivalent atomic sizes of $\rm Li$ and $\rm Ni$ atoms, which lie at close proximity to the honeycomb slabs leads to the interchange of their crystallographic sites (commonly referred to as $\rm Li/Ni$ `cationic mixing').\cite{Choi1996} Consequently, the mobility of $\rm Li$ is obstructed through the collisions with $\rm Ni$ atoms within the 2D honeycomb surface that act as impurities, a process quantum mechanically referred to as scattering. Within our heuristic approach, scattering of $\rm Li$ ions through collisions with $\rm Ni$ costs more (activation) energy than in the case of $\rm Na$ or $\rm K$, $E_{\rm a}^{\rm Li} \gg E_{\rm a}^{\rm Na}, E_{\rm a}^{\rm K}$). Thus, the Einstein-Smoluchowski relation $\mu = D/k_{\rm B}T$ together with Arrhenius equation $D = D_{0}\exp(-E_{\rm a}/k_{\rm B}T)$ leads to a smaller ionic mobility in the case of $\rm {\it A} = Li$ ions. This trend is indeed shown in \textbf{Fig. \ref{Fig_6}} and \red{\textbf{Table \ref{Table_2}}}, wherein $\rm Li$-based honeycomb layered oxides, regardless of the temperature, still show inferior ionic conductivity compared to those with $\rm Na$ or $\rm K$.

\newpage

\section{Electrochemistry of honeycomb layered oxides}

\begin{figure*}[!b]
\centering
  \includegraphics[height=10.8cm]{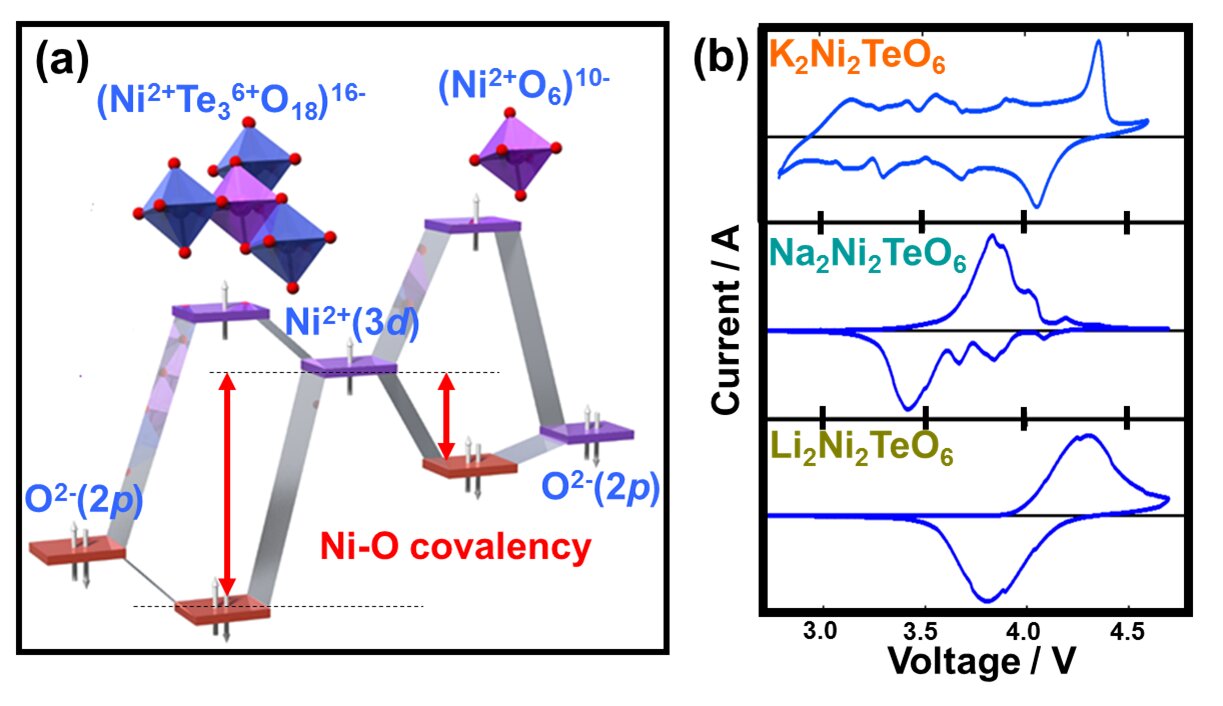}
  \caption{High-voltage electrochemistry of honeycomb layered oxides. (a) Molecular orbital calculations of the voltage increase arising from the `inductive effect' that alters the covalency of Ni--O bonds, due to the presence of more electronegative Te atom surrounded by a honeycomb configuration of Ni atoms. (b) Voltage-response curves (technically referred to as cyclic voltammograms) of honeycomb layered compositions ($A^{+}_2M^{2+}_2D^{6+}\rm O_6$ ($A$ = Li, Na, K)), showing their potential as high-voltage cathode materials for rechargeable alkali cation batteries. Technically, these cyclic voltammograms (voltage-response curves) were plotted under a scan rate of 0.1 millivolt per second. Part of the data in (b) was adapted from ref. \cite{Masese2018}
  under Creative Commons licence 4.0.}
  \label{Fig_9}
\end{figure*}

\subsection{Theoretical basis for high voltages in honeycomb layered oxides}

The layered structure consisting of highly oxidisable 3$d$ transition metal atoms in the honeycomb slabs segregated pertinently by alkali metal atoms, renders this class of oxides propitious for energy storage. In principle, classical battery electrodes rely on the oxidation (or reduction) of constituent 3$d$ metal cations to maintain charge electro-neutrality, thus facilitating the extraction (or reinsertion) of alkali metal cations, a process referred to as `charge-compensation'.\cite{goodenough2013li} In principle, the constituent pentavalent or hexavalent $d^0$ cations (such as $\rm Bi^{5+}$, $\rm Sb^{5+}$, $\rm Te^{6+}$, $\rm W^{6+}$ and so forth) do not participate in the charge-compensation process during battery performance. However, the highly electronegative $\rm WO_6^{6-}$ (or $\rm TeO_6^{6-}$, $\rm WO_{6}^{6-}$, $\rm BiO_6^{7-}$, $\rm SbO_6^{7-}$, $\rm RuO_6^{7-}$, {\it et cetera}) anions lower the covalency of the bonds formed between the oxygen ($\rm O$) atoms and 3$d$ transition metal ($\rm {\it M}$) resulting in an increase the ionic character of $M$-$\rm O$ bonds within the honeycomb layered oxides.\cite{Goodenough2013} Consequently, the energy required to oxidise $\rm {\it M}$ cations increases, inducing a staggering increase in the voltage of related honeycomb layered oxides within the battery. This process is commonly referred to as `inductive effect'.\cite{Padhi1997} For clarity, the electronegativity trend (based on the Pauling scale) is generally as follows: $\rm W > Ru > Te > Sb > Bi$. Honeycomb layered oxides such as $\rm Li_4Ni^{2+}TeO_6$, and more recently, $\rm Li_2Ni^{2+}_2TeO_6$, manifest higher voltages (over $4$ V) in comparison to other layered oxides or compounds containing $\rm Ni^{2+}$.\cite{Sathiya2013, Grundish2019, Zvereva2015a} This is rationalised by considering $\rm Te^{6+}$ as a $\rm TeO_6^{6-}$ moiety, which being more electronegative than anions such as $\rm O^{2-}$, increases the voltage necessary to oxidise $\rm Ni^{2+}$ (or technically as redox potential) through the inductive effect (as succinctly shown in \textbf{Fig. \ref{Fig_9}a}).\cite{Sathiya2013} Voltage increase due to this inductive effect has, in particular, been noted in polyanion-based compounds when $\rm (SO_4)^{2-}$ are replaced either by $\rm (PO_4)^{3-}$ or $\rm (PO_4F)^{4-}$ anion moieties.\cite{Masquelier2013} Therefore, the inductive effect seems to be typical and represents a crucial strategy when tuning the voltages of honeycomb layered oxides. Indeed, besides $\rm Li_2Ni_2TeO_6$ and $\rm Li_4NiTeO_6$, analogues consisting of $\rm Na$ (such as $\rm Na_2Ni_2TeO_6$ and $\rm Na_4NiTeO_6$) and $\rm K$ alkali atoms (such as $\rm K_2Ni_2TeO_6$ and $\rm K_4NiTeO_6$) also exhibit high voltages surpassing those of layered oxides in their respective fields.\cite{Yang2017, Masese2018, Gupta2013}

\subsection{Electrochemical measurements}

Theoretical insights regarding the high voltage innate in the aforementioned honeycomb layered oxides are validated by experimental investigations. Typical electrochemical measurements performed include: cyclic voltammetry, which assesses the voltages during charging and discharging at which the 3$d$ transition metal redox processes occur as well as other structural changes and galvanostatic charge/discharge measurements, which principally determine pivotal battery performance metrics, {\it inter alia}, (i) the amount of alkali atoms electrochemically extracted or inserted ({\it id est}, capacity) during charging and discharging, (ii) how fast the alkali atoms can be extracted or inserted (referred to as rate performance), (iii) voltage regimes where alkali atoms are dominantly being extracted or inserted and (iv) the nature of the extraction or reinsertion process of alkali atoms, for instance, whether it occurs topotactically (referred to as a single-phase, solid-solution or monophasic behaviour) or as a multiple phase (referred to as a two-phase or biphasic behaviour). \textbf{Figure \ref{Fig_9}b} illustrates the cyclic voltammograms of $\rm {\it A}^{+}_2Ni^{2+}_2Te^{6+}O_6$ ($\rm {\it A} = Li$, $\rm Na$ and $\rm K$), depicting voltage peaks/humps around $4$ V where the redox process of $\rm Ni$ (in this case $\rm Ni^{2+}/Ni^{3+}$) occur during electrochemical extraction/insertion of alkali atoms. It is noteworthy that the larger the ionic radius of $\rm {\it A}$ is, the more pronounced the minor voltage humps are seen. This is indicative of structural changes (phase transitions) occurring, details of which shall be elaborated in a later section. Usually the voltage response curves (cyclic voltammograms) should nicely mirror each other (taking the line where the current density is set as zero to be the mirror plane in \textbf{Fig. \ref{Fig_9}b}). However, due to some electrochemical issues (such as inherently slow alkali-ion kinetics), the voltages at which the redox processes or structural changes occur deviate from each other as seen in \textbf{Fig. \ref{Fig_9}b}. This is technically referred to as `voltage polarisation' or `voltage hysteresis'.\cite{Stern1957} Voltage polarisation can significantly be decreased by partial substitution (or doping) of the constituent 3$d$ transition metal atoms with isovalent metals. For instance, partial doping with $\rm Zn$, $\rm Mn$ or $\rm Mg$ in $\rm Na_3Ni_2SbO_6$ leads to lower voltage polarisation compared to that of the undoped $\rm Na_3Ni_2SbO_6$.\cite{Kee2020, Aguesse2016} Furthermore, doped oxides present higher voltages; depicting doping as another feasible route to increase the voltages of these honeycomb layered oxides.\cite{Kee2016}

\subsection{Suitable electrolytes for high-voltage honeycomb layered oxides}

\begin{figure*}[!t]
\centering
  \includegraphics[height=14cm]{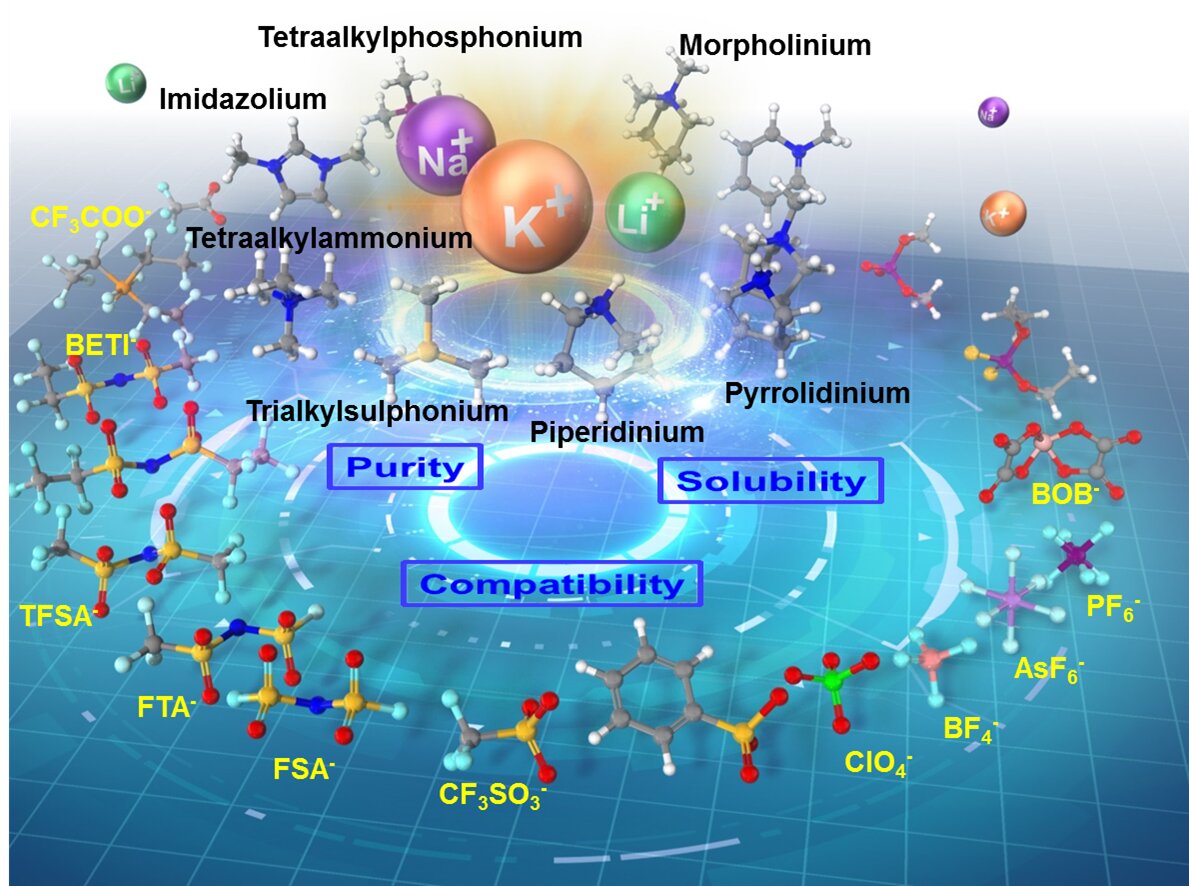}
  \caption{
  Illustration showing a selection of electrolytes (in particular ionic liquids) which guarantee the stable electrochemical performance of honeycomb layered oxides. In principle, ionic liquids consist of organic cations (pyrrolidinium, imidazolium, piperidinium, {\it et cetera}) coupled with organic or inorganic anions (such as $\rm BF_4^{-}$, $\rm PF_6^{-}$, $\rm ClO_4^{-}$, {\it et cetera}.) Organic cations are shown in black, whilst organic or inorganic anions are in yellow. Purity of the salts, solubility and compatibility with the honeycomb layered oxide cathode materials, amongst other factors are necessary to consider when obtaining suitable ionic liquids for high-voltage operation. Readers may further refer to the literature for more details regarding the ionic liquids.\cite{Matsumoto2019, yamamoto2017, hwang2018na, kubota2012b, chen2016b, matsumoto2014, kubota2008, fukunaga2012, ding2013, hagiwara2000, monti2014, wongittharom2014, wang2015rechargeable, chen2014, fukunaga2016, noor2013, chen2013, kalhoff2015, ohno2005, macfarlane2014,Yoshii2019, Masese2018}}
  \label{Fig_10}
\end{figure*}

\begin{figure*}[!t]
\centering
  \includegraphics[height=10.3cm]{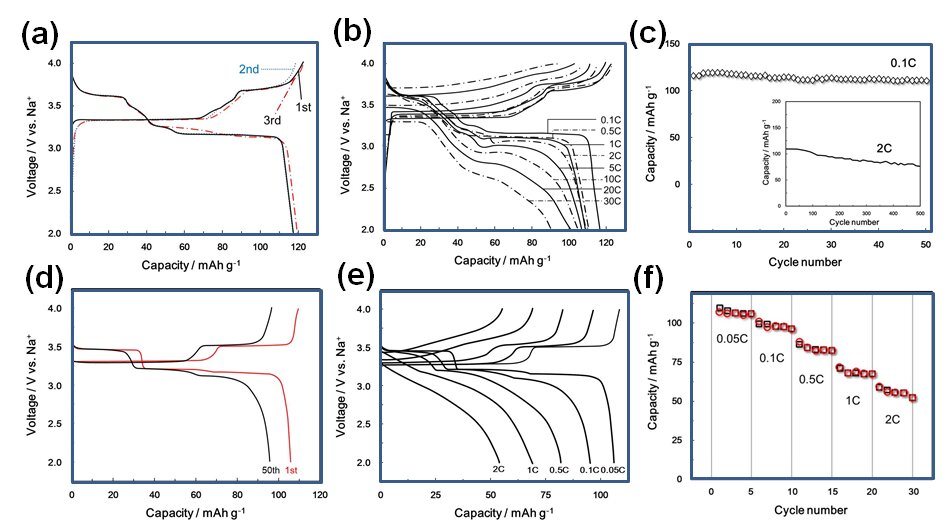}
  \caption{Electrochemical performance of representative honeycomb layered oxide cathode materials. (a) Voltage-capacity plots of $\rm Na_3Ni_2SbO_6$ showing the initial (dis)charge curves under a $\rm Na$-ion (de)insertion rate (current density) commensurate to $0.1 C$. Technically, $n\,C$ rate denotes the number of hours ($1/n$) necessary to (de)insert alkali-ions to the full theoretical capacity (alkali ion occlusion capacity per formula unit). (b) Rate capability of $\rm Na_3Ni_2SbO_6$. (c) Capacity retention of $\rm Na_3Ni_2SbO_6$ at (de)insertion rates of $0.1\,C$ and $2\,C$ (inset). (d) Voltage-capacity plots of $\rm Na_3Ni_2BiO_6$ at a current density equivalent to $0.05\,C$. (e) Corresponding rate performance, showing $\rm Na_3Ni_2BiO_6$ to also sustain fast $\rm Na$-ion kinetics. (f) Capacity retention of $\rm Na_3Ni_2BiO_6$ at various rates. (a), (b) and (c) were reproduced and adapted from ref. \citenum{Yuan2014}
  with permission. (d), (e) and (f) were reproduced and adapted from ref. \citenum{Bhange2017}
  \red{by} permission \red{of the Royal Society of Chemistry}.}
  \label{Fig_10b}
\end{figure*}

\begin{figure*}[!t]
\centering
  \includegraphics[height=14cm]{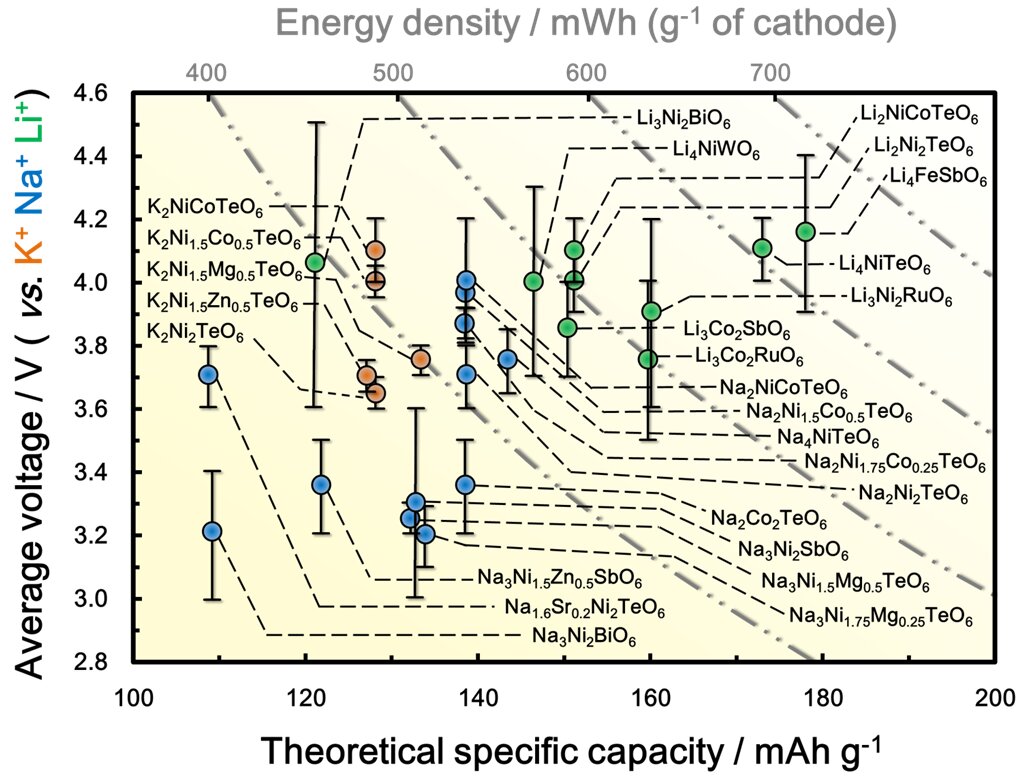}
  \caption{
  Voltage-capacity plots of various honeycomb layered oxides \red{encompassing mainly pnictogen or chalcogen atoms}, showing their potential as high-energy-density contenders for high-voltage alkali-ion batteries.\cite{Sathiya2013, Grundish2019, Yang2017, Yuan2014, Masese2018, Bhange2017, Masese2019, Yoshii2019, Wang2019b, Zheng2016, Ma2015, Kee2016, Aguesse2016, Kee2020, Matsumoto2005, Matsumoto2006, Sakaebe2003, Wang2018, Bao2014, Kim2017, Huang2020, Wang2019a, Paharik2017, Dai2017, Han2016, Zhao2019, McCalla2015a} The error bars represent the upper and lower limits of the voltages attained experimentally. Note also that the theoretical capacities have been calculated based on the change in oxidation states of transition 3$d$ metals as charge-compensation cations.}
  \label{Fig_11}
\end{figure*}

Precise and adequate evaluation of the voltage responses of high-voltage cathode materials, such as the aforementioned $\rm {\it A}_2Ni_2TeO_6$, demands the utilisation of stable electrolytes that can sustain high-voltages. Conventional electrolytes consisting of organic solvents are prone to decomposition during high-voltage operations; rendering them unsuitable for the high-voltage performance innate in such honeycomb layered oxides electrodes. Ionic liquids, which comprise entirely of organic cations and (in)organic anions, are a growing class of stable and safe electrolytes exhibiting a plenitude of distinct properties; pivotal amongst them being their good electrochemical and thermal stability, low volatility and low flammability.\cite{Yoshii2019, Matsumoto2006, Sakaebe2003} These attributes assure improved safety for batteries utilising ionic liquids. Matsumoto and co-workers were amongst the first to show exemplary performance of layered oxides such as $\rm LiCoO_2$ with the use of ionic liquids.\cite{Matsumoto2005, Matsumoto2006, Sakaebe2003} Honeycomb layered oxides, such as $\rm K_2Ni_2TeO_6$ and their cobalt-doped derivatives,\cite{Masese2018, Masese2019, Yoshii2019} have been shown to display stable performance at high-voltage operation in electrolytes comprising ionic liquids plausible candidates of which are depicted in \textbf{Fig. \ref{Fig_10}}.

\subsection{Alkali-ion kinetics and redox processes}

Fast kinetics of alkali ions within an electrode during electrochemical extraction/insertion is a crucial parametric that influences the rate performance of battery performances. For instance, previous reports have attributed the good rate performance and excellent cyclability of $\rm Na_3Ni_2SbO_6$ cathode material  (\textbf{Figs. \ref{Fig_10b}a}, \textbf{\ref{Fig_10b}b} and \textbf{\ref{Fig_10b}c}) to the fast interlayer kinetics of $\rm Na^+$ within the highly-ordered $\rm Na_3Ni_2SbO_6$.\cite{Yuan2014, Wang2018} Honeycomb layered oxides such as $\rm Na_3Ni_2SbO_6$ also exhibit preponderant rate capabilities (as shown in \textbf{Figs. \ref{Fig_10b}d} and \textbf{\ref{Fig_10b}e}) and can sustain fast $\rm Na$-ion kinetics upon successive Na-ion (de)insertion (\textbf{Fig. \ref{Fig_10b}f}).\cite{Bhange2017} Pertaining to structural stability, the manner in which 3$d$ transition metal atoms (for example $\rm Ni$ atoms in $\rm Na_3Ni_2SbO_6$) are arranged in a honeycomb configuration endows it not only with good thermal stability but also structural stability to sustain repeatable alkali atom extraction and insertion ($\rm Na$ atoms in this case).\cite{Wang2019b}

Besides the high-voltage and facile alkali-ion kinetics, this class of honeycomb layered oxides can accommodate ample amounts of alkali atoms depending on the choice of both $d^0$ (4$d$ or 5$d$) cations and 3$d$ transition metal atoms. The increase of the amount of alkali atoms accommodated within the interlayers of the honeycomb slabs implies an increase in the energy storage capacity, indicative of a high energy density. For instance, more of alkali atoms can be extracted from honeycomb layered oxide compositions such as $\rm {\it A}^{+}_3Ni^{2+}_2Sb^{5+}O_6$ ($\rm {\it A} = Li, Na$ and $\rm K$) or $\rm {\it A}^{+}_3Ni^{2+}_2Bi^{5+}O_6$ than in $\rm {\it A}^{+}_2Ni^{2+}_2Te^{6+}O_6$, despite the higher molar mass of $\rm {\it A}^{+}_3Ni^{2+}_2Bi^{5+}O_6$ compared to $\rm {\it A}^{+}_2Ni^{2+}_2Te^{6+}O_6$. The voltage-capacity plots of representative honeycomb layered oxides that can be utilised as cathode materials for rechargeable alkali-ion batteries are shown in \textbf{Fig. \ref{Fig_11}}.\cite{Sathiya2013, Grundish2019, Yang2017, Yuan2014, Masese2018, Bhange2017, Masese2019, Yoshii2019, Wang2019b, Zheng2016, Ma2015, Kee2016, Aguesse2016, Kee2020, Matsumoto2005, Matsumoto2006, Sakaebe2003, Wang2018, Bao2014, Kim2017, Huang2020, Wang2019a, Paharik2017, Dai2017, Han2016, Zhao2019, McCalla2015a} Note that the capacities of these oxides have been calculated based on the manifold oxidation states of the constituent transition states that can be allowed to facilitate maximum extraction of alkali atoms from the layered structures. \blue{In principle, the theoretical capacity ($Q$ (mAh g$^{-1}$)) of a material is determined by the molar mass ($M$ (g mol$^{-1}$)) and the number of electrons ($n$) involved during alkali-ion extraction (charging) or insertion (discharging), in accordance with the following equation\red{,
\begin{align}
    Q = \frac{n \times N \times e}{M} = \frac{n \times F}{3.6\,M} = 2.68 \times 10^4 \times \frac{n}{M},
\end{align}
}
where $N$ is the Avogadro constant ($6.02 \times 10^{23}$ mol$^{-1}$), $e$ is the elementary charge ($1.602 \times 10^{-19}$ C) and $F$ the Faraday constant ($96485.3$ C mol$^{-1}$).} It is apparent that these honeycomb layered oxides exhibit competitive energy storage capacities to justify them as high-energy-density contenders for rechargeable batteries.

Another point of emphasis is the nature of the redox process occurring within these honeycomb layered oxides. During the charge compensation redox process, the constituent $\rm Ni$ cations ({\it videlicet}., $\rm Ni^{2+}/Ni^{3+}$) are completely utilised \red{in} oxides such as $\rm {\it A}^{+}_2Ni^{2+}_2Te^{6+}O_6$ ($\rm {\it A} = Li, Na$ and $\rm K$) ensuring full electrochemical extraction of the alkali atoms. However, for oxides such as $\rm {\it A}^{+}_4Ni^{2+}Te^{6+}O_6$, it is impossible to fully extract all the alkali {\it A} atoms relying on the redox process of constituent Ni atoms ($\rm Ni^{2+}/Ni^{4+}$) alone.

\subsection{Anionic redox processes}

To fully tap the capacity (hence energy density) of such oxide compositions, the redox process of anions such as oxygen also have to be utilised, besides the redox process of 3$d$ transition metal cations. Formation of ligand holes, peroxo- or superoxo-like species are expected to occur in the oxygen orbital when anionic redox processes take place, and sometimes oxygen ($\rm O_2$) may be liberated leading to complete structural collapse; thus affecting the cyclability/performance durability of such oxides when used as battery materials.\cite{Grimaud2016}

Anionic redox processes provide a judicious route to utilise the full capacity of electrode materials and has been a subject that has attracted humongous interest in the battery community in recent years.\cite{Grimaud2016, laha2015oxygen, hong2019, yahia2019, assat2018fundamental, li2017anionic, xu2019review, li2020, foix2016, yabuuchi2016, qiao2018, mccalla2015visual, oishi2016, xie2017, watanabe2020, yamamoto2020, masese2015, oishi2015, oishi2013c, masese2020redox} Apart from facilitating an increase in the redox voltage of honeycomb layered oxides, the presence of $d^0$ cations (such as $\rm W^{6+}, Te^{6+}, Sb^{5+}$, {\it et cetera}.) also helps stabilise the anion-anion bonding that accompanies the oxygen redox chemistry. For example, the existence of highly valent $\rm W^{6+}$ (5$d^0$) cations in $\rm Li_4NiWO_6$ strongly stabilises the $\rm O$-$\rm O$ bonds, thereby averting the formation of gaseous $\rm O_2$ following anion oxidation.\cite{Taylor2019} Just like $\rm Li_4NiWO_6$, other honeycomb layered oxides such as $\rm Li_4FeSbO_6$ have also been found to manifest good oxygen-based redox reversibility, but it generates a large voltage hysteresis in the process.\cite{McCalla2015a, Jia2017} Further investigations on the oxygen-based redox reversibility are still ongoing in this field to uncover the factors underlying the large voltage hysteresis and determine ways to minimise it. What is emerging with these honeycomb layered oxides is that the presence of high-valency $d^0$ (4$d$ or 5$d$) is a crucial condition to produce not only high redox (and in some cases paradoxical) voltages, but also invoke oxygen redox chemistry aside from 3$d$ cationic redox processes. Moreover, the possibility to expand the materials platform of these honeycomb layered compounds through partial substitution with isovalent or even aliovalent 3$d$ transition metals, renders them as apposite model compounds to study numerous electrochemical aspects.

\newpage

\section{Topological phase transitions in honeycomb layered oxides}

Honeycomb layered oxides are susceptible to undergoing structural changes (phase transitions) upon electrochemical alkali-ion extraction. The presence of divalent transition metals ($\rm {\it M}^{2+}$) in the honeycomb slabs plays a major role in inducing these transitions during alkali-ion reinsertion process. In principle, when alkali atoms are electrochemically extracted, the valency state (oxidation state) of the transition metal atoms residing in the honeycomb slabs increase and {\it vice versa} during the reinsertion process; earlier defined as the charge-compensation process. Voids or vacancies created during alkali atom extraction leads to enhanced electrostatic repulsion between the metal atoms residing in different slabs; leading to an increase in the interslab distance.\cite{saadoune1996, croguennec2000, hironaka2017, wang2018vacancy}

\subsection{Structural changes as phase transitions}

Evolution of the structural changes upon alkali-ion extraction and reinsertion can readily be discerned using X-ray diffraction (XRD) analyses. During alkali-ion extraction, ($00z$) Bragg diffraction peaks that reflect the honeycomb interslab planes shift towards lower diffraction angles indicating the expansion of the interslab distance/spacing. The reverse process occurs during alkali-ion reinsertion, as has been exemplified in $\rm K_2Ni_2TeO_6$ upon potassium-ion extraction and reinsertion (as shown in \textbf{Fig. \ref{Fig_12}a)}.\cite{Masese2018} Apart from overall peak shifts observed during topotactic alkali-ion extraction and reinsertion (which technically manifests a single-phase (monophasic/solid-solution) behaviour), peak broadening or asymmetric peaks can be observed along with the disappearance of peaks and the emergence of new ones (reflecting a two-phase/biphasic behaviour).

\subsection{Stacking sequence changes as phase transitions}

The phase transition behaviour of honeycomb layered cathode oxides during alkali-ion extraction (charging) and reinsertion ({\it id est}, discharging), entails intricate structural changes that affects the coordination environment of alkali atoms. For instance, electrochemical sodium ($\rm Na$)-ion extraction from $\rm Na_3Ni_2BiO_6$ and $\rm Na_3Ni_2SbO_6$ during charging process leads to a sequential change in the bond coordination of Na, namely from the initial octahedral ($\rm O$) coordination to prismatic ($\rm P$) and finally to an octahedral ($\rm O$) coordination.\cite{Yuan2014, Liu2016, Bhange2017, Zheng2016, Wang2018} Further, the manner of stacking of repetitive honeycomb slabs per unit cell changes from $3$ to $1$. Thus, the phase transition of $\rm Na_3Ni_2BiO_6$ during charging process can be written in the following Hagenmuller\red{-Delmas'} notation\cite{Delmas1976, delmas1981} as previously described: $\rm O3\rightarrow P3 \rightarrow O1$ stacking mode. However, phase transitions can influence crucial electrochemical performance parametrics such as the rate capabilities of related oxides when used as battery materials. As such, crucial strategies have been sought to suppress the intricate phase transformation processes, for example, through doping or partial substitution of the transition metal atoms in the honeycomb slabs ({\it exempli gratia}, $\rm Na_3Ni_{1.5}{\it M}_{0.5}BiO_6$ (where $\rm {\it M} = Mg$, $\rm Zn$, $\rm Ni$, $\rm Cu$)) or even the alkali atoms in for instance $\rm Na_{1.6}Sr_{0.2}Ni_2TeO_6$.\cite{Gupta2013, Wang2017}

\begin{figure*}[!t]
\centering
  \includegraphics[height=18cm]{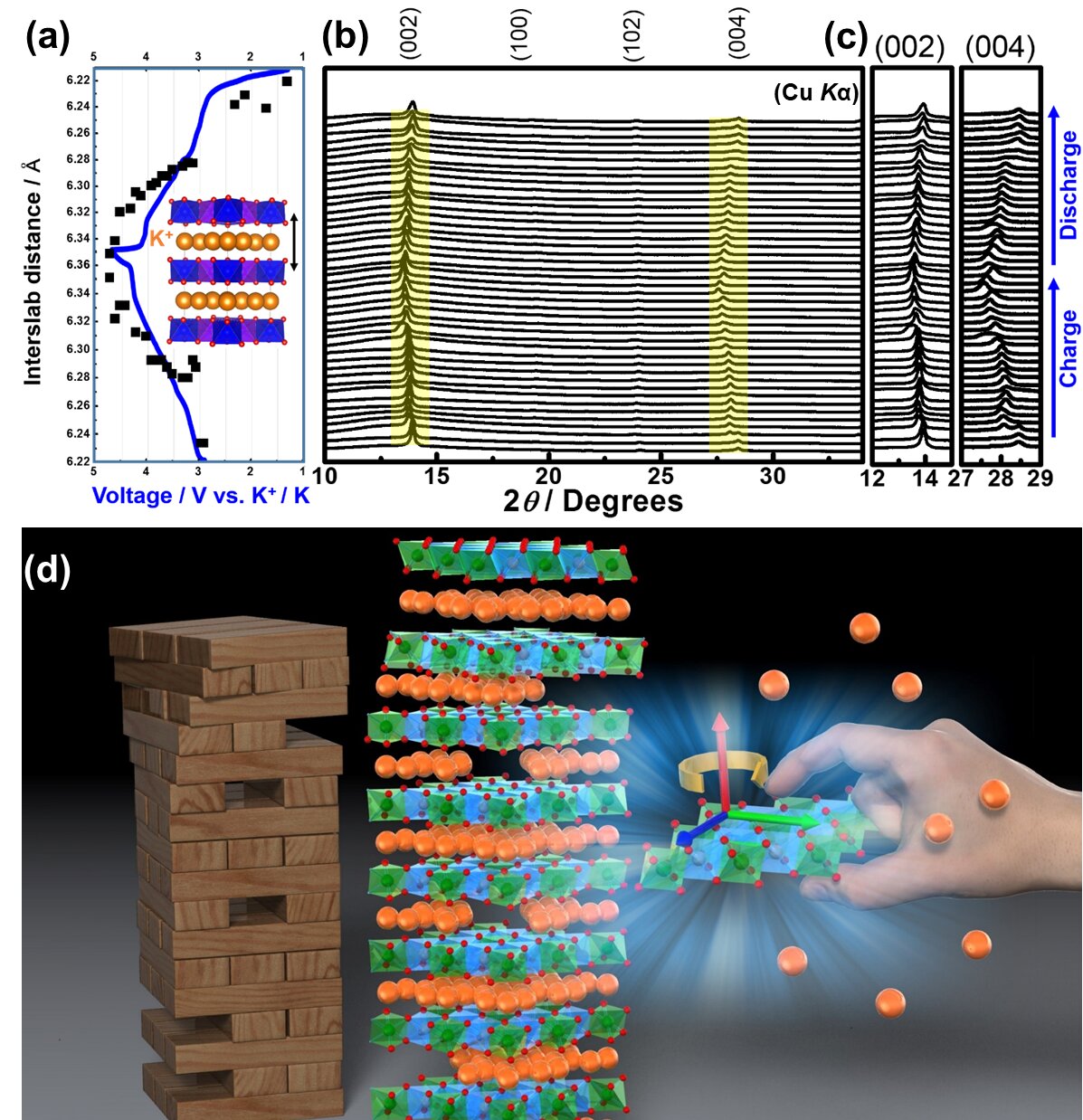}
  \caption{Phase transitions of honeycomb layered oxides. (a) Increase/decrease of the interslab distance ($\Delta z$) of honeycomb layered oxide $\rm K_2Ni_2TeO_6$ with charging (K$^+$ extraction)/ discharging (K$^+$ reinsertion). (b) Crystal structural evolution of $\rm K_2Ni_2TeO_6$ upon charging and discharging, showing the occurrence of intricate phase transition mechanism. (c)  Broadening and shifting of the (002) and (004) Bragg diffraction peaks that are sensitive to alkali-ion extraction/reinsertion during discharging/charging. (d) Rendition of the phase transition in these classes of layered oxides that entails complex phase transitions (mono- and bi-phasic, and amongst others), akin to a process of successively removing blocks from a complete `{\it Jenga} wooden blocks set' which can account for the Devil's staircase-like voltage profiles typically observed for honeycomb layered oxides.\cite{Yuan2014, Masese2018, Kee2016, Aguesse2016, Kee2020, Wang2018, Dai2017, Wang2017} (a-c) Reproduced from ref. \cite{Masese2018}
  under Creative Commons licence 4.0.}
  \label{Fig_12}
\end{figure*}

Multiple phase transformations observed in honeycomb layered oxides during alkali-ion extraction and reinsertion have a profound effect on their electrochemical characteristics such as rate performance and nature of the voltage profiles. These intricate phase transitions lead to the appearance of staircase-like voltage profiles as is often observed in the voltage-capacity profiles of most of the reported honeycomb layered cathode oxide materials.\cite{Yuan2014, Masese2018, Kee2016, Aguesse2016, Kee2020, Wang2018, Dai2017, Wang2017} Shifting of the honeycomb slabs during electrochemical alkali-ion extraction and reinsertion, or what is commonly termed as interslab gliding, has been rationalised to occur as the alkali atoms rearrange their occupying positions (alkali atom ordering). Such a complex phase transformation process can be envisioned through successively removing blocks from a complete `{\it Jenga} wooden blocks set', as shown in \textbf{Fig. \ref{Fig_12}b}. Assuming that the `blocks' are the `alkali atoms', removal of these wooden blocks will lead to rearrangement of the whole {\it Jenga} set to avoid structural collapse either by sliding (gliding) or rotation (shear) of the blocks (slabs). A mechanism akin to this {\it Jenga}-like mechanism, which is further discussed below, can account for the Devil's staircase-like voltage profiles typically observed for honeycomb layered oxides.\cite{Yang2017, Yuan2014, Masese2018, Gupta2013, Liu2016, Bhange2017, Zheng2016, Kee2016, Aguesse2016, Kee2020, Wang2018, Wang2017} 

\newpage

\subsection{Topological order and phase transitions in honeycomb layered oxides}

Phase transformation behaviour observed upon electrochemical alkali-ion extraction and reinsertion can spur enigmatic structural changes, like the aforementioned `{\it Jenga}-like' transitions. A comprehensive analysis of this mechanism calls for a deeper understanding based on a more comprehensive theory. Nonetheless, we hereafter highlight an approach based on heuristics founded on geometry, topology and electromagnetic considerations.\cite{Kanyolo2020} Readers may find it prudent to revise topics on tensor calculus, index notation, Einstein convention\cite{einstein1916} and other widely useful concepts in applied mathematics such as Gaussian curvature (Gauss-Bonnet theorem) in 2D,\cite{dubrovin2012, Allendoerfer1943} the Levi-Civita symbol and Chern-Simons theory.\cite{Allendoerfer1943, Dunne2007, Marino2005, Dunne1990, Witten2011, Alexander2009, Axelrod1993, Jackiw1990, Berry1984, schwarz2005, Alexander2009} Here, we use units where Planck's constant and the speed of the massless photon in the crystal are set to unity: $\hbar = c = 1$.

\subsubsection{An idealised model of topological phase transitions in honeycomb layered oxides}

\begin{figure*}[!b]
\centering
  \includegraphics[height=16cm]{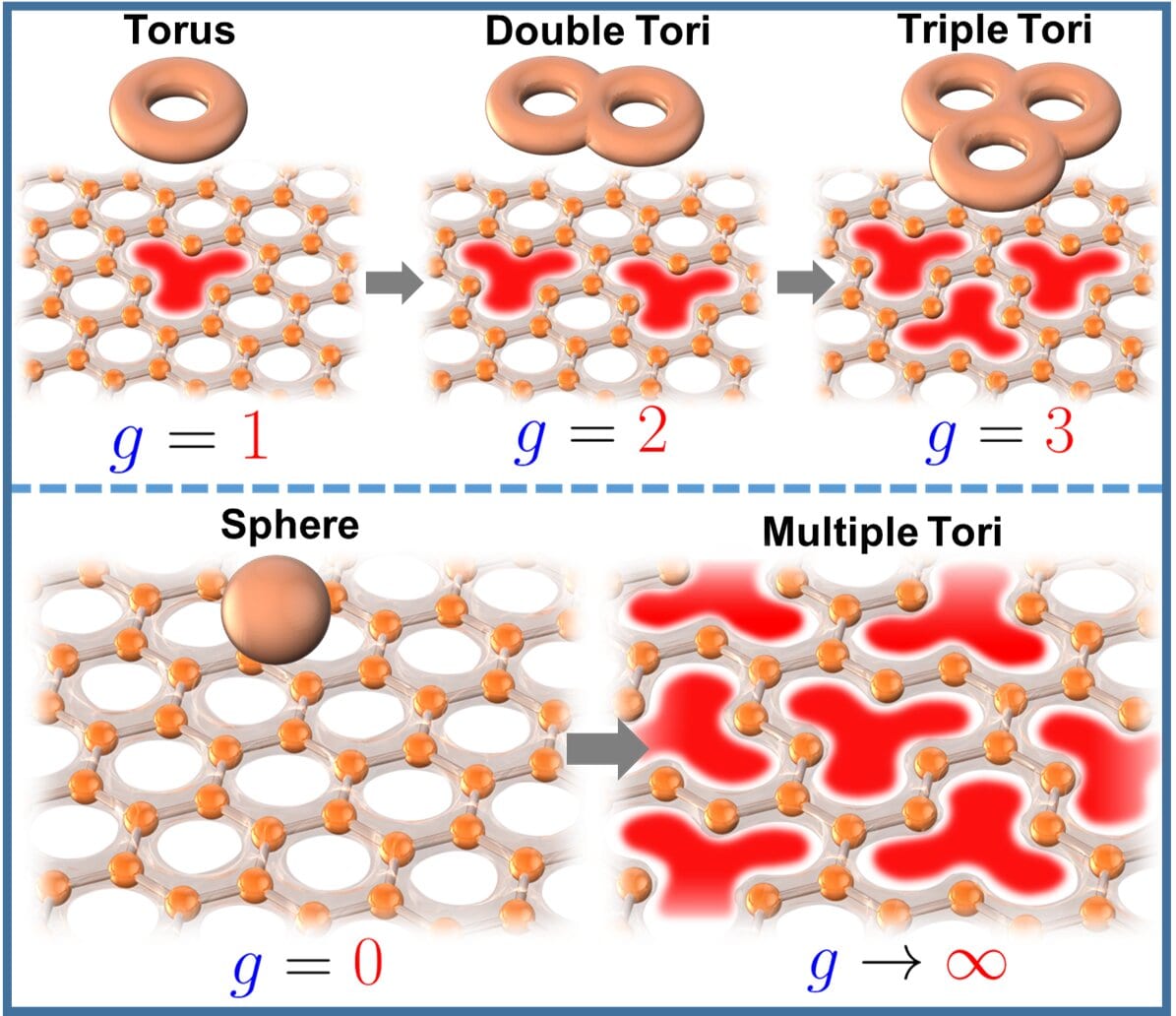}
  \caption{Atomic rearrangement triggered by extraction of alkali-atoms in honeycomb layered oxides such as $\rm K_2Ni_2TeO_6$, where $g \in integer$ is the number of alkali-atoms extracted by applying an external voltage in the $ab$ ($x$--$y$) plane.\cite{Kanyolo2020} The tori denote the various geometrical objects with holes denoted as $g$ (for genus). The tori can be mathematically mapped to the various configurations of the honeycomb lattice with ionic vacancies also denoted as $g$.}
  \label{Fig_13}
\end{figure*}

\begin{figure*}[!t]
\centering
  \includegraphics[height=9.5cm]{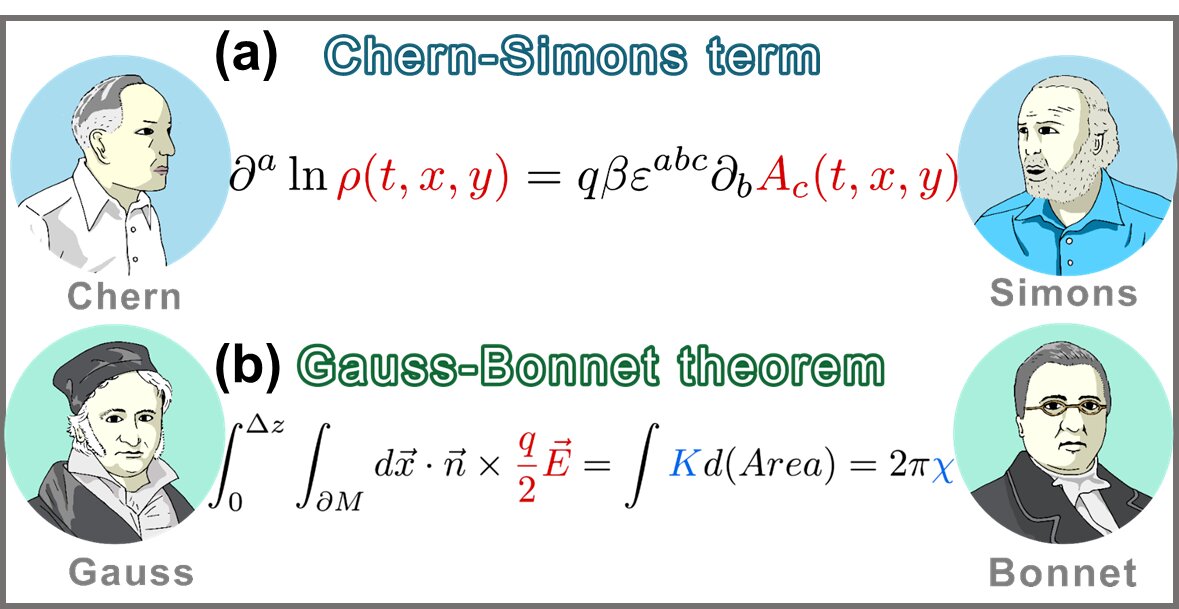}
  \caption{Topological transitions of honeycomb layered oxides.\cite{Kanyolo2020} (a) A two\red{-}dimensional (2D) field theory relating a Chern-Simons term\cite{Zee2010, Jackiw1990} to the ionic concentration $\rho(t,x,y)$ (charge density of the cations) where $q$ is the unit charge of a single cation, $\beta = 1/k_{\rm B}T$ is the inverse temperature and $A_{c} = (V, A_x, A_y)$ is the 2D electromagnetic potential (see \textbf{eq. (\ref{Chern_Simons_eq})}). (b) A Gauss-Bonnet theorem\cite{Allendoerfer1943, dubrovin2012} relating the applied electric-field $\vec{E} \equiv E_i = \varepsilon^{ibc}\partial_bA_c$ to the Euler characteristic $\chi = 2 - 2g$ and Gaussian curvature $K$ of the honeycomb surface $M$, where $\Delta z \sim \lambda_{c} = 2\pi/m$ is taken to be the order of the Compton wavelength of the cations and $\vec{n} = (0,0,1)$ is a vector normal to the honeycomb surface (see eq. \ref{Gauss_Bonnet_eq}). $\chi(M)$ is applied to estimate the transitions from the complete $g = 0$ honeycomb configuration to $g \in integer$ quasi-stable configurations such as the three-clover atomic arrangements depicted in \textbf{Fig. \ref{Fig_13}}.}
  \label{Fig_14}
\end{figure*}

Amongst some of the configuration of alkali atoms in a two-dimensional (2D) lattice of honeycomb layered oxides is shown in \textbf{Fig. \ref{Fig_13}}. Note that such a configuration has also been observed for some potassium atoms in $\rm K_2Ni_2TeO_6$ through XRD\cite{Masese2018} and electron microscopy studies (see \textbf{Fig. \ref{Fig_3_2}}). Potassium extraction (as is the case when a voltage is applied during charging process), for instance, leads to a non-sequential interslab distance increase; rendering the alkali cations to move in an undulating 2D surface (technically exhibiting a Gaussian curvature). Charge conservation in such a 2D undulating surface implies that the charge density vector $j_{a} = (\rho, j_{x}, j_{y})$, satisfies $\partial_{a}j^{a} = 0$ which has the solution $j^{a} = \varepsilon^{abc}\partial_{b}A_{c}$ (where $A_{c} = (V, A_x, A_y)$ is the 2D electromagnetic vector potential and $\varepsilon^{abc}$ is the totally anti-symmetric Levi-Civita symbol), hence leading to a Chern-Simons term, $\varepsilon^{abc}\partial_{b}A_{c}$.\cite{Jackiw1990, Zee2010} In turn, the honeycomb lattice introduces further constraints on the electrodynamics of these cations. In particular, since the cations (absent the applied voltage) form a 2D honeycomb lattice where the (free) alkali cations that contribute to the diffusion current $j^{a}$ are extracted from the 2D lattice by the potential energy $qV$ of the applied voltage, the total number of these free alkali cations ($g \in integer$) are related to the quasi-stable configurations displayed in \textbf{Fig. \ref{Fig_13}} that we shall refer to as 3 (leaf)-clover configurations. We shall consider each configuration as a $g$-torus where $g \in integer$ is the genus of an embedded 2D surface linked to the diffusion heuristics applied earlier in the review. Note that each $g$-torus supplies a unit charge $q$ of a single alkali cation, and thus determines the total charge density $\rho$ of the alkali cations which is related to the diffusion current $j_{i} = q\mu\rho E_{i} = \rho v_{i}$ along $ab$ plane of the honeycomb slabs. Consequently, these ideas can be summarised by a useful set of equations consistent which also contain the diffusion approach already tackled in the previous section of the review (also illustrated in \textbf{Fig. \ref{Fig_14}a} and \textbf{\ref{Fig_14}b}),
\begin{subequations}
\begin{align}\label{Chern_Simons_eq}
 \partial^{a}\ln\rho(t,x,y) = q\beta\varepsilon^{abc}\partial_{b}A_{c},\\
 \frac{q}{2m}\int_{\partial M}d\vec{x}\cdot(\vec{n}\times\vec{E}) = \int_{M}K\,d(Area) = 2\pi \chi = 2\pi (2 - g),
 \label{Gauss_Bonnet_eq}  
\end{align}
\end{subequations}
where $m$ is the mass of the cations, the interlayer (separation) distance, $\vec{n} = (0,0,1)$ is the normal vector to the $ab$ plane, $g$ is (approximately) the number of free cations, $\rho \propto \exp(-\beta E_{\rm a})$ is the ionic charge density with $E_{\rm a}$ the energy of the cations and $\vec{\nabla}\cdot\vec{E} = 8\pi \rho/q^2$, $\beta = 1/k_{\rm B}T$ is the inverse temperature, $K$ is the Gaussian curvature of a curved closed intrinsic surface $M$ and $\partial M$ is the boundary of $M$ representing the diffusion pathways of $g$ number of cations which form honeycomb lattice on $M$ displayed in \textbf{Fig. \ref{Fig_13}}. Thus, the integral equation is simply the well-known Gauss-Bonnet theorem.\cite{Allendoerfer1943}

In the special case of static equilibrium when the ionic density $\rho$ is strictly time-independent $\partial \rho/\partial t = 0$ and the electromagnetic vector potential is given by $A_{c} = (V(x,y), 0, 0)$, the Chern-Simons term reduces to $\partial^{a}\ln \rho(t,x,y) = q\beta \varepsilon^{abc}\partial_{b}A_{c} \rightarrow \partial_{i}\ln \rho(x,y) = q\beta \varepsilon_{ij}\partial_{j}V(x,y)$  which yields $\rho \vec{v} = q\mu\rho \equiv \sigma\vec{E}$ with the ansatz $E_{\rm a}(x,y) = \int_{\partial M} d\vec{x}\cdot(\vec{n}\times\mu^{-1}\vec{v})$, where $\varepsilon_{ij}$ is the 2D Levi-Civita symbol and $\mu = D/k_{\rm B}T$ is the mobility of the cations. Hence, the energy evaluated over a closed loop $E_{\rm a}\left ( \oint_{\partial M} \right ) = m(g - 1)$ over the honeycomb surface conveniently counts the missing mass of cations within the loop, as shown in \textbf{Fig. \ref{Fig_13}}. Equivalently, this corresponds to the (activation) energy $E_{\rm a} = (g - 1)mc^2$ needed to render the cations mobile, where $c = 1$ is the speed of the massless photon in the crystal. This means that the quasi-stable configuration with $g = 1$ requires no activation energy to create and can be considered as a ground state of the system. However, since the other configurations are shifted by a constant energy $E_{\rm a} = mg$ from this ground state, the system contains an additional $g - 1$ number of stable configurations.

\subsubsection{Topological order in honeycomb layered oxides}

On the other hand, according to \textbf{eq. (\ref{Chern_Simons_eq})}, the ionic density is time-dependent, $\partial \rho/\partial t \neq 0$, when a magnetic field $B_z = \partial_xA_y - \partial_yA_x$ is present. Since $g \in integer$ corresponds to the aforementioned 3-(leaf) clover configurations on the honeycomb lattice, magnetic fields drive the system out of one configuration to the next via extraction of cations from the honeycomb lattice. We shall refer to this mechanism of adiabatic extraction of the alkali cations from the honeycomb surface accompanied by introduction of time-varying electromagnetic fields as {\it Jenga} mechanism, in analogy with the game of the same name. 

Similar to the total collapse of the pieces in {\it Jenga} at the end of the game, this process of extraction of alkali cations and subsequent restabilisation cannot continue indefinitely since \textbf{eq. (\ref{Chern_Simons_eq})} and \textbf{eq. (\ref{Gauss_Bonnet_eq})} remain valid only around equilibrium and the conditions of adiabatic perturbations around equilibria $g$ values. Whence, the transformation of the complete honeycomb structure into a predominantly $3$-clover configuration should induce a phase transition. One possible approach to a theoretical treatment of such transitions is to apply the Berezinskii-Kosterlitz-Thouless model\cite{Kosterlitz1973} of phase transitions to the magnetic fields (or fluxes) introduced during this dynamical {\it Jenga} phase. Another approach is to consider the phase transitions that may be triggered by sound waves in the crystal arising from rapid (non-adiabatic) extractions of the cations from the honeycomb surface. Geometrically, this entails periodically time-varying Gaussian (curvature) metric analogous to gravitational waves in the space-time geometry. When quantised, these sound waves are phonons that can mediate a weak attractive force between the positively charged fermionic cations (forming Cooper-pairs) and hence may lead to superconductivity.\cite{Cooper1956, Bardeen1973} In contrast, an idealised approach to the dynamics of the cations has been proposed in ref. \cite{Kanyolo2020}, 
where bosonic cations form a Bose-Einstein condensate\cite{Gross1963, Pitaevskii1961} below the critical temperature and their dynamics are consistent with \textbf{eq. (\ref{Chern_Simons_eq})} and \textbf{eq. (\ref{Gauss_Bonnet_eq})}. Above the critical temperature, unpaired charged vortices appear representing diffusion channels of the cations under small curvature perturbations around $g \simeq 1$. Of course, further research of the physics of the {\it Jenga} mechanism including other non-adiabatic phenomena testing the validity or failures of \textbf{eq. (\ref{Chern_Simons_eq})} and \textbf{eq. (\ref{Gauss_Bonnet_eq})} will certainly be the focus of frontier research in the coming years.

\newpage

\section{High-precision measurement of diffusion and magneto-spin properties
}

In the previous section(s), we discussed the physics and electro-chemistry of the diffusion of cations within the honeycomb layers. However, we neglected their magneto-spin interactions with the inter-layers which in turn can substantially affect their mobility and hence, their solid-state alkali-ion diffusion properties. This approximation is valid since the alkali cations ({\it exempli gratia} $\rm K$) are known to generally possess an inherently weak nuclear magnetic moment which barely interacts with the octahedra ({\it exempli gratia} $\rm TeO_6$) in the inter-layers. In particular, the diffusion dynamics of the cations is resilient to local magneto-spin interactions in the honeycomb layers since the weak magnetic fields originating from the large number of cations in the honeycomb layers tend to randomise and average out according to central limit theorem.\cite{Billingsley1995} This means that even though the Gaussian average (mean) of the magnetic fields vanishes, $\langle B_{z}(t) \rangle = 0$, the mean-square $\langle B_{z}(t)B_{z}(0)\rangle \neq 0$ need not vanish. Hence, the diffusion and magneto-spin properties of the cations are encoded in the mean of the random magnetic fields in the honeycomb layers. However, measuring these properties by applying the Gaussian average over magnetic quantities is an intricate task that often proves elusive to undertake due to a scarcity of effective techniques.

\subsection{Considering effects of alkali-ion diffusion on muon spin-polarisation}


In 2D, the Langevin equation given in \textbf{eq. (\ref{Langevin-Fick_eq}b)} is replaced by,
\begin{subequations}
\begin{align}\label{Langevin_2D_eq}
    \frac{d\vec{p}}{dt} = -\frac{1}{\mu}(\vec{n}\times\vec{v}) + q(\vec{n}\times\vec{E}) + q(\vec{n}\cdot\vec{B})\vec{n},
\end{align}
which together with \textbf{eq. (\ref{Chern_Simons_eq})} form the Langevin-Fick framework of equations (analogous to \textbf{eq. (\ref{Langevin-Fick_eq})}).\cite{Kanyolo2020} Notice that since the magnetic field $\vec{B} \propto \vec{\eta}$ is proportional to the Langevin force, its mean-square is given by the fluctuation-dissipation theorem, \cite{Weber1956} $\langle B_{z}(t)B_{z}(0) \rangle = 2k_{\rm B}T\mu q^2f(t) \neq 0$ with $f(t)$ a function of time. Consequently, the mobility $\mu$ (related to the diffusion coefficient by the Einstein-Smoluchowski relation $\mu = \beta D$) can be determined from the mean-square of the local magnetic fields in the honeycomb layers through the (dynamic) Kubo-Toyabe (KT) function\cite{kubo1967, hayano1979, yamazaki1979muon} given by,
\begin{align}\label{Kubo_Toyabe_eq}
    P_{z}(t) = \frac{1}{3} + \frac{2}{3}\left (1 - \Delta_{v}^2t^2\right )\exp(-\frac{1}{2}\Delta_{\nu}^2t^2),
\end{align}
\end{subequations}
where $P_{z}(t)$ is the spin-polarisation of the particle and $\Delta_{\nu}^2 = \gamma^2\langle B_{z}^2(0)\rangle$ is the decay rate of the particle with $\gamma$ its gyromagnetic ratio. The KT function effectively describes the time evolution of a spin-polarised particle in zero magnetic field with a non-vanishing mean-square. This singles out particles (in the standard model of particle physics) with a strong gyromagnetic moment as ideal for probing such weak magneto-spin and diffusion properties since their spin-polarisation will precess according to the KT function. Notably, muon spin rotation, resonance and relaxation (abbreviated as $\mu^{+}$SR) is a potent measurement technique that avails this univocal information pertaining to alkali-ion diffusion properties of materials to electrochemists and material scientists.\cite{Sugiyama2009a, Mansson2013, Sugiyama2011, Sugiyama2012a, Sugiyama2012c}


\subsection{Rationale and methodology behind applying muon spin rotation and relaxation measurement techniques 
}

At this juncture, it is imperative to explain the rationale for the use of muons in analysis of diffusion and magneto-spin properties of materials. Muons stand out from other members of the lepton family of elementary particles mainly owing to the following reasons:
\begin{itemize}
  \item Muons are abundant and are indeed a product of cosmic radiation (recall the {\it Aurora Borealis} and {\it Aurora Australis}). Muons can also be artificially produced using spallation sources such as ISIS Neutron and Muon Source (UK), Japan Proton Accelerator Research Complex (JPARC), TRIUMF (Canada) and Paul Scherrer Institute (PSI, Switzerland);
  \item The spin configuration of muons are traceable (technically, muons are 100\% spin-polarised since they are produced via the decay of a (positive) pion at rest into a positron and an electron neutrino via the weak interaction, which violates parity),\cite{Lee1956, blundell1999, muon_bible2011} implying that they are easy to detect via their decay products unlike other members of the lepton elementary particles. This aspect endows $\mu^{+}$SR measurements with an upper edge over other resonance techniques such as nuclear magnetic resonance (NMR).  In addition, muons possess a high gyromagnetic ratio ($\gamma_{\mu} = 135.5$ MHz/T) meaning that they are very sensitive to weak magnetic fields;
  \item Unlike electrons, muons have a finite lifetime that is appreciable; thus, $\mu^{+}$SR offers a unique measurement time window that complements conventional techniques such as NMR and neutron diffraction.
\end{itemize}

\begin{figure*}[!b]
\centering
  \includegraphics[height=9cm]{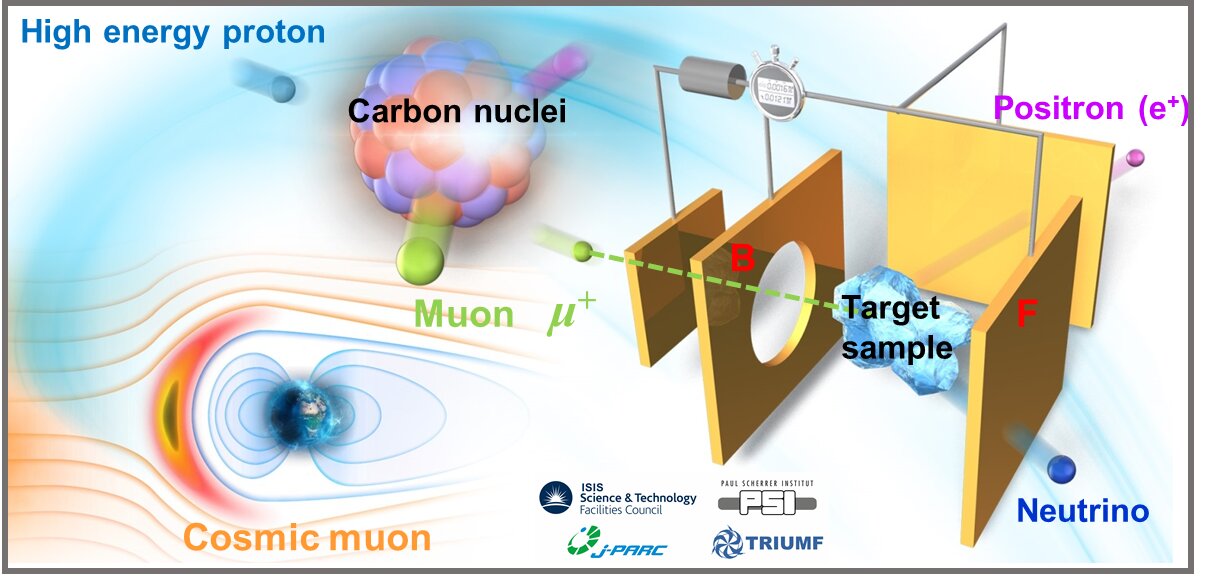}
  \caption{
  Working principle of (anti-)muon spin relaxation ($\mu^{+}$SR) as a potent tool for investigating the diffusive and magnetic properties of target materials. The (anti-)muon is produced when a high energy proton beam is directed onto carbon nuclei which produce (positive) pions. The (positive) pion decays into an (anti-)muon and a muon-neutrino, which subsequently decays to a positron, an anti-muon neutrino and an electron-neutrino which escape the sample. The difference in the positron counts in the forward (F) and backward (B) detectors normalised by the total count, the asymmetry function $A(t)$, gives the spin relaxation of the (anti-)muon in the sample.}
  \label{Fig_15}
\end{figure*}

Detection of alkali-ion diffusion by muons first entails the embedding of muons into a sample (or muon implantation), the sample in this case being the layered oxide material. These muons are artificially produced via the bombardment of high energy protons onto a carbon \red{(graphite)} or beryllium target, as is schematically shown in \textbf{Fig. \ref{Fig_15}}. The muons (in this case, positive muons (anti-muons)) are then focused using a collimator to the sample where they bind with oxygen ions ($\rm O^{2-}$) to form stable $\mu^{+}$--$\rm \,\,O^{2-}$ bonds. The implantation of muons into a honeycomb layered oxide is illustrated in \textbf{Fig. \ref{Fig_16}a}, where muons typically reside at the vicinity of oxygen ions at distances in the ranges of $1 \sim 1.2$ \AA.\cite{Mansson2013} The implanted muons are initially static and are able to sense the local nuclear magnetic field in the layered oxide when it is in a paramagnetic state, a behaviour that can mathematically be expressed using a static Kubo-Toyabe (KT) function,\cite{kubo1967, hayano1979, yamazaki1979muon} as is also shown in \textbf{Fig. \ref{Fig_16}b}. When alkali-ion diffusion occurs, the local nuclear field haphazardly fluctuates and the implanted muons acquire a dynamic contribution in the KT function through the hopping rate of the cations; thus are able to sense the local field that is randomly fluctuating at an average rate. The mobility of alkali cations can be increased by temperature beyond a certain critical temperature $T_{\rm c}$ where the alkali cations become mobile, thus inducing an additional fluctuation in the local mean-square magnetic field leading to a conspicuous increase of the fluctuation (collision) rate $\nu_0 \rightarrow \nu(T)$, where the mean-square magnetic field is given by $\langle B_{Z}(t)B_{z}(0) \rangle = \langle B_{z}^2(0) \rangle \exp(-\nu t)$. Consequently, the self-diffusion coefficient $D_{\rm self} = \sum_{i = 1}^{n}\frac{1}{N_{i}}Z_{i}s_{i}^2\nu_{0}$ related to the diffusion coefficient, $D(T)$ by a Boltzmann factor $D(T) = D_{\rm self}\exp(-E_{\rm a}/k_{\rm B}T) = \sum_{i = 1}^{n}\frac{1}{N_{i}}Z_{i}s_{i}^2\nu(T)$ is accurately determined using the $\mu^{+}$SR measurements by considering the collisions as a Markov process\cite{behrends2000, Billingsley1995} over $n$ paths of the cations in the 2D honeycomb lattice where $N_i$ is the number of cation sites, $Z_i$ the vacancy fraction and $s_i$ the length of the mean-free path between collisions.\cite{borg2012}


\subsection{Muon spin rotation and relaxation diffusion coefficient measurement results}

\begin{figure*}[!b]
\centering
  \includegraphics[height=8.2cm]{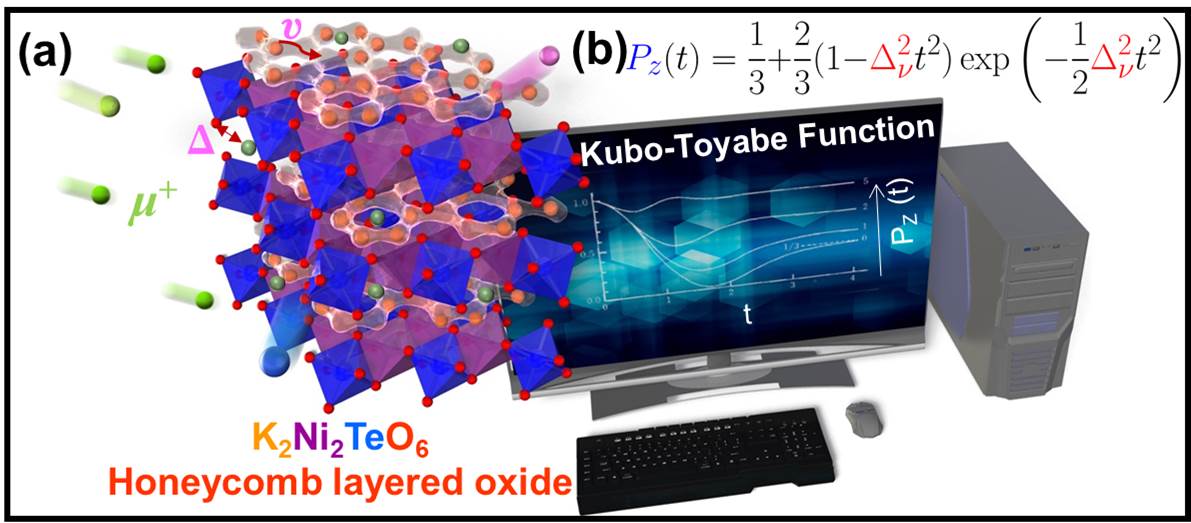}
  \caption{High-precision measurement of magneto-spin and diffusion properties of honeycomb layered oxides relevant to phase transitions. (a) The anti-muon implantation into a honeycomb layered oxide framework with a stoichiometric composition of, for instance, $\rm K_2Ni_2TeO_6$. The anti-muon is expected to bind onto the oxygen ions located in the octahedral ($\rm TeO_6$ and $\rm NiO_6$) structures of the material altering the typical decay rate  of the (anti-)muon. The hopping rate, $\nu$ of the diffusing potassium ($\rm K$) cations along the honeycomb depends on their interaction with the anti-muons through their random nuclear magnetic fields which alters the anti-muon decay rate $\Delta_{\nu}$. (b) The analysis of alkali-ion diffusion using the dynamical Kubo-Toyabe function,\cite{kubo1967, hayano1979, yamazaki1979muon} $P_z(t)$ which describes the relaxation of muon spin polarisation in the presence of a particular (typically Gaussian) distribution of nuclear magnetic fields of the cations in the honeycomb layered oxide material. The total asymmetry function in $\mu^+$SR experiments depends on the Kubo-Toyabe function, which depends on the decay rate of the anti-muons due to transport properties of the cations such as their hopping rate in the material. The hopping rate in turn determines the self-diffusion coefficient of the material.}
  \label{Fig_16}
\end{figure*}

\begin{figure*}[!t]
\centering
  \includegraphics[height=6cm]{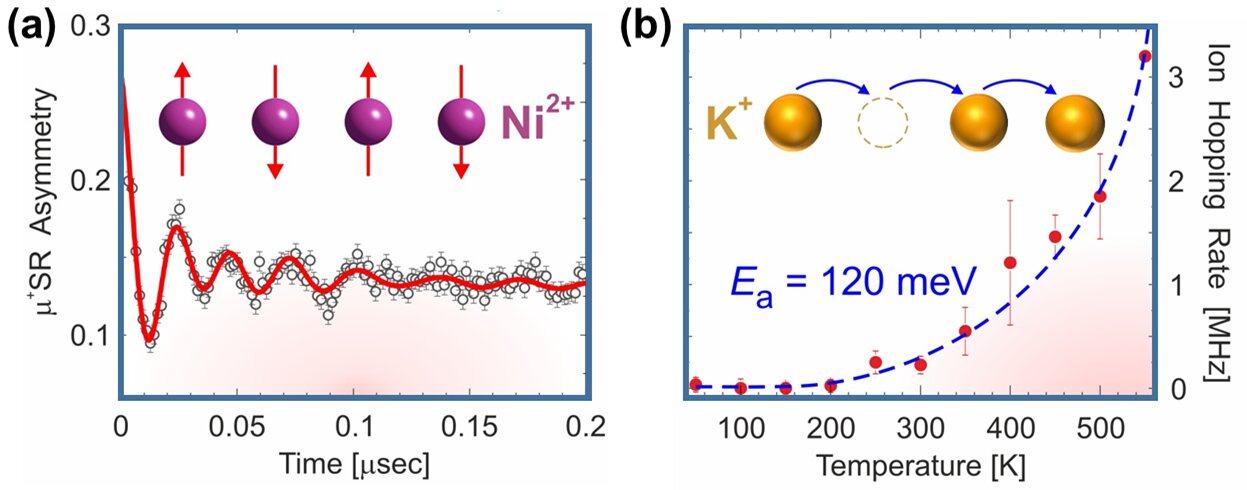}
  \caption{High-precision measurement of diffusion and magneto-spin properties of honeycomb layered oxides relevant to phase transitions. (a) Presence of an antiferromagnetic spin ordering in $\rm K_2Ni_2TeO_6$ below $26$ K revealed by a clear oscillation in the $\mu^+$SR time spectra. (b) The onset and evolution of K-ion diffusion revealed by an exponential increase in field fluctuation rate ($=$ ion hopping rate) from which the activation energy ($E_{\rm a}$) of the diffusion process can be determined.\cite{Matsubara2020}}
  \label{Fig_17}
\end{figure*}

Sugiyama, M{\aa}nsson and co-workers have pioneered the use of $\mu^{+}$SR measurements in the study of both the magneto-spin and alkali-ion diffusive properties in a wide swath of layered materials such as  Li$M \rm O_2$ (where $M = \rm Ni$ and $\rm Co$), $\rm LiCrO_2$, $\rm LiNi_{1/3}Co_{1/3}Mn_{1/3}O_2$,  $\rm Li_2MnO_3$ and even to $\rm NaCoO_2$.\cite{Sugiyama2009c, Sugiyama2009a, Sugiyama2009b, Sugiyama2010, Sugiyama2010a, Sugiyama2010b, Sugiyama2011, Sugiyama2012a, Sugiyama2012b, Sugiyama2012c, Sugiyama2013a, Sugiyama2013b, Sugiyama2014, Sugiyama2015a, Sugiyama2015b, Sugiyama2015c, Sugiyama2018, Sugiyama2018b, Sugiyama2020, Ikedo2010, Mukai2010, Ohta2010, Umegaki2019, Forslund2019, Sugiyama2018, Sassa2018, Forslund2018, Nozaki2018, Umegaki2018, Mansson2018, Forslund2017, Nozaki2018a, Umegaki2017, Mukai2013a, Mukai2013b, Nozaki2014, Mansson2014, Ikedo2010a, Mansson2013, Ofer2010, Chow2012, Ofer2012, Wikberg2012} Investigation of potassium-ion ($\rm K^{+}$) dynamics in related layered materials is particularly unwieldy, due to the innately weak nuclear magnetic moment of potassium relative to other ions such as lithium ($\rm Li$) and sodium ($\rm Na$). This renders it difficult to capture the dynamics of $\rm K^+$ in layered materials using standard techniques such as NMR spectroscopy. As discussed above, the fact that spin-polarised muons possess a strong gyromagnetic moment, makes $\mu^{+}$SR measurements particularly ideal for capturing the dynamics of cations such as the $\rm K^{+}$ with an extremely weak nuclear magnetic moment in materials. For clarity, the nuclear magnetic moment/gyromagnetic ratio of $\rm K$ ($\mu[^{39}\rm K] = 0.39\,\, \mu$N, $12.50$ MHz/T) is much smaller than for $\rm Li$ ($\mu[^{7}\rm Li] = 3.26\,\,\mu$N, $108.98$ MHz/T) and $\rm Na$ ($\mu[^{23}\rm Na] = 2.22\,\, \mu$N, $70.81$ MHz/T). The $\mu^{+}$SR asymmetry function time spectrum of honeycomb layered oxide $\rm K_2Ni_2TeO_6$ (or equivalently as $\rm K_{2/3}Ni_{2/3}Te_{1/3}O_2$) measured below the antiferromagnetic transition temperature ($26$ K) as shown in \textbf{Fig. \ref{Fig_16}a}, is shown in \textbf{Fig. \ref{Fig_17}a}, where precession of the muon (spin-polarisation) occurs. It is evident that the muon precesses due to the emergence of a spontaneous internal magnetic field, resulting in a clear oscillation of the time spectrum. This is a response that is typically observed from a muon ensemble in a magnetically ordered state (in this case, antiferromagnetic Ni spin ordering in $\rm K_2Ni_2TeO_6$).\cite{Matsubara2020}

The dependency plot of the fluctuation rate, which is dynamically related to the hopping rate of $\rm K^+$ with temperature, $\nu(T)$ is shown in \textbf{Fig. \ref{Fig_17}b}. Between $250$ K and $500$ K, this fluctuation rate increases with temperature signifying the onset of diffusive motion of $\rm K^{+}$ in $\rm K_2Ni_2TeO_6$. The hopping rate nicely obeys a trend akin to Arrhenius equation $\nu(T) = \nu_{0}\exp(-E_{\rm a}^{\rm K}/k_{\rm B}T)$ from where an activation energy commensurate to approximately $E_{\rm a}^{\rm K} \simeq 120$ meV is obtained.\cite{Matsubara2020} The diffusion coefficient can be calculated using the above hopping rate assumptions to yield a diffusion coefficient value of $D^{\rm K}(T) = 1.2\times 10^{-10} \rm cm^2/s$, which is an order of magnitude lower than that of layered materials such as $\rm LiCoO_2$.\cite{park2010review, Sugiyama2009a}

Caution needs to be taken when interpreting the $\mu^{+}$SR measurement data, as muons {\it per se} can also be mobile in inactive materials. $\rm K_2Ni_2TeO_6$ indeed shows reversible $\rm K^{+}$ extraction and insertion behaviour (thus, electrochemically active) at room temperature; thus, the onset of $\rm K^{+}$ diffusive motion that arises at $T > 250$ K is irrefutable. The feasibility of utilising $\mu^{+}$SR measurements to further unveil the intricacies of the dynamics of such cations as $\rm K^{+}$, which tend to possess low nuclear magnetic moments, will expand the pedagogical scope of cationic intercalation (insertion) and deintercalation (extraction) dynamics within honeycomb layered oxides and other layered materials.

\newpage

\section{Summary and future challenges for honeycomb layered oxides}

This review provides an elaborate account of the exceptional chemistry and the physics that make honeycomb layered oxides a fledgling class of compounds. We explore the prospects that would result \blue{in} myriads of compositions expected to be reported in the coming decades. \blue{Majority of the honeycomb layered oxides reported typically engender $\rm Li^+$ or $\rm Na^+$ as resident cations. However, the further adoption of honeycomb layered oxides with large-radii alkali ions, such as $\rm K^+$,$\rm Rb^+$ and $\rm Cs^+$ or even coinage metal ions such as $\rm Ag^+$, $\rm H^+$, $\rm Au^+$, $\rm Cu^+$, {\it et cetera}, is expected to further increase the compositional space of related compounds, unlocking manifold functionalities amongst this class of materials. In fact, preliminary theoretical computations have affirmed the feasibility of preparing honeycomb layered oxides encompassing cations such as $\rm Rb^+$, $\rm Cs^+$, $\rm Ag^+$, $\rm H^+$, $\rm Au^+$, $\rm Cu^+$, {\it et cetera} to adopting, for instance, a chemical composition of $\rm {\it A}_2Ni_2TeO_6$, where $\rm {\it A} = Rb$, $\rm Cs$, $\rm Ag^+$, {\it et cetera}. By the same token, synthesis of honeycomb layered materials that encompass alkaline-earth metals such as $\rm {\it A} = Ba$, $\rm Sr,$ {\it et cetera} has also been proposed as another avenue of augmenting the various combinations of these materials. Indeed, studies pertaining compositions such as  $\rm {\it A}Ru_2O_6$, where $\rm {\it A} = Ba$, $\rm Sr,$ {\it et cetera} have recently been reported.\cite {pearce2020} Undoubtedly, this class of materials offers an extensive platform worthy of pursuit in the coming years (\textbf{Fig. \ref{Fig_3_0}a}).}  

\blue{To predict or interpret possible emergent features of honeycomb layered oxides, it is crucial to understand their unique structural frameworks and the local and bulk atomistic changes they undergo. A combination of crystallography techniques that include transmission electron microscopy (TEM), neutron diffraction and X-ray diffraction are expected to offer a holistic view of the arrangement of atoms within the honeycomb lattice and the global order of atoms within the honeycomb structure of the new materials. Moreover high-resolution TEM at low temperatures, as can be availed by cryogenic microscopy, is a possible route to discern the arrangement of transition metal atoms in the honeycomb lattice at low temperatures where transitions tend to occur.\cite{DeYoreo2016} Experimental and theoretical reports on the structures of these materials has already revealed some stacking disorders with honeycomb layered species such as $\rm Na_3Ni_2SbO_6$, $\rm Na_2Zn_2TeO_6$ and $\rm Na_2Ni_2TeO_6$, associating them with emergent functionalities such as phase transitions, magnetic ground states and ionic diffusion.\cite{Xiao2020, Bianchini2019, Li2019b, Kurbakov2020} Although defects have been known to have both prolific and detrimental effects on honeycomb layered oxides, they still remain vastly underexplored. Nonetheless, {\it de novo} computational and experimental techniques are expected to uncover new defect physics and chemistry that will expand their uses into multiple fields.}

On another front, doping offers a prospective route to availing more possibilities with a broader scope of chemical compositions that display improved electrochemical and additional magnetic properties. 
\blue{From an electrochemical perspective}, doping with non-magnetic atoms such as magnesium or strontium generally reinforces the crystalline structure by suppressing electrochemically\red{-}driven phase transitions, whilst increasing the thermodynamic entropy of the materials. Increased entropy has added advantages that include raising the working voltage as well as facilitating multiple redox electrochemistry during battery operations as elucidated in \textbf{Fig. \ref{Fig_9}} and \textbf{Fig. \ref{Fig_11}}. Relating to ionic conductivity, partial doping of the transition metal atoms in the honeycomb slab with aliovalent or isovalent atoms is a pertinent strategy to increase the ionic conductivity of honeycomb layered oxides. For instance, partial substitution of $\rm Zn^{2+}$ with $\rm Ga^{3+}$ in $ \rm Na_2Zn_2TeO_6$ solid-state electrolyte (with a wide voltage tolerance) aids to increase the $\rm Na^+$-ion mobility (conductivity) due to increased formation of $\rm Na^+$ vacancies.\cite{Li2018Chemistry} Theoretical investigations done by Sau and co-workers have accentuated the profound effect of transition metal substitution in $\rm Na_2{\it M}_2TeO_6$ ({\it M} = Ni, Zn, Co, Cu). \cite {sau2016ion1, sau2015ion1, sau2016ion} In contrast, doping with magnetic atoms, as shown in \textbf{Fig. \ref{Fig_5}}, reveals fascinating magnetic behaviour that places honeycomb layered oxides among\red{st} the exotic quantum materials.

\begin{figure*}[!b]
\centering
  \includegraphics[height=13.5cm]{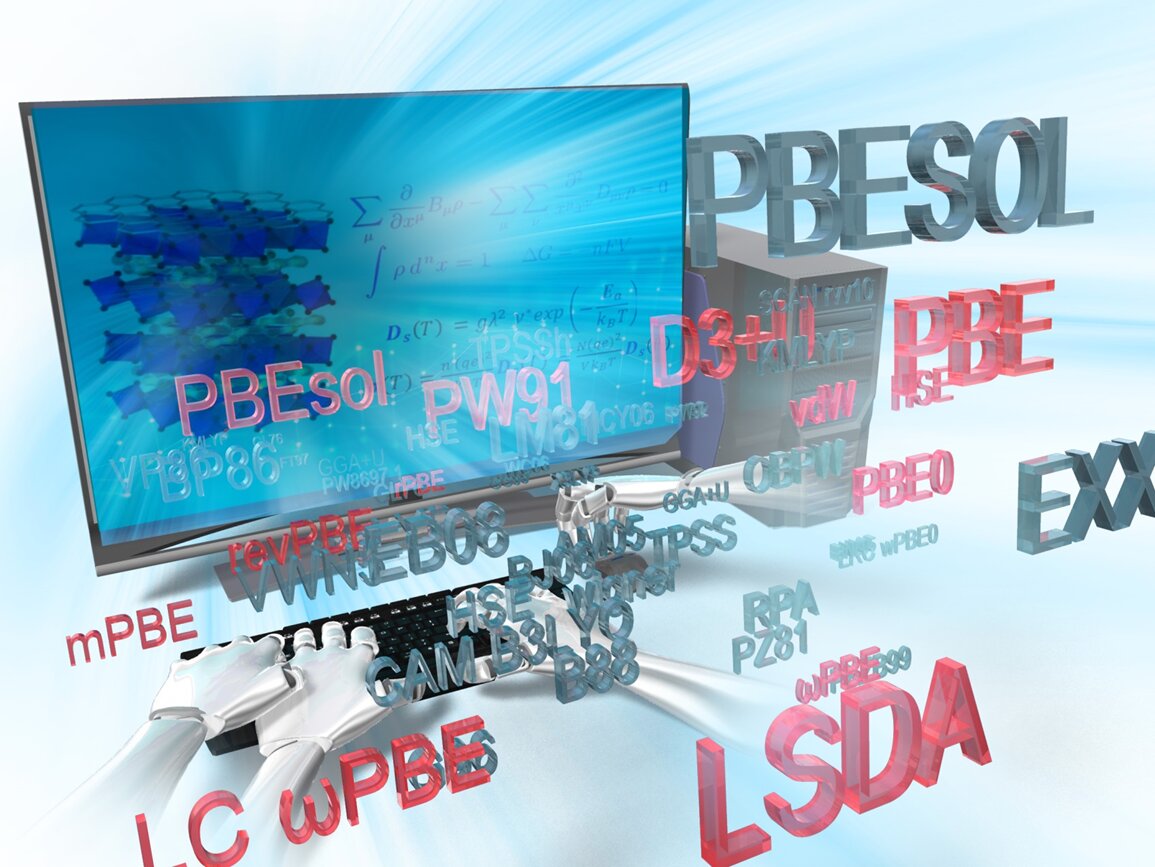}
  \caption{Computational design techniques\cite{gonze2002, setyawan2010, jain2011, hautier2012} that typically can be applied to simulate various physicochemical properties of honeycomb layered oxide frameworks. The schematic illustrates the potential of using these computational techniques for the designing of new chemical compositions of honeycomb layered oxide materials.}
  \label{Fig_18}
\end{figure*}

Moreover, topochemical reactions, for instance, chemical ion-exchange of $\rm Rb$ or silver ($\rm Ag$) with potassium ($\rm K$) in $\rm K_2Ni_2TeO_6$ can aid build an entire host of new oxide materials with a wide swath of physicochemical properties. Indeed, such a design strategy has proven effective in the synthesis of $\rm Ag_3Ni_2BiO_6$, for instance, via topochemical ion-exchange of $\rm Li_3Ni_2BiO_6$.\cite{Berthelot2012} Additionally, the introduction of alkali cations with differing ionic radii makes the tuning of the distance between the honeycomb layers (interslab/interlayer distance) possible; thus presenting avenues to tune the interlayer magnetic couplings as discussed in \textbf{Fig. \ref{Fig_8}}. This guarantees the feasibility to not only adjust the electrochemical properties but also to tweak the physicochemical aspects such as the magnetic dimensionality of the honeycomb lattice. This calls for further exploratory synthesis to be augmented with computational protocols, as schematically adumbrated in \textbf{Fig. \ref{Fig_18}} to expedite the design of new honeycomb layered oxide compositions.

There has been significant progress in the physics entailing topological states, for which honeycomb layered oxides play a pivotal role in advancing this topical field. In this review, we have discussed the Kitaev and Haldane magnetic (spin) interactions within the honeycomb lattice that offer a path to the experimental realisation of Kitaev quantum spin liquid and a quantum anomalous Hall insulator (Chern insulator) respectively.\cite{Kitaev2006, zhou2017, zhong2020, liu2018, sano2018, xu2020, Regnault2011, Thonhauser2006} In addition, higher-order magnetic interactions induced by the angle between the spins of the magnetic cations, introduces other interactions: mainly, the Heisenberg and asymmetric / Dzyaloshinskii-Moriya (DM) interactions.\cite{Moriya1960, Dzyaloshinsky1958} Due to the additional angular space-time dependent degree of freedom, these interactions are considered of higher order and thus very elusive to realise without the presence of, for instance, single-crystals of target honeycomb layered oxide materials. Irradiating circularly-polarised oscillating electric fields on preferably single crystals within a Floquet model (theory) is a plausible route to realising DM interactions within honeycomb layered oxide materials, as has been suggested by several authors.\cite{Owerre2016, Owerre2017, Owerre2019} The primary significance of these interactions is the evaluation of magnetic skyrmions\cite{felser2013, finocchio2016, fert2017, nagaosa2013, fert2013} - quasi-particles that have been predicted to exist in certain magnetic condensed matter systems such in magnetic thin films either as dynamic excitations or stable/metastable configurations of spin; which shows great promise in topological quantum computing applications.\cite{Nayak2008, Haldane1988, Plekhanov2017, Wright2013}

\begin{figure*}[!b]
\centering
  \includegraphics[height=13.5cm]{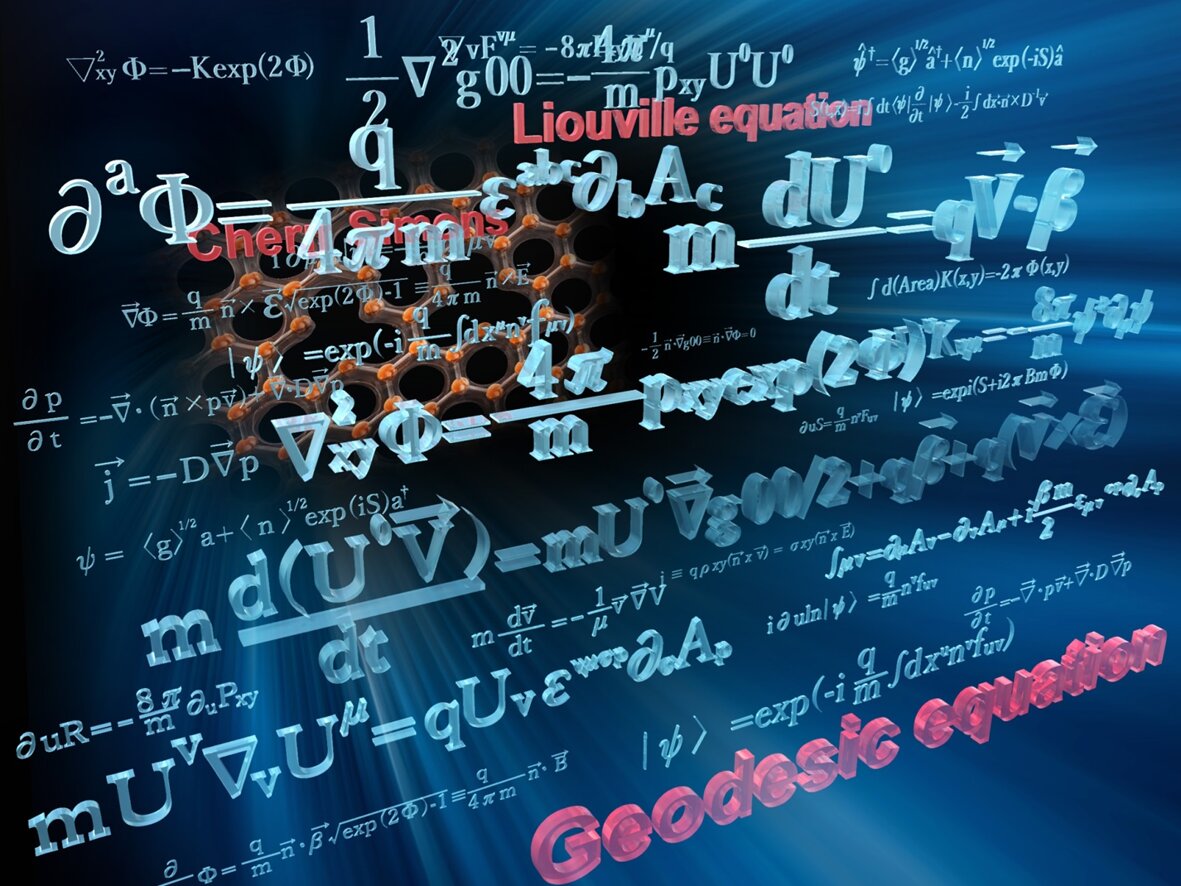}
  \caption{Schematic of the rich science anticipated in honeycomb layered oxide materials relating to topology, curvature, geodesics, Chern-Simons theory, Liouville's equation, Brownian motion and Bose-Einstein condensates.\cite{Kanyolo2020, Zee2010, Jackiw1990, dubrovin2012, Gross1963, Pitaevskii1961}}
  \label{Fig_19}
\end{figure*}

Regarding single-crystal growth, the high thermal stability of honeycomb layered oxides, such as tellurates, bismuthates and antimonates, makes them suitable for crystallisation at high temperatures conducive for their preparation. In fact, the possibility of growing single crystals in honeycomb layered oxides (such as $\rm Na_2Ni_2TeO_6$, $\rm Na_2Cu_2TeO_6$, $\rm Na_3Cu_2SbO_6$ and $\rm Na_2Co_2TeO_6$) using high-temperature solid-state reactions has already been achieved.\cite{Sankar2014, Xiao2019, Yao2020} Another fascinating pursuit will be the design of thin films from honeycomb layered oxide materials, either using molecular beam epitaxy (MBE), atomic laser deposition (ALD) or pulsed laser deposition (PLD), which will aid to accurately visualise the presence of magnetic skyrmions or any emergent topological physics that covers, \textit{inter alia}, superconductivity, magneto-resistance and ferro-electricity. Moreover, an extension of the $\mu^+$SR technique, namely low energy $\mu^+$SR (LEM) offers the unique possibility to tune the muon implantation depth into samples from 5-500 nm. Thus, synthesising thin film battery cells, \textit{exempli gratia} from honeycomb layered oxide materials, allows one to gain unique access to depth-resolved studies of the dynamics of the cations at and across the buried interfaces, \textit{isto es} the solid-state cathode / electrolyte / anode, respectively.\cite{morenzoni2004, sugiyama2015d}

A plethora of unprecedented amazing phenomena may also be found when honeycomb layered oxides are subjected under high-pressure (stress) conditions. This has, amongst other things, the effect of making higher-order interactions finite and thus non-negligible. In particular, exerting pressure perpendicular to the honeycomb slabs bring into play 3D interactions that may have been otherwise negligible. Experimentally, $\rm Na_2Cu_2TeO_6$ shows new bond coordination (dimerisation of $\rm Cu$ bonds) at high pressure, leading to a change in the magnetic properties technically referred to as magnetic phase transitions.\cite{Koo2008, gazizova2018} Generally, high pressure exerted in these layered oxides can introduce defects or microstructure evolutions that may show great potential for novel functional materials. Although the global topology of honeycomb layered materials is robust against local defects, whenever these defects are related to topological invariants ({\it exempli gratia} Berry's phase,\cite{Cohen2019, Kanyolo2019, liu2018berry, ao1993berry, loss1990berry, tomita1986, zak1989berry, samuel1988, Chang1996, Aronov1993, Xiao2010} they will affect global properties of the material such as phase transitions, as exemplified in \textbf{Fig. \ref{Fig_12}}, \textbf{Fig. \ref{Fig_13}} and \textbf{Fig. \ref{Fig_14}}. Phase transition phenomena inherent in these classes of honeycomb layered oxide materials will certainly necessitate the use of spectroscopic techniques such as muon spin relaxation ($\mu^{+}$SR), as well as computational and theoretical techniques as displayed in \textbf{Fig. \ref{Fig_18}} and \textbf{Fig. \ref{Fig_19}} respectively, to discern the nature of the spin interactions innate at high-pressure regimes. Moreover, resolution at the atomic-scale of related functional materials when subjected to ultra-high pressure will attract tremendous research interests in the coming years. 

\begin{figure*}[!b]
\centering
  \includegraphics[height=13.8cm]{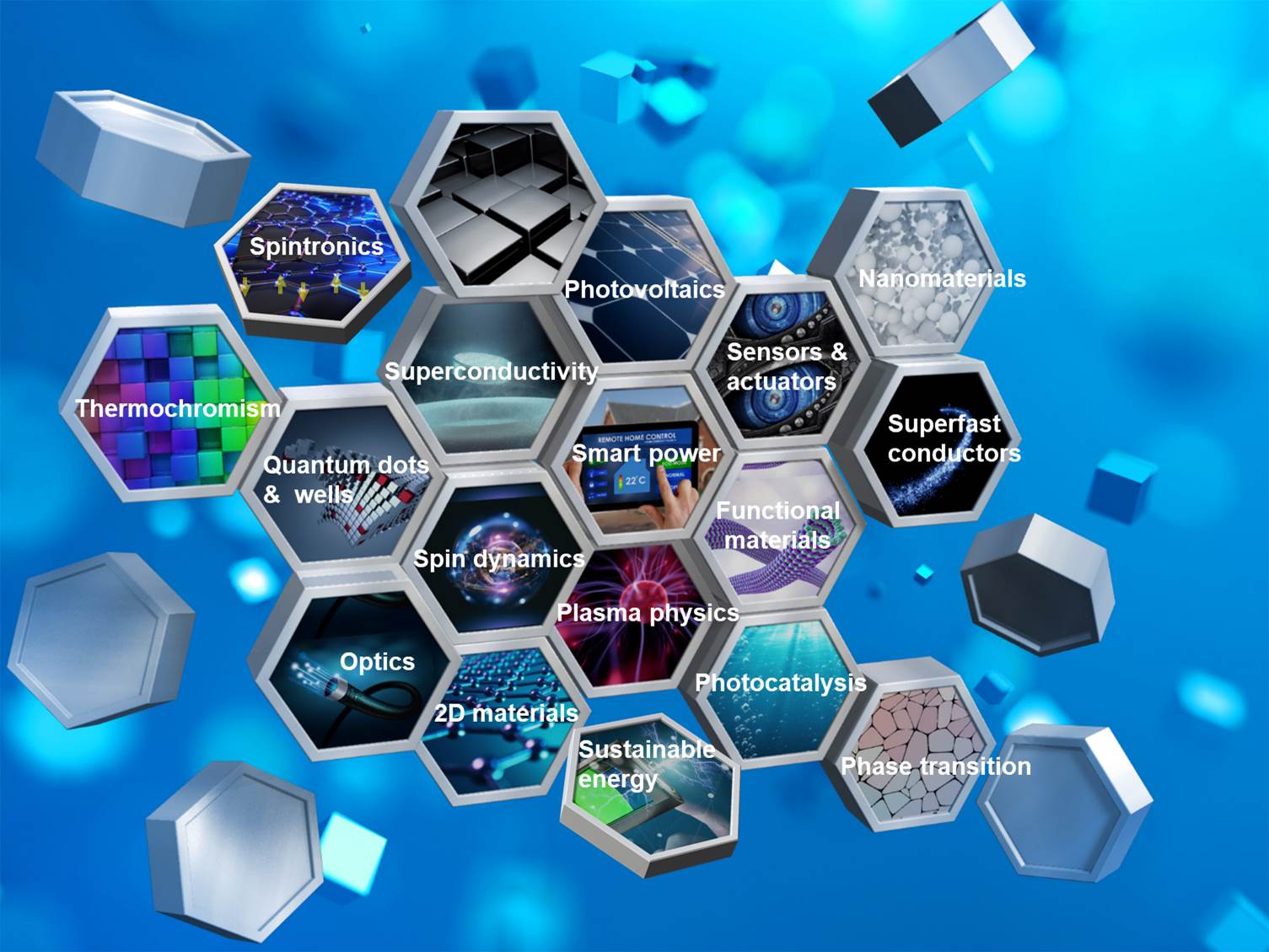}
  \caption{Diversity of honeycomb structures in various realms of science and technology. The schematic highlights the future perspectives of honeycomb layered oxides that can be envisioned such as superconductivity, phase transitions, photocalysis, thermochromism, spin dynamics, photovoltaics and optics.}
  \label{Fig_20}
\end{figure*}

The heightened interest in oxide materials based on honeycomb layers is expected to spearhead the design of a new generation of materials that promise to make remarkable contributions in the fields of energy, electronic devices, catalysis, and will ultimately benefit the scientific community in a broad swath of fields in the coming decades, as can be envisaged in \textbf{Fig. \ref{Fig_20}}. Recent reports are also emerging on honeycomb layered oxides as photocatalysts, optical materials, superfast ionic conductors, and so forth.\red{\cite{Kadari2016, Pu2018, Roudebush2015, Bera2019, Bera2020, Li2018ACSApplMater, Wu2018, Deng2019, Wu2020, Dubey2020, pei2019, li2020c, shi2020}} A grand challenge with most of these materials lies in their handling. Particularly for honeycomb layered oxides comprising alkali ions with large radii such as potassium and rubidium, handling demands the presence of a controlled atmosphere ({\it videlicet}, storage in argon-purged glove boxes) as they are sensitive to moisture (hygroscopic) and air. Future work should also focus on the improvement of the stability of related honeycomb layered oxides, for instance, when exposed to air; to enable handling and mass production of these materials in ambient conditions. Their instability can be contained and controlled, for example, by tuning their chemical composition. Partial substitution of the constituent transition metal atoms is a possible route, as has been noted when $\rm Na_3Ni_2SbO_6$ is partially substituted even with a minuscule amount with $\rm Mg$, $\rm Mn$ or even $\rm Ru$.\cite{Kee2016, Kee2020, Xiao2020a} \blue{On another front, the use of multiple transition metals in equivalent amounts has also been presented as a new route for the design of stable layered oxides (often referred to as `high-entropy oxides') with unique physicochemical properties.\cite {sarkar2019, sarkar2018, zhao2020} Although this concept presents new possibilities for the design and application of honeycomb layered oxides, it is still in infancy with a lot of growth potential.}  

\blue{In brief},
partial substitution also induces a change in the \red{phase} transitions observed when alkali cations are electrochemically extracted, as is the case when they are used as battery materials. Hygroscopicity presents another avenue for tuning the interslab distance and editing electrochemical profiles in some materials bringing forth several advantages such as superconductive phase transitions, as has been noted in layered $\rm NaCoO_2$ when hydrated.\cite{Takada2003} 

Honeycomb layered oxides \red{(particularly for compositions incorporating pnictogen or chalcogen atoms that have been delved in this review)} can serve as high-voltage cathode materials for rechargeable batteries, as summarised in \textbf{Fig. \ref{Fig_10}}, exhibiting theoretically high capacities. A challenge is their safe and stable operation at high-voltage regimes; warranting the adoption of stable electrolytes that can tolerate high-voltage battery operation. Ionic liquids, which consist of organic or inorganic anions and organic cations, manifest a plenitude of desirable properties. Paramount amongst them is their low flammability, good chemical stability and excellent thermal stability.\cite{MacFarlane2016, Matsumoto2019} In particular, the inherently large voltage tolerance makes ionic liquids propitious when matched to high-voltage layered cathodes during battery operation. Stable performance of high-voltage layered cathode materials using piperidinium-based ionic liquids has been shown;\cite{Matsumoto2005, Sakaebe2003} likewise, assessment of high-voltage honeycomb layered oxides using stable electrolytes (such as ionic liquids) is a plausible route for harnessing their high electrochemical performance. A schematic list of the choice of stable electrolytes, especially ionic liquids, for honeycomb layered oxide cathode materials is furnished in \textbf{Fig. \ref{Fig_9}}. On another note, exotic redox chemistry can be manifested in honeycomb layered oxides. For instance, $\rm Li_4FeSbO_6$ is currently amongst model materials to study oxygen anion redox chemistry; a topical area in battery research nowadays.\cite{McCalla2015a, Grimaud2016, Taylor2019, Jia2017} Much room still exists in the search for related honeycomb layered oxides.

\red{At this juncture it begs the question: \textit{quo  vadis}, honeycomb layered oxides?} The rich electrochemical, magnetic, electronic, topological and catalytic properties generally innate in layered materials, indubitably present a conducive springboard to break new ground of unchartered quantum phenomena and the coexistent electronic behaviour in two-dimensional (2D) systems. It is our expectation that this will unlock unimaginable applications in the frontier fields of computing, quantum materials and internet-of-things (IoT). 

Finally, the vexing question of why magnetic atoms in the slabs of these layered oxides conveniently align in a honeycomb architecture, to our knowledge, remains to be addressed; an attestation that the landscape of honeycomb layered oxide materials still remains broad and uncharted, moving forward into this new age of avant-garde innovation. An eminent mathematician has elegantly posited a solution to why bees prefer the honeycomb architecture in what now is emerging as `the Honeycomb conjecture'.\cite{Hales2001} Presumably, it is through a review of the materials found in nature that we can glean insights for future design in this universe of honeycomb layered oxide materials.

\newpage

\section*{Acknowledgements}
This work was conducted under the auspices of the National Institute of Advanced Industrial Science Technology (AIST), Japan Society for the Promotion of Science (JSPS KAKENHI Grant Number 19 K15685) and Japan Prize Foundation. Part of this research was also supported by the European Commission through a Marie Sk\l{}odowska-Curie Action and the Swedish Research Council - VR (Dnr. 2014-6426, 2016-06955 and 2017-05078), the Carl Tryggers Foundation for Scientific Research as well as the Swedish Foundation for Strategic Research (SSF) within the Swedish national graduate school in neutron scattering (SwedNess). The authors would also like to thank Dr Minami Kato, Dr Kohei Tada, Dr Ola Kenji Forslund, Dr Elisabetta Nocerino, Dr Konstantinos Papadopoulos, Mr Anton Zubayer and Dr Keigo Kubota for fruitful discussions on this manuscript. We acknowledge that this review paper was proofread and edited through the support provided by Edfluent services. \red{T. M. gratefully acknowledges Natsumi Ishii and Rei Ishii for the unwavering support in conducting this work.}

\newpage

\section*{Conflicts of interest}
There are no conflicts to declare.





\newpage

\bibliography{rsc} 
\bibliographystyle{rsc} 

\end{document}